\documentclass[%
reprint,
superscriptaddress,
nofootinbib,
amsmath,amssymb,
aps,
]{revtex4-1}

\usepackage{graphicx}
\usepackage{grffile}
\usepackage{dcolumn}
\usepackage{bm}
\usepackage{epsfig}
\usepackage{mathrsfs}  
\usepackage{subfigure}
\usepackage{multirow}
\usepackage{epstopdf}
\usepackage{amsmath}
\usepackage{algorithmicx}
\usepackage{amssymb}
\usepackage{tensor}
\usepackage[math]{cellspace}
\usepackage{bookmark}
\usepackage[usenames,dvipsnames]{xcolor}
\usepackage{hyperref}
\usepackage{slashed}
\usepackage{youngtab}
\usepackage{physics}
\usepackage{tensor}
\usepackage[capitalise]{cleveref}

\newcommand{\MPl}{M_{\text{Pl}}}

\newcommand{\autoCheckMissing}[1]{#1}

\crefformat{equation}{Eq.~(#2#1#3)}
\crefformat{table}{Tab.~(#2#1#3)}
\crefformat{figure}{Fig.~(#2#1#3)}
\crefformat{appendix}{App.~(#2#1#3)}
\crefformat{section}{Sec.~(#2#1#3)}
\crefformat{subsection}{Subsec.~(#2#1#3)}

\makeatletter
\renewcommand*\env@matrix[1][\arraystretch]{%
  \edef\arraystretch{#1}%
  \hskip -\arraycolsep
  \let\@ifnextchar\new@ifnextchar
  \array{*\c@MaxMatrixCols c}}
\makeatother

\allowdisplaybreaks
\begin{document}

\title{Nonlinear Evolution of Quadratic Gravity in 3+1 Dimensions}

\author{Aaron Held}
\email{aaron.held@uni-jena.de}
\affiliation{
Theoretisch-Physikalisches Institut, Friedrich-Schiller-Universit\"at Jena,
Max-Wien-Platz 1, 07743 Jena, Germany
}
\affiliation{
The Princeton Gravity Initiative, Jadwin Hall, Princeton University,
Princeton, New Jersey 08544, U.S.
}

\author{Hyun Lim}
\email{hyunlim@lanl.gov}
\thanks{\\Both authors contributed equally. The names are listed alphabetically.}
\affiliation{
Computational Physics and  Methods (CCS-2), 
Los Alamos National Laboratory, 
Los Alamos, NM  87545 USA
}
\affiliation{Center for Theoretical Astrophysics,
Los Alamos National Laboratory,
Los Alamos, NM 87545 USA 
}

\begin{abstract}
We present a numerically stable system of (3+1) evolution equations for the nonlinear gravitational dynamics of quadratic-curvature corrections to General Relativity (Quadratic Gravity). We also report on the numerical implementation of these evolution equations. We recover a well-known linear instability and gather evidence that -- aside from said instability -- Quadratic Gravity exhibits a physically stable Ricci-flat subsector. In particular, we demonstrate that Teukolsky-wave perturbations of a Schwarzschild black hole as well as a full binary inspiral (evolved up to merger) remain Ricci flat throughout evolution. This suggests that, at least in vacuum, classical Quadratic Gravity can mimic General Relativity, even in the fully nonlinear strong-gravity regime.
\end{abstract}

\maketitle

\section{Motivation}
\label{sec:motivation}

The dynamics of General Relativity (GR) is governed by terms at linear order in (Riemann) curvature. As we gain access to the strong gravity regime~\cite{LIGOScientific:2016aoc, EventHorizonTelescope:2019dse}, we probe potential new physics which becomes relevant at higher order in curvature. Such new physics is suggested by the cosmological riddles of dark matter~\cite{Bertone:2004pz, Barack:2018yly} and dark energy~\cite{Peebles:2002gy}. Moreover, GR predicts its own breakdown, as singularity theorems~\cite{Penrose:1964wq, Hawking:1965mf, Geroch:1966ur, Hawking:1969sw} imply geodesic incompleteness in the interior of black holes. In the context of new physics at strong curvature, such a breakdown is not surprising: Close to the formation of a singularity, curvature scales grow (arbitrarily) large, hence, potential higher-order curvature corrections are no longer negligible, and the dynamics of GR needs to be modified to account for these corrections. 
If curvature corrections are present, the respective new-physics scale may occur anywhere between the largest currently accessible curvature scales and the Planck scale.
\\

Taking a step beyond GR, we focus on dynamics at quadratic order in curvature.
Such quadratic-curvature corrections are widely expected to arise from quantum fluctuations, see~\cite{tHooft:1974toh,Stelle:1976gc,Goroff:1985sz,Avramidi:1985ki,vandeVen:1991gw,Ohta:2018sze} for perturbative quantum gravity,
\cite{Loll:2019rdj}~for lattice approaches to quantum gravity,
\cite{Ashtekar:1988sw}~for loop-quantum gravity,
\cite{Boulware:1985wk, Zwiebach:1985uq}~for string theory,
and~\cite{Benedetti:2009rx, Ohta:2015zwa, Knorr:2021slg, Baldazzi:2021orb}~for
asymptotically safe gravity.

Quadratic curvature corrections occur in the form of gravitational self-interactions~\cite{Stelle:1976gc, Stelle:1977ry} and in the form of non-minimal couplings of curvature to other fields~\cite{Kanti:1995vq,Alexander:2009tp}. Both sectors can be unified in the context of an effective field theory of gravity and matter, see, e.g.,~\cite{Ruhdorfer:2019qmk}.

Field redefinitions can mix between the pure-gravity and the non-minimal sector and, moreover, between different orders in curvature, see, e.g.,~\cite{Burgess:2003jk,Endlich:2017tqa}. Several different terms may thus be physically equivalent if the field redefinitions do not impact physical conclusions.

In the following, we will focus on gravitational self-interactions, we will not perform field redefinitions, and we will neglect any potential non-minimal couplings of curvature to other fields. We abbreviate the respective theory as Quadratic Gravity (QG) -- sometimes also called Stelle-gravity~\cite{Stelle:1976gc, Stelle:1977ry}.
\\

General Relativity tends to hide singularities, and thus regions of diverging curvature, behind horizons~\cite{Penrose:1969pc,Wald:1997wa}, see~\cite{Choptuik:1992jv,Gundlach:2007gc} for the potential exception of critical collapse. 
Experimental probes of horizon-scale physics~\cite{LIGOScientific:2016aoc,EventHorizonTelescope:2019dse} thus provide the most promising way to constrain potential new physics at large curvature.
Here, we are motivated, in particular, by the rapidly growing catalog of gravitational-wave events~\cite{LIGOScientific:2018mvr,LIGOScientific:2020ibl,LIGOScientific:2021djp}. 

Utilizing said data to constrain new physics~\cite{LIGOScientific:2021sio} will eventually require predictions for gravitational wave forms in theories beyond GR. The key tool to predict the respective nonlinear dynamics close to merger is well-posed numerical evolution, see~\cite{Shibata:1995we,Baumgarte:1998te,Pretorius2005,Pretorius:2005gq} for pioneering work in numerical relativity and~\cite{Sarbach2012,Isenberg2014} for reviews of the well-posed initial value problem in GR.

Beyond GR, numerical evolution in the presence of non-minimally coupled scalar degrees of freedom has received much attention~\cite{Witek2019,Okounkova2019,Okounkova2020,Witek2020,Ripley2020a,Ripley2020b,Okounkova:2020rqw,East2021,Silva:2020omi,Figueras:2020dzx} and (for a specified set of theories) well-posedness has been established at weak non-minimal coupling~\cite{Kovacs:2020pns,Kovacs:2020ywu}. See also~\cite{Cayuso:2020lca, Cayuso:2023aht} for evolution including pure-gravity operators at quartic order~\cite{Endlich:2017tqa} and by means of damped high-frequency modes~\cite{Cayuso:2017iqc}.

In previous work~\cite{Held2021}, we verified stable numerical evolution in the spherically-symmetric sector of Quadratic Gravity.
Here, we report on an extension of the evolution equations to (3+1) dimensions, following, in particular, the pioneering work of Noakes~\cite{Noakes:1983}, see also~\cite{Morales:2018imi}.

In \cref{sec:QG}, we start by reviewing QG, its equations of motion, and the propagating degrees of freedom. In~\cref{sec:3+1derivation}, we perform a $(3+1)$ decomposition and derive our key analytical result: a set of $1^\text{st}$-order evolution equations. In~\cref{sec:numerical}, we describe our specific numerical implementation and verify numerical stability. In~\cref{sec:results}, we present first physical results which suggest that QG exhibits a nonlinearly stable Ricci-flat subsector which is fully equivalent to GR. In~\cref{sec:conclusion}, we conclude with a discussion and an outlook on future work. Several technical details are relegated into appendices.

We use the $(-,+,+,+)$ signature and use Latin letters as spacetime indices. Moreover, we work in Planck units, i.e., setting the speed of light $c=1$. For clarity, we keep Newton's constant $G$ explicit.
Round (square) brackets denote full \mbox{(anti-)symmetrization} of the enclosed indices.

\section{Setup: Quadratic Gravity}
\label{sec:QG}

The action of Quadratic Gravity (QG) is given by
\begin{align}
\label{eq:action}
    S_\text{QG} = \int_x
    \left[
        \mathcal{L}_\text{mat}[\Phi]
        +\frac{1}{16\pi G}R
        +\alpha R_{ab}R^{ab}
        -\beta R^2
    \right],
\end{align}
where $\int_x$ is shorthand notation for $\int d^4x \sqrt{\text{det}(-g)}$.
In the following, the first term is taken to be independent of the curvature and depends solely on minimally coupled matter fields (and on the cosmological constant). The matter fields are collectively denoted by $\Phi$. The second term is linear in the curvature and corresponds to GR, parameterized by Newton's constant $G=1/(8\pi \MPl)$ (or, equivalently, by the Planck mass $\MPl$). The third and fourth term are quadratic in the curvature and are parameterized by couplings $\alpha$ and $\beta$. In four dimensions, $\alpha$ and $\beta$ are dimensionless and all other (vacuum) terms at quadratic order in the curvature can be rewritten into linear combinations of the included ones by means of the Gauss-Bonnet identity.
We neglect boundary terms and non-minimal couplings between matter and curvature.
\\

The theory of QG, as defined in \cref{eq:action}, propagates (i)~the usual graviton, i.e., a massless spin-2 mode;
(ii)~a massive spin-0 mode; and (iii)~a massive spin-2 mode. The massive spin-2 mode has an opposite-sign kinetic term (in comparison to the other two modes) and is thus an Ostrogradski ghost. The massive spin-0 and spin-2 mode have respective masses
\begin{align}
\label{eq:masses}
    m_0^2 = -\frac{1}{32\pi G(3\beta - \alpha)}\;,
    \quad\quad\quad
    m_2^2 = -\frac{1}{16\pi G\alpha}\;.
\end{align}
In the following, we express the dimensionless couplings $\alpha$ and $\beta$ in terms of the masses $m_0$ and $m_2$.

Due to the inclusion of quadratic-curvature terms, the dynamics of QG is governed by fourth-order equations of motion. Nevertheless, the full theory can be described in terms of the same degrees of freedom~\cite{Noakes:1983,Hindawi:1995an,Hinterbichler:2015soa} as the linearized theory. To make this explicit, the Ricci scalar $\mathcal{R}$ and the traceless Ricci tensor $\widetilde{\mathcal{R}}_{ab}=R_{ab} - 1/4g_{ab}R$ can be promoted to independent evolution variables, as indicated by the calligraphic notation. This allows to write the equations of motion, obtained by varying the action in \cref{eq:action}, as follows\footnote{While \cite{Stelle:1977ry,Lu:2015psa,Held:2022abx} use different definitions of the couplings (related by the Gauss-Bonnet identity), the respective equations of motion are all equivalent. Some signs in~\cite{Noakes:1983} differ which, however, does not affect conclusions about a well-posedness.
}~\cite{Stelle:1977ry,Noakes:1983,Hindawi:1995an,Lu:2015psa,Held:2022abx}:
\begin{widetext}
\begin{alignat}{2}
\label{eq:metric}
    &\text{massless spin-2:}&
    \quad\quad\quad
    G_{ab}(\Box g) &=\;\widetilde{\mathcal{R}}_{ab} - \frac{1}{4} g_{ab}\mathcal{R}\equiv \frac{1}{\MPl^2}\widetilde{T}_{ab}\;,
    \\[0.1em]
\label{eq:trace}
    &\text{massive spin-0:}&
    \Box\,\mathcal{R}&=\;m_0^2\,\mathcal{R}
    +\frac{m_0^2}{\MPl^2}\,T^c_{\phantom{c}c}\;,
    \\[0.1em]
\label{eq:traceless}
    &\text{massive spin-2:}&
    \Box\,\widetilde{\mathcal{R}}_{ab}&=\;m_2^2\,\widetilde{\mathcal{R}}_{ab}
    - \frac{m_2^2}{\MPl^2}\,T^\text{(TL)}_{ab}
    + 2\,\widetilde{\mathcal{R}}_{a}^{\phantom{a}c}\widetilde{\mathcal{R}}_{bc}
    - \frac{1}{2}g_{ab}\widetilde{\mathcal{R}}^{cd}\widetilde{\mathcal{R}}_{cd}
    + \frac{1}{3}\left(\frac{m_2^2}{m_0^2}+1\right)\mathcal{R}\,\widetilde{\mathcal{R}}_{ab}
    \nonumber\\
    &&
    &
    - \frac{1}{3}\left(
        \frac{m_2^2}{m_0^2}-1
    \right)\left[
        \nabla_a\nabla_b \mathcal{R} 
        - \frac{1}{4}g_{ab}\left(
            m_0^2 \mathcal{R}
            + \frac{m_0^2}{\MPl^2}\,
            T^c_{\phantom{c}c}
        \right)
    \right]
    - 2\,\widetilde{\mathcal{R}}^{cd}C_{acbd}\;.
\end{alignat}
\end{widetext}
For reasons detailed below, we will refer to these equations as the metric equation, the trace equation, and the traceless equation, respectively.

The metric equation, i.e., \cref{eq:metric}, is nothing but the definition of the Einstein tensor: in terms of the metric on the left-hand side (LHS); and in terms of the fiducial variables on the right-hand side (RHS). It provides a second-order evolution equation for the metric. The fiducial variables $\mathcal{R}$ and $\widetilde{\mathcal{R}}_{ab}$, appearing on the RHS, are effectively equivalent to matter source terms, for which we have defined a fiducial stress-energy tensor $\widetilde{T}_{ab} \equiv \MPl^2(\widetilde{\mathcal{R}}_{ab} - \frac{1}{4}g_{ab}\mathcal{R})$. Hence, the metric equation can be treated as in GR.
For instance, one can make use of harmonic gauge to diagonalize the metric equation~\cite{Noakes:1983}. Alternatively, one may use the BSSN formalism~\cite{Shibata:1995we, Baumgarte:1998te}, as we do in \cref{sec:numerical}.

The trace equation, i.e., \cref{eq:trace}, provides a $2^\text{nd}$-order evolution equation for $\mathcal{R}$. 
The traceless equation, i.e., \cref{eq:traceless}, provides a $2^\text{nd}$-order evolution equation for $\widetilde{\mathcal{R}}_{ab}$.
Herein, we split the actual matter sources into a trace ($T^c_{\phantom{c}c}$) and a traceless ($T^\text{(TL)}_{ab}$) part which, in turn, source the respective fiducial variables.
\\

To keep the equations as concise as possible, we have also introduced the Weyl-tensor $C_{acbd}$. The latter can be expressed in terms of $R_{abcd}$, $\widetilde{\mathcal{R}}_{ab}$, and $\mathcal{R}$ as
\begin{align}
    C_{acbd} = R_{acbd} 
    + g_{b[c}\widetilde{\mathcal{R}}_{a]d}
    + g_{d[a}\widetilde{\mathcal{R}}_{c]b}
    +\frac{1}{6} g_{b[a} g_{c]d} \mathcal{R}
    \;.
\end{align}
In the evolution equations of $\mathcal{R}$ (\cref{eq:trace}) and $\widetilde{\mathcal{R}}_{ab}$ (\cref{eq:traceless}), derivatives of the metric only enter via double covariant derivatives as well as in $R_{abcd}$.

\section{Derivation: (3+1)-decomposition of the evolution equations}
\label{sec:3+1derivation}

The evolution system, as given in \cref{eq:metric,eq:trace,eq:traceless}, is a good starting point to perform the (3+1)-decomposition. Herein, we decompose the metric, i.e.,
\begin{align}
    g_{ab} = \gamma_{ab} - n_an_b
    \label{eq:decomp_metric}
\end{align}
into the spatial metric $\gamma_{ab}$ and the normal vector $n^a$ orthogonal to the spatial hypersurface. (The normal vector is chosen such that $n^an_a = -1$.)
Covariant derivatives $\nabla_a$ are projected onto spatial and normal part via
\begin{align}
    \label{eq:decomp_covd}
    \nabla_a = 
    (\gamma\indices{^b_a} - n_a n^b)\nabla_b \equiv 
    D_a - n_a n^b \nabla_b
    \;,
\end{align}
where we have defined the usual spatial covariant derivative $D_a \equiv \gamma\indices{_a^b}\nabla_b$.

Moreover, we introduce the usual geometric definition\footnote{The extrinsic curvature can be defined as the symmetric part of the spatial projection of the gradient of the normal vector, i.e., as $K_{ij} \equiv - \gamma\indices{_i^a} \gamma\indices{_j^b} \nabla_{a} n_{b}$, but if the normal vector is rotation free, the antisymmetric part vanishes and the strict definition reduces to the one in \cref{eq:def-extrinsic}} of the extrinsic curvature $K_{ij}$ and the acceleration $a_i$, respectively, as the mixed and the spatial projection of the gradient of the normal vector, i.e.,
\begin{align}
    \label{eq:def-acceleration}
    a_{i} &\equiv \gamma\indices{_i^b} n^a \nabla_{a} n_{b}\;,
    \\
    \label{eq:def-extrinsic}
    K_{ij} &\equiv - \gamma\indices{_i^a} \gamma\indices{_j^b} \nabla_{a} n_{b}\;.
\end{align}
The purely temporal projection of $\nabla_{a} n_{b}$ vanishes such that one may abuse notation and also write $a_{b} = n^a \nabla_{a} n_{b}$. In this case, $n^ba_b=0$. In complete equivalence to the above geometric definition, one can give a dynamical definition of the extrinsic curvature as a 1st-order variable for the metric, i.e., as $K_{ij}\equiv-\frac{1}{2}\mathcal{L}_n\gamma_{ij}$, where $\mathcal{L}_n$ denotes the Lie derivative along $n^a$. Both definitions are fully equivalent and imply each other.
\\

In the following, we reduce the remaining $2^\text{nd}$-order derivatives in the time-direction, i.e., along $n^a$, to $1^\text{st}$-order derivatives. In anticipation of that, we define additional $1^\text{st}$-order variables
\begin{align}
\label{eq:defV}
    \widetilde{V}_{ab}\equiv&
    -n^c\nabla_c\widetilde{\mathcal{R}}_{ab}\;,
    \\
\label{eq:defRhat}
    \hat{\mathcal{R}}\equiv&
    -n^c\nabla_c\mathcal{R}\;,
\end{align}
for the fiducial Ricci variables.

Furthermore, we decompose the fiducial traceless-Ricci tensor $\widetilde{\mathcal{R}}_{ab}$ and its $1^\text{st}$-order variable $\widetilde{V}_{ab}$ such that
\begin{align}
    \mathcal{A} &\equiv \gamma^{cd}\widetilde{\mathcal{R}}_{cd}\;,
    &\mathcal{B} &\equiv \gamma^{cd}\widetilde{V}_{cd}\;,
    \nonumber\\
    \mathcal{A}_{ab} &\equiv \gamma_a^{c}\gamma_b^d\widetilde{\mathcal{R}}_{cd} - \frac{1}{3}\gamma_{ab}\mathcal{A}\;,
    &\mathcal{B}_{ab} &\equiv \gamma_a^{c}\gamma_b^d\widetilde{V}_{cd} - \frac{1}{3}\gamma_{ab}\mathcal{B}\;,
    \nonumber\\
    \mathcal{C}_a &\equiv 
    n^c\gamma_a^d\widetilde{\mathcal{R}}_{cd}\;,
    &\mathcal{E}_a &\equiv 
    n^c\gamma_a^d\widetilde{V}_{cd}\;,
    \nonumber\\[0.4em]
    \Rightarrow\quad
    \mathcal{A} &= n^an^b\widetilde{\mathcal{R}}_{ab}\;,
    &\mathcal{B} &= n^an^b\widetilde{V}_{ab}\;,
    \label{eq:traceless3+1split}
\end{align}
where the last two relations are enforced by the tracelessness of $\widetilde{\mathcal{R}}_{ab}$ and $\widetilde{V}_{ab}$.
Equivalently, one may write this (3+1) split as
\begin{align}
    \widetilde{\mathcal{R}}_{ab} &= 
    \mathcal{A}_{ab} 
    + \frac{1}{3}\,\gamma_{ab}\,\mathcal{A}
    - 2\,n_{(a}\mathcal{C}_{b)}
    +n_an_b\,\mathcal{A}
    \;,
    \nonumber\\
    \widetilde{V}_{ab} &= 
    \mathcal{B}_{ab} 
    + \frac{1}{3}\,\gamma_{ab}\,\mathcal{B}
    - 2\,n_{(a}\mathcal{E}_{b)}
    +n_an_b\,\mathcal{B}
    \;.
    \label{eq:decomp_RabVab}
\end{align}
The remaining metric-dependent quantities can be decomposed using the conventional Gauss-Codazzi and Ricci equations, as collected in \cref{app:GaussCodazzi}.

We decompose the actual matter sources following the usual convention, i.e.,
\begin{align}
    \rho &= n_{a} n_{b}\,T^{ab}
    \;,
    \notag\\
    S_i &= - \gamma_{ia} n_{b}\,T^{ab}
    \;,
    \notag\\
    S_{ij} &= \gamma_{ia} \gamma_{jb}\,T^{ab}
    \;.
    \label{eq:matter-decomp}
\end{align}
Similarly, we decompose the fiducial matter sources, i.e.,
\begin{align}
    \widetilde{\rho} &= 
    n_{a} n_{b}\,\widetilde{T}^{ab} = 
    \MPl^2\left(\mathcal{A} + \frac{1}{4}\mathcal{R}\right)
    \;,\notag\\*
    \widetilde{S}_i  &= 
    - \gamma_{ia} n_{b}\,\widetilde{T}^{ab}=
    -\autoCheckMissing{\MPl^2}\,\mathcal{C}_i
    \;,\notag\\*
    \widetilde{S}_{ij} &= 
    \gamma_{ia} \gamma_{jb}\,\widetilde{T}^{ab}=
    \autoCheckMissing{\MPl^2}\left(
        \mathcal{A}_{ij}
        +\frac{1}{3}\gamma_{ij}\mathcal{A}
        \autoCheckMissing{-}\frac{1}{4}\gamma_{ij}\mathcal{R}
    \right)
    \label{eq:fid-matter-decomp}
    \;.
\end{align}
For the actual matter sources, we note that $T^\text{(TL)}_{ab}$, appearing in \cref{eq:traceless}, is traceless in 4D but the 3D projections do not vanish, i.e.,
\begin{align}
    n^an^b\,T^\text{(TL)}_{ab} &= \frac{1}{4}(S+3\rho)\;,
    \\*
    \gamma^{ab}\,T^\text{(TL)}_{ab} &= \frac{1}{4}(S+3\rho)\;,
    \\*
    \gamma_i^a\gamma_j^b\,T^\text{(TL)}_{ab} &= S_{ij} - \frac{1}{4}\gamma_{ij}(S-\rho)\;.
\end{align}
With these definitions at hand, the decomposition of the three evolution equations (metric equation, trace equation, and traceless equation, cf. \cref{eq:metric,eq:trace,eq:traceless}) is tedious but essentially straightforward. After the decomposition, we also identify which of the decomposed equations correspond to constraints, constraint evolution, or physical evolution equations. The busy reader may skip to \cref{sec:summary-of-evolution-eqs} where we summarize the result.

\subsection{(3+1) decomposition of the metric equation}

\cref{eq:metric}, determines the evolution of the metric $g_{ab}$. The fiducial variables $\mathcal{R}$ and $\widetilde{\mathcal{R}}_{ab}$ can be treated as fiducial matter sources. The actual matter sources $T^{ab}$ do not appear in the metric evolution. They will only affect the other evolution equations.
As for most numerical efforts in GR, our starting point for the metric sector is the York-variant of the ADM equations \cite{York:2014}, i.e.,
\begin{align}
	\left(n^c\nabla_c\gamma_{ij}\right) =& 
	-2\,D_{(i}n_{j)}
	-2\,K_{ij}
	\:,
    \label{eq:evol-gamma-York}
     \\[1em]
     \left(n^c\nabla_c K_{ij}\right) =&
     - a_ia_j
     -2\,D_{(i}a_{j)}
     - 2\,K_{m(i}D_{j)}n^m
     \nonumber\\*[0.5em]&
     - 2K_{im}K^{m}_{j}
     + K\,K_{ij}
     +{}^{(3)}\!R_{ij}
     \nonumber\\*&
     - \frac{1}{\MPl^2}\left(
        \widetilde{S}_{ij} 
        - \frac{1}{2}\gamma_{ij}(\widetilde{S} - \widetilde{\rho})
     \right)
     \;,
     \label{eq:evol-extrinsic-York}
     \\[0em]
     0 =&\;D_jK^{j}_{i} - D_i K - \autoCheckMissing{\frac{1}{\MPl^2}}\widetilde{S}_i \;,
     \label{eq:shift-constraint-York}
     \\[0.5em]
     0 =&\;{}^{(3)}\!R - K_{ij}K^{ij} + K^2 - \autoCheckMissing{\frac{2}{\MPl^2}}\widetilde{\rho}\;.
    \label{eq:Hamiltonian-constraint-York}
\end{align}
The first equation (evolution of the spatial metric) is a definition, used to reduce the equations from $2^\text{nd}$-order to $1^\text{st}$-order in time. It is the metric equivalent of our definitions in \cref{eq:defV,eq:defRhat}. However, the fiducial variable, that one has introduced in the metric sector, i.e., $K_{ij}$, also carries direct geometric meaning -- it is the extrinsic curvature of the spatial hypersurface.
For the second equation (evolution of the extrinsic curvature), one has used the lapse constraint in \cref{eq:Hamiltonian-constraint-York} to simplify the evolution equation. Hence the appearance of $\widetilde{\rho}$ in \cref{eq:evol-extrinsic-York}.
The $3^\text{rd}$ and $4^\text{th}$ equation correspond to the momentum and Hamiltonian constraint, respectively.
\\

In summary, including fiducial matter sources, the metric equation decomposes in complete equivalence to GR. The spatial projections result in evolution equations for $\gamma_{ij}$ and $K_{ij}$, i.e., for 12 pieces of initial data\footnote{Here, we already assume that hypersurfaces are chosen such as to fix $g_{00}$ and $g_{0i}$ by an appropriate choice of lapse and shift as well as $n^c\nabla_cg_{00}$ and $n^c\nabla_cg_{0i}$ such as to obey a specified gauge choice, e.g., harmonic gauge. This choice of gauge/coordinates already fixes 8 out of 20 pieces of initial data in the second-order evolution of $g_{\mu\nu}$.}.
The mixed and the temporal projections of the metric equation result in 4 constraints -- the Hamiltonian and the momentum constraint. Moreover, there remains coordinate freedom within the spatial hypersurface: We are free to choose the spatial coordinates as well as the initial time, hence removing 4 further pieced of initial data. Overall, as in GR, one finds $12-4-4=4$ independent pieces of initial data, i.e., 2 degrees of freedom, in the metric sector. We will come back to the overall counting of degrees of freedom in \cref{sec:summary-of-evolution-eqs}.

\subsection{(3+1) decomposition of the trace equation}

\cref{eq:trace} determines the evolution of the fiducial Ricci scalar $\mathcal{R}$. Since the only derivatives appear in $\Box\mathcal{R}$, it is of quasi-linear form. We (3+1)-decompose the covariant derivatives on the left-hand side (LHS) as
\autoCheckMissing{
\begin{align}
    \Box \mathcal{R} = 
    n^a \nabla_a \hat{\mathcal{R}} 
    +(D_i + a_i)D^i\mathcal{R}
    - K \hat{\mathcal{R}}\;.
\end{align}
}
Combining the above result with the RHS of \cref{eq:trace} provides two $1^\text{st}$-order (in time) equations
for $\mathcal{R}$ and $\hat{\mathcal{R}}$, i.e.,
\begin{align}
\label{eq:evol-trace}
    n^a \nabla_a \mathcal{R} &= 
    - \hat{\mathcal{R}}\;,
    \\
\label{eq:evol-dttrace}
    n^a \nabla_a \hat{\mathcal{R}} &= 
    - (D_i + a_i)D^i\mathcal{R}
    \nonumber\\*&\quad\;
    + K \hat{\mathcal{R}} 
    + m_0^2 \mathcal{R}
    \autoCheckMissing{+ \frac{m_0^2}{\MPl^2}(S - \rho)}
    \;,
\end{align}
where $\rho$ and $S = \gamma^{ab}T_{ab} = \gamma^{ab}S_{ab}$ correspond to the trace of the actual matter source terms, decomposed analogously to \cref{eq:fid-matter-decomp}.
\\

In summary, we find two evolution equations and no constraints in the trace sector.

\subsection{(3+1) decomposition of the traceless equation}

\cref{eq:traceless} evolves the traceless fiducial Ricci tensor $\widetilde{\mathcal{R}}_{ab}$. Without recasting, this equation is not of quasi-linear form. Therefore, while performing the (3+1) decomposition, we can expect to have to use the previous evolution equations to remove all second order time derivatives on the RHS. This procedure is reminiscent of the order reduction in~\cite{Noakes:1983}.
\\

Before doing so, we consider the (3+1) decomposition of the LHS, i.e.,
\autoCheckMissing{
\begin{align}
    \label{eq:boxRab}
    \Box \widetilde{\mathcal{R}}_{ab} = 
    n^c \nabla_c \widetilde{V}_{ab}
    + (D_c + a_c)D^c\widetilde{\mathcal{R}}_{ab}
    - K \widetilde{V}_{ab}
    \;,
\end{align}
}
\\
where, as for the fiducial Ricci scalar, we have introduced the first-order fiducial variable $\widetilde{V}_{ab} = - n^c \nabla_c \widetilde{\mathcal{R}}_{ab}$, cf.~\cref{eq:defV}.
Herein, the spatial covariant derivatives should strictly be understood as a shorthand notation, i.e.,
$
D_cD^c\widetilde{\mathcal{R}}_{ab} \equiv 
\gamma\indices{^d_c} \nabla_d (\gamma^{ce} \nabla_e \widetilde{\mathcal{R}}_{ab})
$
and
$
a^c D_c\widetilde{\mathcal{R}}_{ab} \equiv
a^e\gamma_e^c\nabla_c\widetilde{\mathcal{R}}_{ab}
$.
This subtlety is important since $\widetilde{\mathcal{R}}_{ab}$ is not yet projected and thus contains temporal components.
The derivation is made explicit in \cref{app:boxRab}. 
Overall, this renders the LHS manifestly $1^\text{st}$-order in time.

On the RHS of \cref{eq:traceless}, the only derivative terms are contained in $\nabla_a\nabla_b \mathcal{R}$ and in the Riemann tensor $R_{acbd}$. The Riemann tensor can be decomposed in the usual way, cf.~\cref{app:GaussCodazzi}.
Regarding $\nabla_a\nabla_b \mathcal{R}$, we find
\autoCheckMissing{
\begin{align}
    \nabla_a\nabla_b \mathcal{R} &=
    D_a D_b \mathcal{R}
    + 2\,n_{(a}D_{b)}\hat{\mathcal{R}}
    - 2\,K_{ab} \hat{\mathcal{R}}
    \nonumber\\*&\quad
    - n_a n_b \left(n^c \nabla_c \hat{\mathcal{R}}\right)
    + n_a n_b\,a_c D^c\mathcal{R}\;,
\end{align}
}
which is, as expected, symmetric in $(a,b)$. Here, we have used (i) the projection of the covariant derivative, cf.~\cref{eq:decomp_covd}; (ii) the geometric definitions of acceleration and extrinsic curvature, i.e., \cref{eq:def-acceleration,eq:def-extrinsic}; and (iii) the identity $0=\nabla_d g^c_a = \nabla_d(\gamma^c_a - n_an^c) = \nabla_d\gamma^c_a - n_a\nabla_dn^c -n^c\nabla_dn_a$.
The calculation is made fully explicit in \cref{app:DDR}.
\\

Collecting everything, we find $1^\text{st}$-order evolution equations for the fiducial variables $\widetilde{\mathcal{R}}_{ab}$ and $\widetilde{V}_{ab}$, i.e.,
\begin{widetext}
\begin{align}
    \label{eq:evol-Rab}
    n^c \nabla_c \widetilde{\mathcal{R}}_{ab} =&
    -\widetilde{V}_{ab}\;,
    \\[1em]
    \label{eq:evol-Vab}
    n^c\nabla_c\widetilde{V}_{ab} =&
    - (D_c + a_c)D^c\widetilde{\mathcal{R}}_{ab}
    + K \widetilde{V}_{ab}
    + m_2^2\widetilde{\mathcal{R}}_{ab}
    \autoCheckMissing{
        - \frac{m_2^2}{\MPl^2}\,T^\text{(TL)}_{ab}
    }
    +2\,\widetilde{\mathcal{R}}_{a}^{\phantom{a}c}\widetilde{\mathcal{R}}_{bc}
    -\frac{1}{2}g_{ab}\widetilde{\mathcal{R}}^{cd}\widetilde{\mathcal{R}}_{cd}
    +\frac{1}{3}\left(\frac{m_2^2}{m_0^2}+1\right)\mathcal{R}\,\widetilde{\mathcal{R}}_{ab}
    \nonumber\\*&
    - \frac{1}{3}\left(\frac{m_2^2}{m_0^2}-1\right)
    \Bigg[
        \left(
            D_a D_b
            + n_a n_b\,a_c D^c
            -\frac{1}{4}\,g_{ab}\,m_0^2
        \right)\mathcal{R}
        -\frac{1}{4}\frac{m_0^2}{\MPl^2}\,g_{ab}(S-\rho)
        \nonumber\\*&\quad\quad\quad\quad
        + 2\left(
            n_{(a}D_{b)}
            - K_{ab}
        \right)\hat{\mathcal{R}}
        - n_a n_b \left(n^c \nabla_c \hat{\mathcal{R}}\right)
    \Bigg]
    \nonumber\\*&
    - 2\,\widetilde{\mathcal{R}}^{cd}\Bigg[
        g_{b[c}\widetilde{\mathcal{R}}_{a]d}
        + g_{d[a}\widetilde{\mathcal{R}}_{c]b}
        +\frac{1}{6}g_{b[a}g_{c]d}\mathcal{R}
        +{}^{(3)}\!R_{a{cb}d} 
        +2\,K_{a[{b}}K_{d]{c}}
        + 4\,a_{[a}n_{c]}a_{[b}n_{d]}
        - 4\,n_{[b}\left(D_{d]}a_{[a}\right)n_{c]}
        \nonumber\\*&\quad\quad\quad\quad
        +4\left(D_{[a}K_{{c][b}}\right)n_{d]}
        +4\left(D_{[{b}}K_{d][a}\right)n_{{c}]}
        +4\,n_{[a}\,K^e_{{c]}}\,K_{e[{b}}n_{d]}
        +4\,\left(
            \gamma^f_{[a}n_{{c}]}\gamma^g_{[{b}}n_{d]}
        \right)
        \left(
            n^e\nabla_eK_{fg}
        \right)
    \Bigg]
    \;.
\end{align}
\end{widetext}
The terms involving $(n^c \nabla_c \hat{\mathcal{R}})$ and $(n^e\nabla_eK_{fg})$ can be writted in terms of the other evolution equations such that no time derivatives remain on the RHS. It remains to project all the non-derivative terms onto spatial and temporal parts and thereby decompose the above two $1^\text{st}$-order traceless equations into spatial and temporal parts.

\subsection{Projection of the traceless equations}

We can explicitly project the traceless equations in order to separate constraint data from initial data. 
In all cases, we will obtain four different projections, i.e., we can obtain (i) the spatial trace with $\gamma^{ab}$; (ii) the spatial projection with $\gamma^a_c\gamma^b_d$; (iii) the temporal projection with $n^an^b$; and (iv) the mixed projection with $n^a\gamma^b_c$ (or equivalently $n^b\gamma^a_c$). To obtain these, we have to commute the projection operators through the covariant derivative on the left-hand side of each respective equation which generates further terms. We present the explicit derivation in \cref{app:LHSprojections} and find
\begin{align}
    \gamma^{ab}\left(n^c\nabla_c\widetilde{\mathcal{R}}_{ab}\right) &=
    \left(n^c\nabla_c\mathcal{A}\right) - 2\,a^c\mathcal{C}_c
    \;,\label{eq:trace-projection_Rab}
    \\
    \gamma_i^a\gamma_j^b\left(n^c\nabla_c\widetilde{\mathcal{R}}_{ab}\right) &=
    \left(n^c\nabla_c\mathcal{A}_{ij}\right) 
    + \frac{1}{3}\gamma_{ij}\left(n^c\nabla_c\mathcal{A}\right)
    \nonumber\\*&\quad\;
    - \frac{2}{3}\mathcal{A}\left(D_{(i}n_{j)} + K_{ij}\right)
    - 2a_{(i}\mathcal{C}_{j)}
    \nonumber\\*&\quad\;
    - 2\,a^c\left(\mathcal{A}_{c(i}n_{j)} + \frac{1}{3}\gamma_{c(i}n_{j)}\mathcal{A}\right) 
    \;,\label{eq:spatial-projection_Rab}
    \\
    n^an^b\left(n^c\nabla_c\widetilde{\mathcal{R}}_{ab}\right) &=
    \left(n^c\nabla_c\mathcal{A}\right)
    - 2\,a^c\mathcal{C}_c
    \;,\label{eq:temporal-projection_Rab}
    \\
    n^a\gamma^b_d\left(n^c\nabla_c\widetilde{\mathcal{R}}_{ab}\right) &=
    \left(n^c\nabla_c \mathcal{C}_d\right) 
    - n_d a^c \mathcal{C}_c
    \nonumber\\*&\quad\;
    - a^a\left(
        \mathcal{A}_{ad} 
        + \frac{2}{3}\gamma_{ad}\mathcal{A}
    \right)
    \;.\label{eq:mixed-projection_Rab}
\end{align}
The analogous projections for the fiducial first-order variables are obtained by the replacements $\widetilde{\mathcal{R}}_{ab}\rightarrow\widetilde{V}_{ab}$, $\mathcal{A}_{ij}\rightarrow\mathcal{B}_{ij}$, $\mathcal{A}\rightarrow\mathcal{B}$, and $\mathcal{C}_a\rightarrow\mathcal{E}_a$.
These left-hand-side projections separate the covariant equations into 
evolution equations (\cref{eq:trace-projection_Rab,eq:spatial-projection_Rab})
and constraint evolution (\cref{eq:mixed-projection_Rab}).
In line with the $(3+1)$ conventions chosen in~\cref{eq:traceless3+1split}, the temporal projection in~\cref{eq:temporal-projection_Rab} is redundant with the spatial trace projection in \cref{eq:trace-projection_Rab}.

The RHS projections are tedious, and we check them in the ancillary files\footnote{See the \texttt{GitHub} repository (\url{https://github.com/aaron-hd/QG-sphSymm-ancillary}). Parts of the derivation make use of the \texttt{xAct} package \cite{Martin-Garcia:2007bqa} (\url{http://www.xact.es/}).}.
Crucially, the RHS terms do not impact the character of the respective projections since they only involve spatial derivatives.
\\

The trace and spatial projection result in the following set of evolution equations for the spacial variables $\mathcal{A}$, $\mathcal{A}_{ij}$, $\mathcal{B}$, and $\mathcal{B}_{ij}$:
\begin{widetext}
\begin{align}
    \label{eq:evol-A}
    n^c\nabla_c\mathcal{A}_{\phantom{ij}}
    &= 
    2\,a^k\mathcal{C}_k
    -\mathcal{B}\;.
    \\[1em]
    \label{eq:evol-Aij}
    n^c\nabla_c\mathcal{A}_{ij} &=
    \frac{2}{3}\,\mathcal{A}\left(
        D_{(i}n_{j)} 
        + K_{ij}
    \right)
    +2\,a^c\left(
        \mathcal{A}_{c(i}n_{j)}
        +\frac{1}{3}\gamma_{c(i}n_{j)}\mathcal{A}
        +\gamma_{c(i}\mathcal{C}_{j)}
    \right)
    - \mathcal{B}_{ij}
    - \frac{2}{3}\gamma_{ij}a^k\mathcal{C}_k
    \;,
    \\[1em]
    \label{eq:evol-B}
    n^c\nabla_c\mathcal{B}_{\phantom{ij}}
    &=
    2\,a^k\mathcal{E}_k
    - \frac{1}{4}\frac{m_2^2}{\MPl^2}(S+3\rho)
    - \left(
        D_iD^i + a_iD^i - m_2^2 +\frac{1}{6}\,\mathcal{R}
    \right)\mathcal{A}
    + K\,\mathcal{B}
    \nonumber\\*&\quad
    +\frac{1}{3}\left(\frac{m_2^2}{m_0^2}+1\right)\mathcal{R}\,\mathcal{A}
    -\frac{1}{3}\left(\frac{m_2^2}{m_0^2}-1\right)\left[
        \left(
            D_iD^i - \frac{3}{4}m_0^2
        \right)\mathcal{R}
        -\frac{3}{4}\frac{m_0^2}{\MPl^2}(S-\rho)
        -2\,K\,\hat{\mathcal{R}}
    \right]
    \nonumber\\*&\quad
    +\frac{3}{2}\left(
        \mathcal{A}_{ij}\mathcal{A}^{ij}
        + \frac{4}{3}\,\mathcal{A}^2
        - 2\,\mathcal{C}^i\mathcal{C}_i
    \right)
    - 2\,\mathcal{C}^i\left(
        D^j K_{ij}
        + a^j K_{ij}
    \right)
    - 4\,K^{ij}D_i\mathcal{C}_j
    \nonumber\\*&\quad
    - 2\left(
        \mathcal{A}^{ij}
        +\frac{1}{3}\gamma^{ij}\mathcal{A}
    \right)\left(
        {}^{(3)}\!R_{ij} 
        + 2\,K_{i[j}K^k_{k]}
    \right)
    + 4\,\mathcal{C}^j\left(
        D_jK - D^iK_{ij}
    \right)
    \nonumber\\*&\quad
    -2\,\mathcal{A}\left(
        a_i a^i 
        + D_{i} a^{i}
        - K^{ij} K_{ij}
        + \gamma^{ij} \left(n^c\nabla_c K_{ij}\right)
    \right)
    \;,
    \\[1em]
    \label{eq:evol-Bij}
    n^c\nabla_c\mathcal{B}_{ij}
    &= 
    \frac{2}{3}\,\mathcal{B}\left(
        D_{(i}n_{j)} 
        + K_{ij}
    \right)
    +2\,a^c\left(
        \mathcal{B}_{c(i}n_{j)}
        +\frac{1}{3}\gamma_{c(i}n_{j)}\mathcal{B}
        +\gamma_{c(i}\mathcal{E}_{j)}
    \right)
    - \frac{m_2^2}{\MPl^2}\left(
        S_{ij} - \frac{1}{4}\gamma_{ij}(S-\rho)
    \right)
    - \frac{1}{3}\gamma_{ij} \left(n^c\nabla_c\mathcal{B}\right)
    \nonumber\\*&\quad 
    -\left(
        D_kD^k + a_kD^k 
        - m_2^2 + \frac{1}{6}\mathcal{R}
    \right)\left(
        \mathcal{A}_{ij} 
        + \frac{1}{3}\gamma_{ij}\mathcal{A}
    \right)
    +K\left(
        \mathcal{B}_{ij}
        +\frac{1}{3}\gamma_{ij}\mathcal{B}
    \right)
    \nonumber\\*&\quad 
    +\frac{1}{3}\left(\frac{m_2^2}{m_0^2}+1\right)
    \mathcal{R}\left(
        \mathcal{A}_{ij}
        +\frac{1}{3}\gamma_{ij}\mathcal{A}
    \right)
    -\frac{1}{3}\left(\frac{m_2^2}{m_0^2}-1\right)\left[
        \left(
            D_iD_j
            -\frac{1}{4}\gamma_{ij}m_0^2
        \right)\mathcal{R}
        -\frac{1}{4}\frac{m_0^2}{\MPl^2}\gamma_{ij}(S-\rho)
        -2\,K_{ij}\hat{\mathcal{R}}
    \right]
    \nonumber\\*&\quad 
    +\frac{1}{2}\,\gamma_{ij}\left(
        \mathcal{A}^{kl}\mathcal{A}_{kl}
        +\frac{4}{3}\mathcal{A}^2
        -2\,\mathcal{C}^k\mathcal{C}_k
    \right)- 2\,\mathcal{C}_{(i}\left(
        D^k K_{j)k}
        + a^k K_{j)k}
    \right)
    - 4\,K_{k(i} D^k \mathcal{C}_{j)}
    \nonumber\\*&\quad 
    -2\left(
        \mathcal{A}^{kl}
        +\frac{1}{3}\gamma^{kl}\mathcal{A}
    \right)\left(
        {}^{(3)}\!R_{ikjl}
        +K_{i[j}K_{l]k}
    \right)
    +4\,\mathcal{C}^k\left(
        D_k K_{ij} 
        - D_{(i}K_{j)k}
    \right)
    \nonumber\\*&\quad 
    -2\,\mathcal{A}\left(
        a_i a_j 
        + D_{(i} a_{j)}
        - K_i^k K_{kj}
        + \gamma_i^k\,\gamma_j^l \left(n^c\nabla_c K_{kl}\right)
    \right)
    \;.
\end{align}
\end{widetext}
Here, we have used the evolution equation for $(n^c\nabla_c\mathcal{A})$ (cf.~\cref{eq:evol-A}) on the RHS of the evolution equation for $(n^c\nabla_c\mathcal{A}_{ij})$ (cf.~\cref{eq:evol-Aij}). It can be verified explicitly that the latter equation is (spatially) traceless.
Analogously, the last term in the first line of \cref{eq:evol-Bij} ensures that the evolution equation for $\mathcal{B}_{ij}$ is (spatially) traceless. We refrain from plugging in \cref{eq:evol-B} (as well as the evolution equation for $K_{ij}$) explicitly to keep the expressions as concise as possible.

\subsection{Bianchi constraints}
\label{sec:bianchi-constraints}

The fiducial variables $\mathcal{R}$ and $\widetilde{\mathcal{R}}_{ab}$ are not physical. Their only purpose is to reduce the order of the system. Naturally, in order for the reduced evolution to capture the physics of the original evolution, and the correct degrees of freedom in particular, we have to ensure that the fiducial variables evaluate to the proper metric quantities~\cite{Noakes:1983}, i.e., that 
\begin{align}
    0=\Delta_{ab} \equiv
    G_{ab}(g) 
    - \widetilde{\mathcal{R}}_{ab}
    +\frac{1}{4}g_{ab}\mathcal{R}
    \;.
\end{align}
However, this equation is nothing but the metric equation itself which we already added to the system of evolution equations. Hence, projection will only reproduce the Hamiltonian constraint (temporal), the momentum constraint (mixed), and the metric evolution equation (spatial). It seems like there are no novel constraints.

Crucially, since we have been replacing $2^\text{nd}$-order variables, we also have to ensure that their $1^\text{st}$ and $2^\text{nd}$ derivatives match the original metric quantities. The simplest such constraint is nothing but the Bianchi identity expressed in terms of the fiducial variables, i.e.,
\begin{widetext}
\begin{align}
\label{eq:constraint-Bianchi}
    0=
    \nabla_b\Delta_a^b =&
    \underbrace{\left[
        - \mathcal{E}_a
        - K_a^b\mathcal{C}_b - K\mathcal{C}_a
        - D^b\mathcal{A}_{ab} - \frac{1}{3}D_a\mathcal{A}
        + \frac{1}{4}D_a\mathcal{R}
    \right]}_\text{spatial}
    + n_a \underbrace{\left[
        \mathcal{B}
        + D^b\mathcal{C}_b
        + \frac{1}{4}\hat{\mathcal{R}}
        + K_{bc}\mathcal{A}^{bc}
        + \frac{4}{3}K\mathcal{A}
    \right]}_\text{temporal}
    \;.
\end{align}
\end{widetext}
Following Noakes~\cite{Noakes:1983}, we refer to these 4 constraints as ``Bianchi constraints''. Similarly, the normal derivative of the Bianchi constraint
\begin{align}
    \label{eq:constraint-Bianchi-dot}
    0 = n_c\nabla^c\left(\nabla_b\Delta_a^b\right)
\end{align}
generates 4 further constraints which we refer to as ``Bianchi-dot constraints'' but which need not be written explicitly for our purposes.

\subsection{Constraint evolution}
\label{sec:fiducial-constraints}

We recall that, in the ADM formalism, the purely temporal and the mixed projection of the Einstein equations result in the Hamiltonian and the momentum constraint. Similarly, the temporal and mixed projection of the higher-derivative equations are not propagating physical degrees of freedom. 
\\

As mentioned before, the temporal projections (cf.~\cref{eq:temporal-projection_Rab}) are fully redundant and merely reproduce the spatial trace projections.
\\

The mixed projections correspond to evolution equations for $\mathcal{C}_i$ and $\mathcal{E}_i$. For instance, the mixed projection of $(n^c\nabla_c\widetilde{\mathcal{R}}_{ab})= -\widetilde{V}_{ab}$ (cf.~\cref{eq:evol-Rab}) with $n^a\gamma^b_i$ results in
\begin{align}
\label{eq:constraint-Rab-mixed}
    n^c\nabla_c\mathcal{C}_i =
    a^k\left[
        \mathcal{A}_{ki}
        + \frac{2}{3}\gamma_{ki}\mathcal{A}
    \right]
    + n_ia^k\mathcal{C}_k
    - \mathcal{E}_i
    \;,
\end{align}
which corresponds to an evolution equation for $\mathcal{C}_i$.
\\

Similarly, the mixed projection of \cref{eq:evol-Vab} corresponds to evolution of $\mathcal{E}_i$.
We refrain from showing the full expansion of the latter projection since (see \cref{sec:summary-of-evolution-eqs}) we can remove $\mathcal{C}_i$ and $\mathcal{E}_i$ by use of the momentum constraint and the spatial projection of the Bianchi constraint, respectively.
Hence, there is no need to explicitly evolve the variables $\mathcal{C}_i$ and $\mathcal{E}_i$. Instead, $\mathcal{C}_i$ and $\mathcal{E}_i$ can be understood as constraint variables and the mixed projections can be interpreted as constraint evolution.

\subsection{Summary of evolution equations and constraints}
\label{sec:summary-of-evolution-eqs}

Overall, the (3+1) decomposition is now phrased in terms of 32 free functions of initial data\footnote{
The components of $n^a$ and its derivatives in the time direction can be seen as the remaining 8 free functions of initial data in order to match with the 40 free functions of initial data expected from the reduction of a $4^\text{th}$-order evolution of a symmetric tensor in 4D. However, here, and in GR, they are fully constrained/determined by gauge/coordinate choice. For instance, one can choose harmonic gauge, in which case $F^a=0$ and $(n^c\nabla_c F^a)=0$ give the respective 8 harmonic constraints.
}, i.e., the spatial metric $\gamma_{ij}$ as well as its $1^\text{st}$-order variable $K_{ij}$; the (fiducial) Ricci scalar $\mathcal{R}$ as well as its $1^\text{st}$-order variable $\hat{\mathcal{R}}$; and the (3+1) components of the (fiducial) traceless Ricci tensor $\mathcal{A}$, $\mathcal{A}_{ij}$, and $\mathcal{C}_i$ as well as their $1^\text{st}$-order variables $\mathcal{B}$, $\mathcal{B}_{ij}$, and $\mathcal{E}_i$.
\\

For these 32 free functions of initial data, only 16 correspond to physical initial data: In the metric sector, the Hamiltonian and momentum constraints in \cref{eq:Hamiltonian-constraint-York,eq:shift-constraint-York} as well as 4 coordinate choices in the initial-data surface reduce from 12 to 4 pieces of initial data. Hence, the metric sector still propagates the expected 2 degrees of freedom of a massless spin-2 mode. While the constraints are modified, the constraint structure remains as in GR.

In the fiducial sector, the 4 Bianchi (cf. \cref{eq:constraint-Bianchi}) and the 4 Bianchi-dot (cf. \cref{eq:constraint-Bianchi-dot}) constraints reduce from 20 to 12 pieces of initial data. Hence, the fiducial sector contains 6 propagating degrees of freedom, corresponding to one massive spin-0 and one massive spin-2 mode. 
\\

While not all of the constraints are algebraic, it is convenient that there are sufficiently many algebraic constraints in order to fully determine and thus remove the initial data for $\mathcal{C}_i$ (by use of the momentum constraint in \cref{eq:shift-constraint-York}) and $\mathcal{E}_i$ (by use of the spatial projection of the Bianchi constraint in \cref{eq:constraint-Bianchi}). In practice, we thus only need to evolve $\gamma_{ij}$ and $K_{ij}$ (see~\cref{eq:evol-gamma-York,eq:evol-extrinsic-York}), $\mathcal{R}$ and $\hat{\mathcal{R}}$ (see~\cref{eq:evol-trace,eq:evol-dttrace}, as well as $\mathcal{A}$, $\mathcal{A}_{ij}$, $\mathcal{B}$, and $\mathcal{B}_{ij}$ (see~\cref{eq:evol-A,eq:evol-Aij,eq:evol-B,eq:evol-Bij}), i.e., only 26 variables.

\section{Method: Numerical evolution}
\label{sec:numerical}

The (3+1) evolution equations derived in the previous section are fully general: we expect them to be compatible with all the state-of-the-art evolution schemes~\cite{Baumgarte:1998te,Shibata:1995we,Alic2012,Alic2013} and numerical code frameworks~\cite{Loffler2011ay,Clough2015,SpecCode,Fernando2019}. In the following, we will focus on the vacuum case, specify to the BSSN formulation~\cite{Shibata:1995we,Baumgarte:1998te}, and numerically evolve the system using the \texttt{Dendro-GR}~\cite{Fernando2019} code.

The purpose of our numerical efforts is twofold. The first purpose is of technical nature: We demonstrate that the evolution system is numerically stable, even in the nonlinear regime. The second purpose is physical: Given numerical stability, we then use the numerical evolution to investigate stability of the Ricci-flat subsector of QG.
\\

From here on, we specify to the usual $(3+1)$ coordinate conventions, in which
\begin{align}
    \beta^a &= (0,\,\beta^i)\;,
    \notag\\
    n^a &= (1/\alpha,\,-\beta^i/\alpha)\;,
    \notag\\
    ds^2 &= 
    -\alpha^2\,dt^2
    +\gamma_{ij}\left(
        dx^i + \beta^i\,dt
    \right)\left(
        dx^j + \beta^j\,dt
    \right)\;,
    \label{eq:coordinate-conventions}
\end{align}
with the lapse function $\alpha$ and the shift vector $\beta^i$. For the evolution of $\alpha$ and $\beta^i$ we choose a standard $(1+\text{log})$ slicing and a $\Gamma$-driver, respectively~\cite{Alcubierre2003}.
\\

Regarding the metric dynamics, the BSSN formulation~\cite{Shibata:1995we,Baumgarte:1998te} (see also~\cite{Beyer:2004sv} for subsequent proof of its strong hyperbolicity), proceeds exactly as in GR. For completeness, we summarize the BSSN evolution equations in \cref{app:BSSN}. Together with the evolution equations for $\mathcal{R}$, $\hat{\mathcal{R}}$, $\mathcal{A}$, $\mathcal{A}_{ij}$, $\mathcal{B}$, and $\mathcal{B}_{ij}$ (see \cref{sec:summary-of-evolution-eqs}), these form the system of partial differential equations (PDEs) that we implement numerically.

\subsection{Numerical setup}
\label{sec:Dendro}

We implement the evolution equations in the \texttt{Dendro-GR}~\cite{Fernando2019} framework. \texttt{Dendro-GR} combines a parallel octree-refined adaptive mesh with a wavelet adaptive multiresolution. An additional Quadratic-Gravity module is built on top of this framework\footnote{See the \texttt{GitHub} repository \url{https://github.com/lanl/Dendro-GRCA}.}. We use a fourth-order finite-difference scheme to evaluate spatial derivatives and a fourth-order Runge-Kutta method to evolve in time. The Courant–Friedrichs–Lewy condition~\cite{Courant:1967} which relates the temporal and spatial disretization is set to 0.25. Therefore, as we increase $N_{x,y,z}$ (or decrease $\Delta (x,y,z)$), the time discretization $\Delta t$ decreases. Other conditions are varied with respect to the test problems. 

\subsection{Numerical stability}
\label{sec:numerical-stability}

%
\begin{figure}[!t]
    \centering
    \includegraphics[trim={0cm 0cm 0cm 0cm},clip,width=\linewidth]{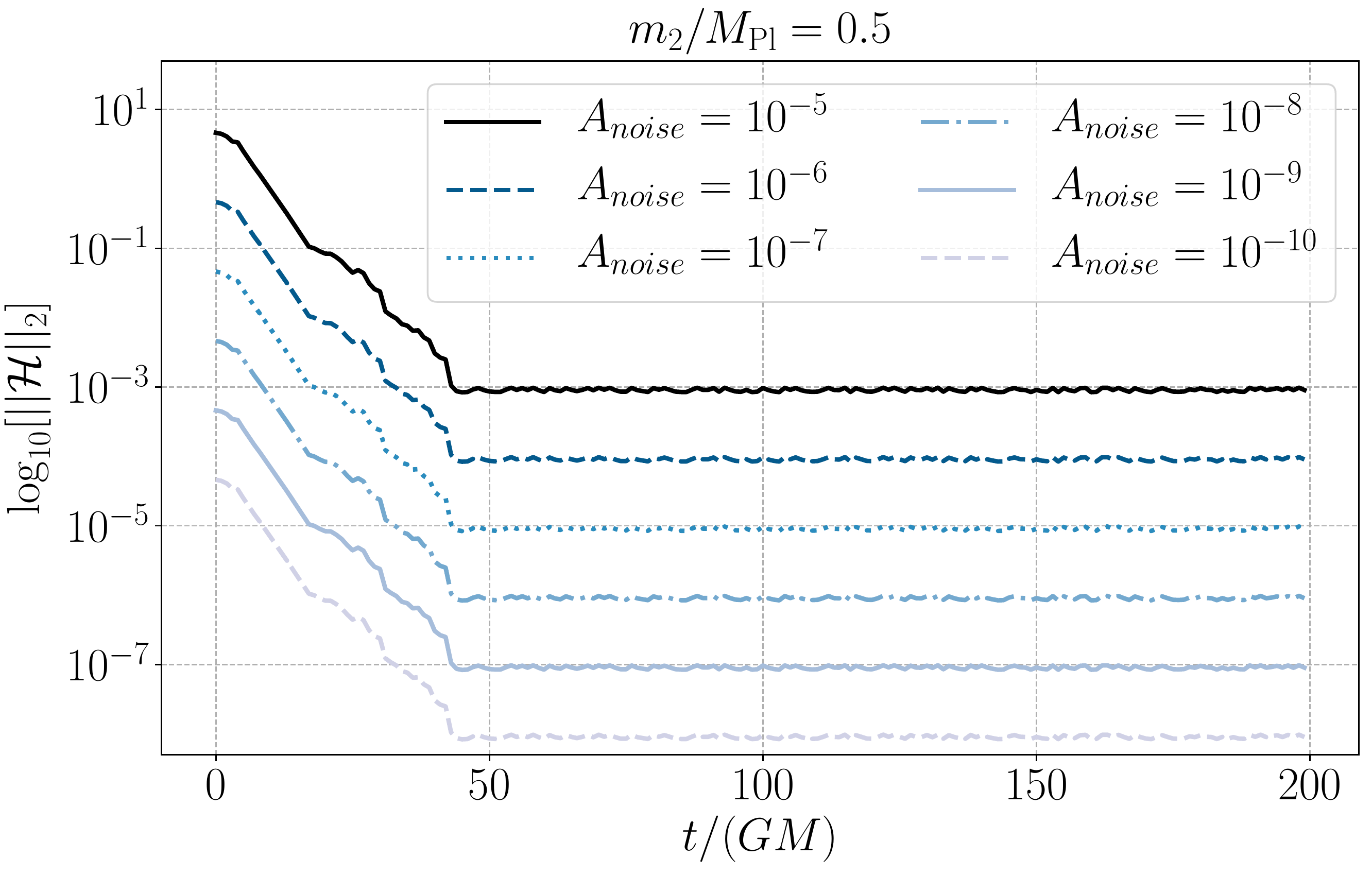}
    \caption{
    \label{fig:single-BH-noise-test}
    Constraint plot ($l2$-norm of the Hamiltonian constraint in \cref{eq:Hamiltonian-constraint-York}) for Kerr initial data with mass $M=1$ and spin $a=0.01$, performing a noise test with different noise amplitudes, ranging from $10^{-5}$ to $10^{-10}$, top to bottom. Each curve represents an increase by a factor of ten in the initial amplitude over the curve below.
    }\end{figure}
To confirm numerical stability, we evolve a single Kerr black hole, perturbed only by numerical noise. We test with puncture initial data~\cite{Steven1997} for a Kerr black hole. Note that the additional QG variables are vanishing since Kerr is a Ricci-flat vacuum solution. 
The Kerr black hole is expressed in Kerr-Schild coordinates such that
\begin{align}
    \label{eqn:kerr}
    ds^2 = (\eta_{ab} + 2H k_a k_b) dx^a dx^b
\end{align}
where $\eta_{ab}$ is usual Minkowski spacetime and 
\begin{align}
    \label{eqn:kerr:component}
    H &= \frac{G\,M\,r}{r^2 + a^2 (z/r)^2}\;, \\
    k_a dx^a &= -dt - \frac{r(xdx + ydy) - a(xdy -ydx)}{r^2+a^2} - \frac{zdz}{r}\;.
\end{align}
Here, $M$ is the black-hole mass and $a$ is the spin parameter. In 3+1 form, we also have
\begin{align}
    \label{eqn:kerr:3p1}
    \alpha &=1/\sqrt{1+2Hk_0 k_0}\;,\\
    \beta_i &= 2H k_0 k_i\;,\\
    \gamma_{ij} &= \delta_{ij} + 2H k_i k_j\;,
\end{align}
and the extrinsic curvature can be obtained as
\begin{align}
    \label{eqn:kerr:Kij}
    K_{ij} = \frac{D_i \beta_j + D_j \beta_i}{2\alpha}\;.
\end{align}
The Kerr-Schild form is a horizon penetrating coordinate system such that there are no coordinate singularities in $\gamma_{ij}$ and $K_{ij}$ at the horizon. Kerr-Schild coordinates cover both the outside and the inside of the black hole. Since Kerr spacetime is Ricci flat, $\mathcal{R}$, $\hat{\mathcal{R}}$, $\mathcal{A}$, $\mathcal{A}_{ij}$, $\mathcal{B}$, and $\mathcal{B}_{ij}$ are initialized as zero. 

We aim to test for numerically (un)stable behavior of the time evolution. In anticipation of the presence of a linear instability in part of the parameter space, cf.~\cref{sec:Gregory-Laflamme}, we choose mass values for which the instability is not relevant. 

To perform a numerical stability test, we add random noise to all components of the initial data such that
\begin{align}
    \label{eqn:id_noise}
    \mathbf{u}(t=0) = \mathbf{u}_0 + A_\textrm{noise} \textrm{RAND}(x)
\end{align}
where $\mathbf{u}=(\gamma_{ij},K_{ij},\mathcal{R},\hat{\mathcal{R}},\mathcal{A},\mathcal{A}_{ij},\mathcal{B},\mathcal{B}_{ij})$ is the state vector for all the evolution variables, $A_\textrm{noise}$ is a noise amplitude which we vary from $10^{-10}$ to $10^{-5}$, and $\textrm{RAND}(x)$ is a random function that generates random values between $-1$ and $1$.
The result is summarized in \cref{fig:single-BH-noise-test}. We find no indication for numerical instability in our evolution scheme. The same holds for all subsequent simulations. The respective constraint plots are presented in \cref{app:convergence}.

\section{Results: Stability of the Ricci-flat subsector of Quadratic Gravity}
\label{sec:results}

In this section, we present our results on the Ricci-flat subsector of Quadratic Gravity. The physical upshot is twofold: first, we recover a well-known linear instability associated to massive spin-2 excitations; second, we demonstrate that -- aside from this linear instability -- even fully dynamical, Ricci-flat solutions like a binary merger seem to be nonlinearly stable.

\subsection{Recovering the linear instability in nonlinear evolution}
\label{sec:Gregory-Laflamme}

%
\begin{figure}[!t]
    \centering
    \includegraphics[trim={0cm 0cm 0cm 0cm},clip,width=\linewidth]{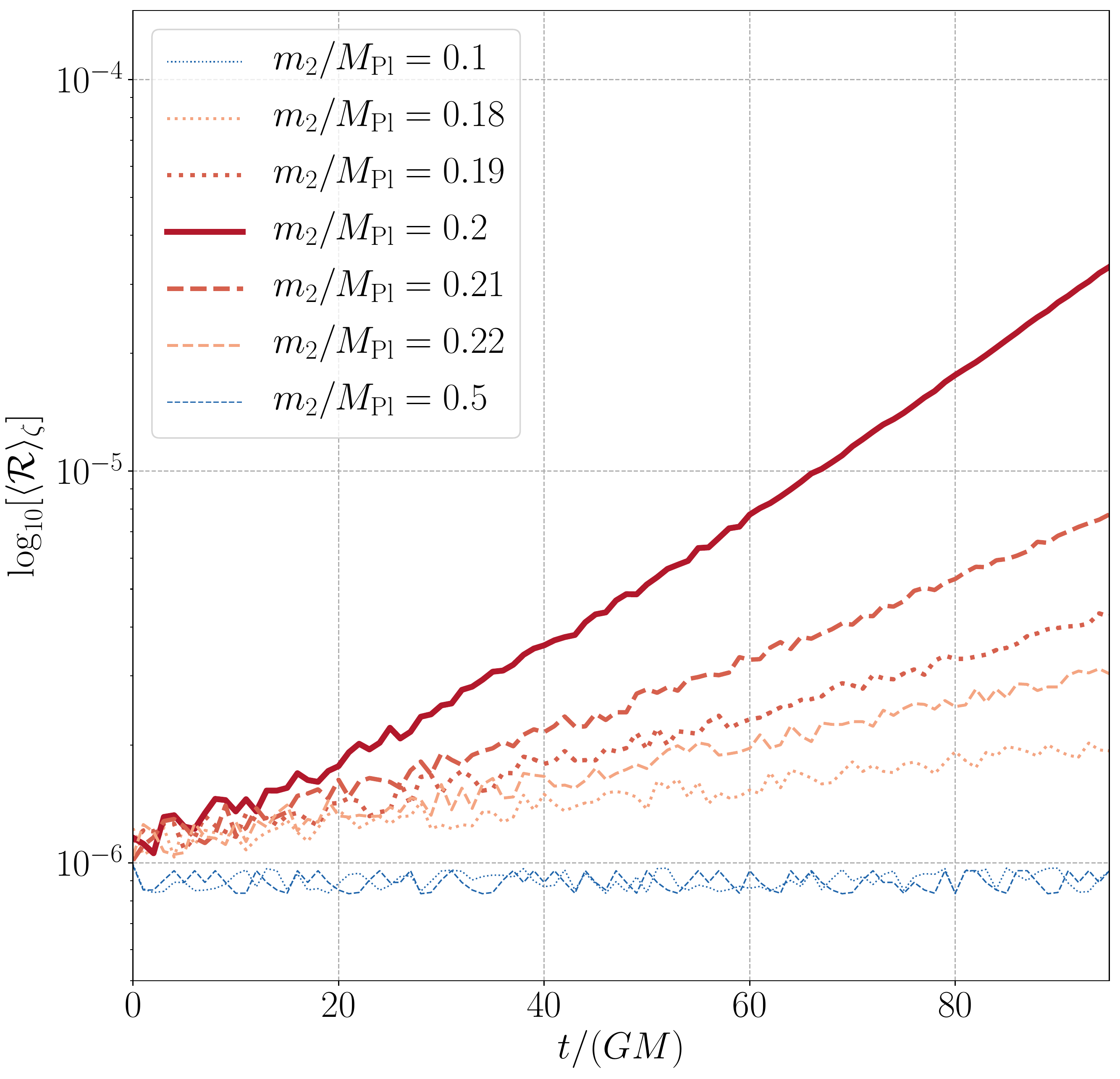}
    \caption{
        To verify the presence of the linear instability, we show the evolution of the spatially averaged Ricci scalar $\log_{10}[\langle \mathcal{R}_\zeta\rangle]$ as a function of evolution time. The spatial average is taken within a cube of edge length $2\zeta$.
    }
\label{fig:GL}
\end{figure}

It is known from the linearized dynamics that a single Schwarzschild black hole can be subject to a linear
instability~\cite{Brito:2013wya,Held:2022abx}, akin to (i.e., linearly equivalent with) the long-wavelength Gregory-Laflamme instability of higher-dimensional black strings~\cite{Gregory:1993vy}. In QG, the onset and the timescale of this instability are determined by the mass $m_2$ of the massive spin-2 degree of freedom and by the gravitational radius $r_g=2\,GM$ of the Schwarzschild black hole~\cite{Brito:2013wya,Held:2022abx}.

In particular, the instability occurs whenever
\begin{align}
    2\,GM\,m_2 \equiv p < p_\text{crit} \approx 0.87\;.
\end{align}
If this inequality is fulfilled, then there exists a 
linear mode which grows like $\sim e^{\text{Im}(\omega) t}$. The exponential growth rate is set by
\begin{align}
    \text{Im}(\omega) = 
    \frac{q(p)}{2\,GM} = 
    m_2\,\frac{q(p)}{p}\;,
\end{align}
where $q(p)$ is a concave function which has been determined numerically (see, e.g.,~\cite[Fig.~2]{Held:2022abx}) and is bounded by
\begin{align}
   q(p) < q_\text{max} = q(p_\text{max}) \approx 0.1\;,
\end{align}
with $p_\text{max}\approx0.4$. Moreover, $\lim_{p\rightarrow p_\text{crit}}q(p) = 0$ and the numerical results indicate that also $\lim_{p\rightarrow 0}q(p) = 0$. Equivalently, the instability timescale in units of the black-hole mass is given by
\begin{align}
    \frac{t_\text{GL}}{GM} \sim
    \frac{1}{GM\,\text{Im}(\omega)} = 
    \frac{2}{q(p)} \gtrsim 20\;.
\end{align}
This means that, with regards to the 
linear instability, there are three different regimes:
\begin{itemize}
    \item If $2\,GM\,m_2 > p_\text{crit}$, no 
linear instability is present.
    \item If $2\,GM\,m_2 \ll p_\text{max}$, the single Schwazschild black hole exhibits a 
linear instability but the exponential growth rate is comparatively slow.
    \item At $2\,GM\,m_2 \approx p_\text{max}$, the exponential growth rate of the 
linear instability is maximized, growing $e$-fold roughly every $t_\text{GL} \approx 20\,GM$.
\end{itemize}
We probe and recover this instability within our numerical evolution. As in \cref{sec:numerical-stability}, we initialize a single Schwarzschild black hole. We detect the instability by calculating the spatially averaged 
Ricci scalar
$\langle\mathcal{R}\rangle_\zeta$ where the spatial average is taken over a cube with $x,\,y,\,z\in[-\zeta,+\zeta]$ and $\zeta=200\,GM$ extends across the full computational domain.

If the instability is present, a non-vanishing  $\langle\mathcal{R}\rangle_\zeta$ is excited by the numerical noise floor in the initial data.

We probe the three different regimes identified above, cf.~\cref{fig:GL}, and find agreement with the expectation from the linear analysis. 
In particular, at $2\,GM\,m_2 \approx p_\text{max}$, we recover the expected timescale of the linear instability.
This also means that we can exclude the presence of further growth modes with a faster timescale.
We thus conclude that the unstable monopole mode identified in the linear analysis is indeed the dominant unstable mode. 

As we demonstrate in \cref{fig:GL}, the linear instability breaks Ricci flatness. Nevertheless, we find that the evolution remains numerically stable, cf.~\cref{fig:constraint_GL} in \cref{app:convergence}. In particular, the constraint violations remain small, even in the presence of a substantial breaking of Ricci flatness. We thus find no indication that well-posed evolution is restricted to the Ricci-flat sector.

We also note that exponential growth -- as expected from the linear analysis -- corresponds to straight lines, given the log-scale in \cref{fig:GL}. Hence, our numerical simulations are in agreement with the linear analysis. Prolonged nonlinear evolution will allow us to clarify the nonlinear fate of the instability. We plan to report on this in future work.

Moreover, the numerical evolution can straightforwardly be extended to rotating Kerr initial data. This allows to numerically explore a potential onset of physical instability for spinning black holes, where only partial results are known in the linearized regime~\cite{Brito:2013wya, Brito:2020lup}.
\\

Having recovered the Gregory-Laflamme-type instability, from here on, we work in the regime in which this linear instability does not occur. In this regime, we expect that a single (Schwarzschild) black hole is stable. In the following two sections, we investigate physical Ricci-flat perturbations. First, in \cref{sec:Teukolsky}, we perturb a single black hole by a gravitational (Teukolsky) wave. Then, in \cref{sec:binary}, we investigate a full binary merger.

\subsection{Physical perturbations: Teukolsky waves}
\label{sec:Teukolsky}
%
\begin{figure}[!t]
    \centering
    \includegraphics[trim={0cm 0cm 0cm 0cm},clip,width=\linewidth]{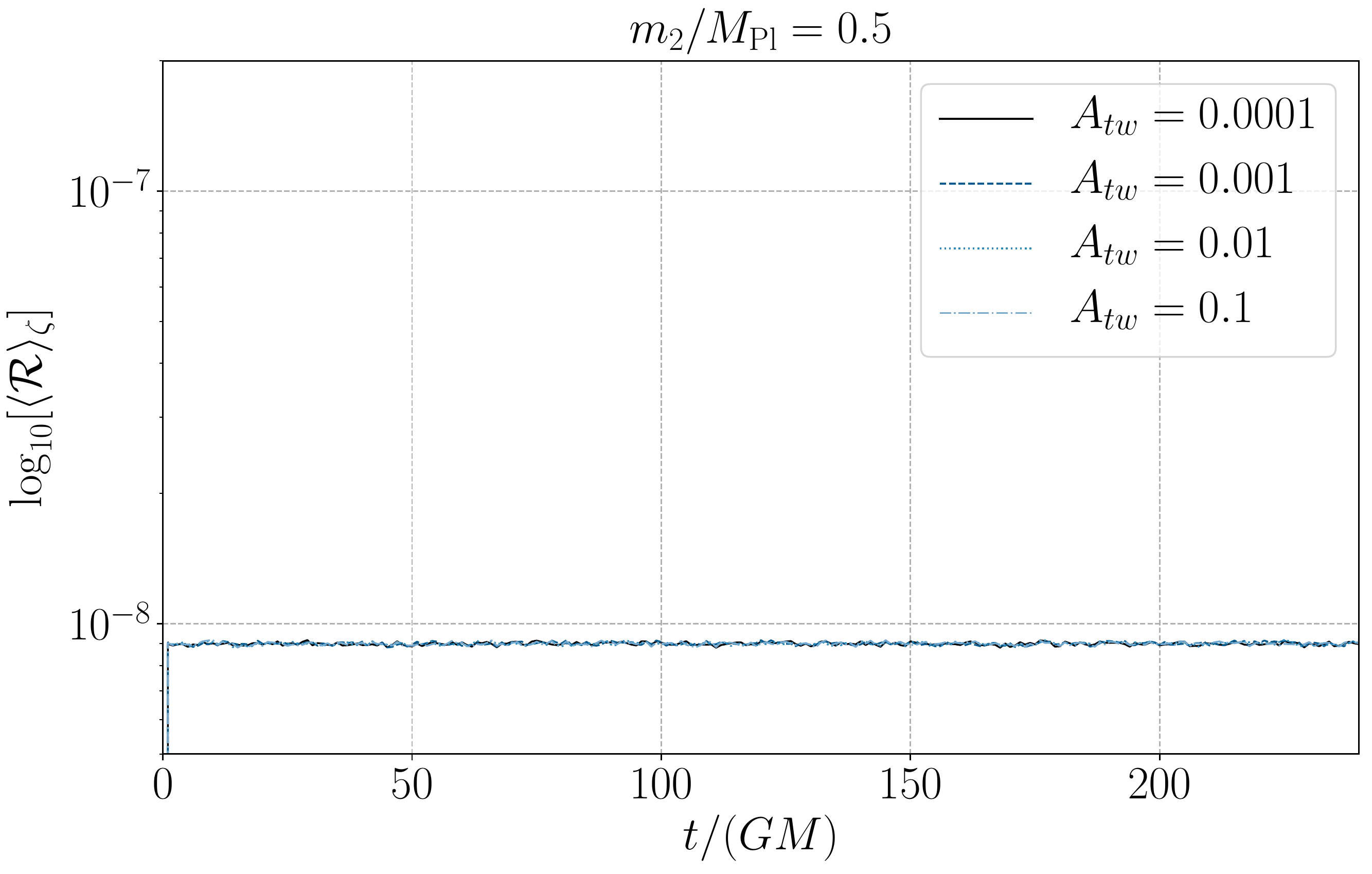}
\caption{
    We verify that Schwarzschild initial data subject to an incident Teukolsky wave remains Ricci flat. Different lines show different magnitudes $A_{tw}$ of the incident Teukolsky wave (cf. legend). The black hole is placed at $x=y=z=0$ and without initial velocity. The Teukolsky wave is initialized at $x=50M$, $y=z=0$ such that it interacts with the black hole at roughly $t/M=50$. We plot the spatially averaged Ricci scalar $\log_{10}[\langle \mathcal{R}\rangle_\zeta]$.
    }
\label{fig:tw_bh_ricci}
\end{figure}

In the previous section, we have recovered the well-known 
linear instability of Schwarzschild black holes in QG. In particular, we have demonstrated how the instability -- if present and with sufficiently fast growth rate -- is excited by the numerical noise floor. 
In the present section, we now separate Ricci-flat physical perturbations from the noise floor. We emphasize that while we consider small perturbations, we nevertheless solve the nonlinear evolution. Constructing initial data which corresponds to physical excitations of modes that break Ricci flatness (as, e.g., the mode that excites the 
linear instability in the previous section) is thus nontrivial since it requires to solve the modified nonlinear constraints. In contrast to the previous section, we, therefore, focus on Ricci-flat perturbations only. For the latter, we can construct initial data just like in GR, once more, making use of the fact that every Ricci-flat solution to GR is also a solution to QG.

There are various ways to construct gravitational-wave initial data, see~\cite{Teukolsky:waves} for Teukolsky waves which correspond to purely quadrupolar gravitational-wave excitations and~\cite{1959AnPhy...7..466B} for the nonlinear construction of Brill waves which correspond to a tower of multipole modes. We specify to Teukolsky waves and adopt Cartesian coordinates in the following.

By construction, Teukolsky waves satisfy the nonlinear momentum constraint.
We follow the standard procedure~\cite{Teukolsky:waves,Baumgarte:2010ndz,Leo2017,Fernandez:2021ckg} to ensure that initial data also satisfies the nonlinear Hamiltonian constraint, i.e., we employ the spatial part of the metric as a conformally related metric in the Hamiltonian constraint and then solve this equation for the conformal factor, i.e., for $\phi$ in our case (cf.~\cref{app:BSSN}). More details of the Teukolsky wave initial data can be found in~\cite{Leo2017,Fernandez:2021ckg}.
\\

We initialize the black hole at the origin of the computational domain and without initial velocity. The Teukolsky wave perturbation is initialized at $50\,GM$ distance to the black hole, from where it propagates radially in all directions. We evolve the resulting simulation up to $t=250\,GM$ such that the evolution time encompasses how the Teukolsky wave interacts with the black hole. In order to confirm physical stability of the Ricci-flat subsector, we show the spatially averaged Ricci scalar $\langle\mathcal{R}\rangle_\zeta$ in \cref{fig:tw_bh_ricci}. Clearly, the Ricci scalar remains vanishing up to numerical noise fluctuations. In particular, the latter noise floor is well separated from the amplitude of physical Teukolsky-wave perturbations $A_\text{tw}$ which are up to $10^7$ times larger, cf. the legend in~\cref{fig:tw_bh_ricci}.

We conclude that, even with significant Ricci-flat perturbations, QG exhibits a stable subsector which mimics vacuum GR. To probe this conclusion further, we now proceed to the fully nonlinear regime of a binary merger.

\subsection{Stability during nonlinear binary evolution}
\label{sec:binary}

%
\begin{figure}[!t]
    \centering
    \includegraphics[trim={0cm 0cm 0cm 0cm},clip,width=\linewidth]{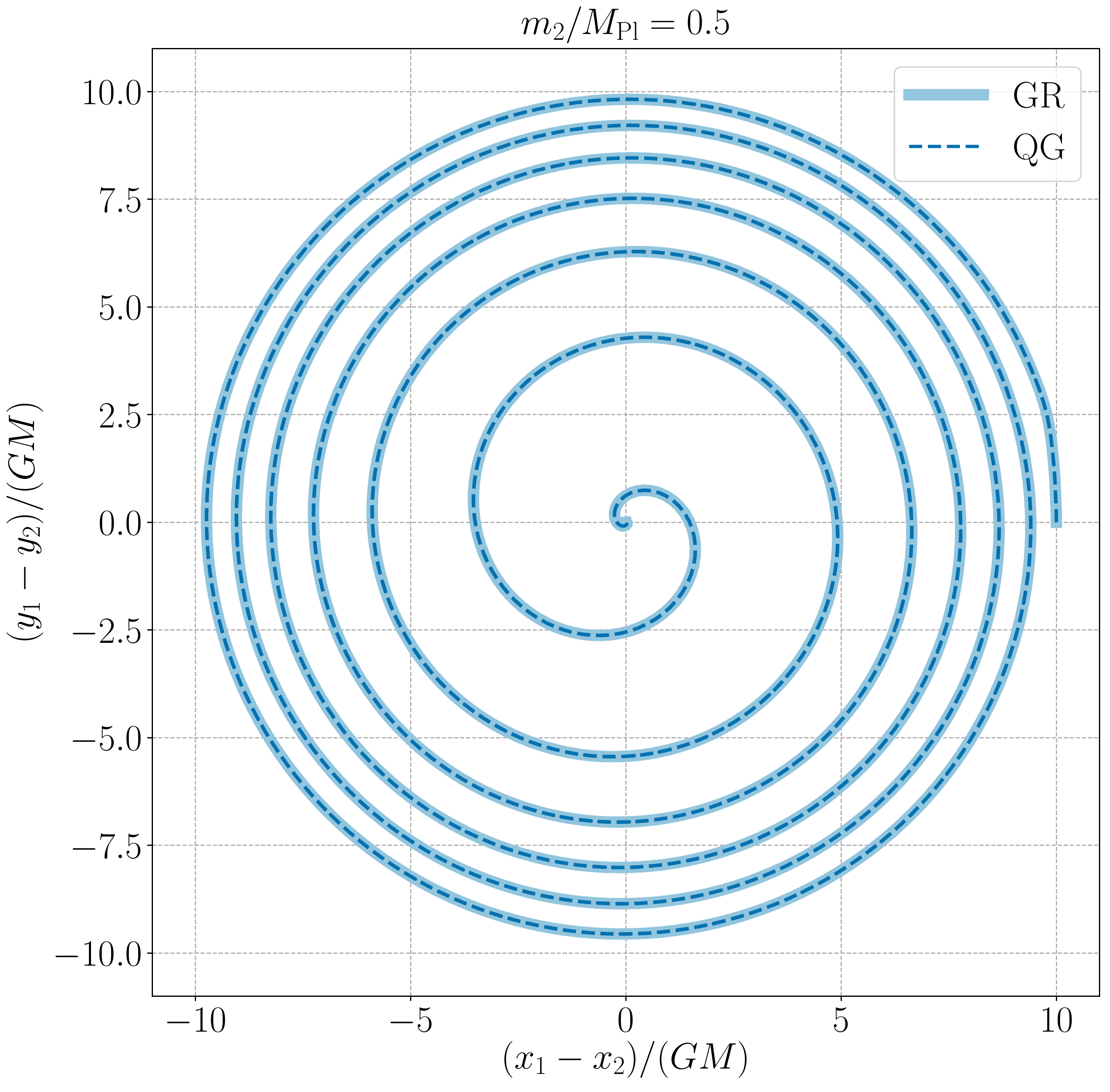}
\caption{We show the trajectory comparison between GR and QG. The evolution captures the last 6 orbits and the merger of a binary-black-hole system for which the initial data matches the one inferred from GW150914.}
\label{fig:bbh}
\end{figure}

From the astrophysical perspective, one of the most interesting questions is to study the evolution of binary systems and the resulting gravitational wave emission. A continuously growing catalog~\cite{LIGOScientific:2018mvr,LIGOScientific:2020ibl,LIGOScientific:2021djp} of gravitational-wave events is being detected by the LIGO/Virgo collaboration. At the same, when binary systems come close to merger, they probe the fully nonlinear regime of the theory and may thus reveal otherwise hidden deviations from GR. One of the possible deviations are the quadratic-curvature corrections investigated in this work, see also~\cite{Witek2019,East2021b,Corman2023,Cayuso:2023aht} for the evolution of binary systems with the inclusion of other (related) deviations from GR. 

Eventually, one would like to compare the theoretical predictions for the extracted gravitational-wave form in GR and in QG (or beyond-GR more generally). However, the previous section suggests that the vacuum sector of QG is fully equivalent to the vacuum sector of GR. If this holds true in the fully nonlinear regime, QG can mimic any binary black-hole (BBH) system and, in particular, the respective gravitational-wave forms obtained in GR. Indeed, this is what we find (see below). Hence, the relevant constraints on QG will likely come from non-vacuum systems and we plan to address this in future work.
\\

As a specific binary example, we use Bowen-York initial data~\cite{bowen1979general,1989fnr..book...89Y}, approximating a binary system which has been matched to the GW150914 LIGO/Virgo event~\cite{Abbott2016}. The respective binary parameters are taken from the \texttt{EinsteinToolkit} library~\cite{Loffler2011ay,Wardell2016}. Since the physical initial data is Ricci-flat, we initialize all the additional QG variables with vanishing values. 

We then track the lapse function to extract the motion of the respective black holes. The trajectory comparison in \cref{fig:bbh} confirms our expectation that the two evolutions are fully equivalent. Once more, we find evidence that QG exhibits a physically stable Ricci-flat subsector which is fully equivalent to GR.

As mentioned above, the obvious next physical question concerns an extension to non-vacuum (and hence non-Ricci-flat) binary systems. In contrast to the present initial data, the fiducial Ricci variables $\mathcal{R}$, $\hat{\mathcal{R}}$, $\mathcal{A}$, $\mathcal{A}_{ij}$, $\mathcal{B}$, and $\mathcal{B}_{ij}$ (see \cref{sec:summary-of-evolution-eqs}), corresponding to the massive spin-0 and the massive spin-2 degrees of freedom, will then, presumably, be excited. We thus expect non-vacuum binary systems, e.g., neutron stars, to show appreciable differences to GR and, therefore, expect the respective waveforms to constrain the quadratic-curvature deviations from GR. All of this comes with the question whether new instabilities arise in the non-vacuum sector of QG. We will address the non-vacuum sector in a separate publication.

\section{Discussion}
\label{sec:conclusion}

We derive a (3+1) evolution system for the nonlinear gravitational dynamics of quadratic-curvature corrections to General Relativity (GR), i.e., for Quadratic Gravity (QG). After verifying numerical stability, we use the nonlinear evolution to verify the nonlinear stability of a Ricci-flat subsector of QG which can mimic GR.

\subsection{Key results}

The key to well-posed nonlinear evolution is based on Noakes' insight~\cite{Noakes:1983} that the Ricci scalar and traceless Ricci tensor can be treated as fiducial variables representing the additional degrees of freedom.
\\

We find that it is possible to solve part of the constraint system algebraically such that we reduce the number of redundant evolution variables.

As for GR, in the metric sector, we evolve twelve $1^\text{st}$-order variables, i.e., the spatial metric $\gamma_{ij}$ and the extrinsic curvature $K_{ij}$, which represent the two degrees of freedom associated with the massless spin-2 graviton.

In the trace sector, the Ricci scalar $\mathcal{R}$ (and its $1^\text{st}$-order variable $\hat{\mathcal{R}}$) correspond directly to an additional massive spin-0 degree of freedom.

In the traceless sector, the spatial part of the traceless Ricci tensor -- which we decompose into a 3-trace and 3-traceless part $\mathcal{A}$ and $\mathcal{A}_{ij}$, respectively -- and the respective $1^\text{st}$-order variables $\mathcal{B}$ and $\mathcal{B}_{ij}$ all-together propagate another twelve pieces of initial data. Two of these are redundant but we do not find an obvious way to remove this redundancy analytically. Overall, these variables correspond to the 5 degrees of freedom of the massive spin-2 mode. 

The respective evolution system, summarized in \cref{sec:summary-of-evolution-eqs}, can be understood as the QG equivalent of the ADM equations for GR, cf.~\cite{York:2014}. In fact, the evolution system contains the standard ADM equations in which the higher-derivative variables appear as fiducial matter sources. Minimally coupled physical matter sources enter the evolution system via the higher-derivative sector. 
\\

We then treat the metric sector as in the BSSN formalism~\cite{Shibata:1995we,Baumgarte:1998te} and verify that the evolution of the resulting system of PDEs is numerically stable. After verifying numerical stability (which we also continue to check throughout all subsequent numerical evolutions, cf.~\cref{app:convergence}), we investigate the physical stability of the Ricci-flat (GR vacuum) subsector of the theory, and find:
\begin{itemize}
    \item
    Our nonlinear results recover a well-known linear instability of Schwarzschild black holes~\cite{Brito:2013wya,Held:2022abx}. At the linear level, this instability is fully equivalent to the Gregory-Laflamme instability~\cite{Gregory:1993vy}. It occurs only if both the spin-2 mass $m_2$ and the black-hole mass $M$ are sufficiently small (in comparison to the Planck mass), i.e., if $\frac{1}{4\pi}\frac{m_2}{\MPl}\frac{M}{\MPl}<0.87$.
    \item
    Aside from this linear instability, we find that, both, physical metric perturbations (e.g., Teukolsky waves as presented in \cref{sec:Teukolsky}) and the fully nonlinear Ricci-flat evolution (e.g., a binary merger as the one presented in \cref{sec:binary}) are physically stable.
\end{itemize}
The latter result is quite nontrivial and suggests that -- at least in parameter ranges for which the Gregory-Laflamme-type instablity is either not present or negligibly small -- QG exhibits a physically stable Ricci-flat subsector. In particular, this suggests that QG can mimic all of the vacuum physics of GR.

\subsection{Outlook}


The presence of a linear instability raises the question of its nonlinear endpoint and the relation to cosmic censorship. (See~\cite{Choptuik:2003qd,Lehner:2010pn,Figueras:2022zkg} for numerical investigation of the nonlinear fate of the Gregory-Laflamme instability for higher-dimensional black strings.) 

More generally, the global stability (i.e., the absence of runaway solutions) and the local stability (i.e., the identification of Lyapunov stable vacua) of Quadratic Gravity are yet to be determined, see also~\cite{Hindawi:1995cu}. We note that stable motion and ghost-like degrees of freedom may not be mutually exclusive~\cite{Deffayet:2021nnt,Deffayet:2023wdg}. With the nonlinear evolution system at hand, we are well-equipped to numerically investigate these questions in future work.
\\

The apparent nonlinear stability of the Ricci-flat sector raises the question how the theory behaves if minimally coupled matter is added to the system. Are there also stable regimes of the non-vacuum theory? If so, is there a stable sector of the theory which deviates appreciably from General Relativity?
As our evolution system already includes matter terms, we plan to also address this question in future work. The key difficulty will be to construct consistent (as in obeying all of the modified constraint equations) initial data for the non-Ricci-flat sector of Quadratic Gravity.
\\

Overall, the numerical stability of the presented evolution system gives access to the fully nonlinear sector of Quadratic Gravity. Moreover, the presented treatment of quadratic-curvature corrections may also inform how to achieve fully stable nonlinear evolution when curvature corrections of yet higher order are present. In particular, any gravitational theory constructed only from Riemann curvature scalars (i.e., scalars formed solely from contractions of the Riemann curvature tensor, in particular, not involving additional covariant derivatives) still maintains fourth-order equations of motion~\cite{Hindawi:1995cu,Bueno:2016ypa}. This suggests that similar techniques to the ones presented here may also apply to a much wider class of gravitational theories, for instance, to the cubic and/or quartic theory~\cite{Endlich:2017tqa,Cayuso:2023aht}.
\\

\paragraph*{Acknowledgements.}
We thank Pau Figueras and Frans Pretorius for many helpful discussions. The work leading to this publication was supported by the PRIME programme of the German Academic Exchange Service (DAAD) with funds from the German Federal Ministry of Education and Research (BMBF). AH acknowledges support by the Deutsche Forschungsgemeinschaft (DFG) under Grant No 406116891 within the Research Training Group RTG 2522/1.
HL is supported by the LANL ASC Program and LDRD grant 20230555ER. This work used resources provided by the LANL Darwin testbed. Darwin is a research testbed/heterogeneous cluster funded by the Computational Systems and Software Environments subprogram of ASC program. LANL is operated by Triad National Security, LLC, for the National Nuclear Security Administration of the U.S.DOE  (Contract No. 89233218CNA000001). This work is authorized for unlimited release under LA-UR-23-23440

\appendix

\section{Gauss-Codazzi-Ricci equations}
\label{app:GaussCodazzi}

The Gauss-, Codazzi-, and Ricci equations are of purely geometric nature. They determine the foliation and are therefore independent of the dynamics, i.e., valid both in GR and QG. They follow from the (3+1) decomposition of the Riemann tensor, i.e., from
\begin{align}
    R_{a{cb}d} =&\;
    {}^{(3)}\!R_{a{cb}d} 
    +2\,K_{a[{b}}K_{d]{c}}
    +4\,n_{[a}\,K^e_{{c}]}\,K_{e[{b}}n_{d]}
    \nonumber\\*&
    +4\left(D_{[a}K_{{c][b}}\right)n_{d]}
    +4\left(D_{[{b}}K_{d][a}\right)n_{{c}]}
    \nonumber\\*&
    +4\,\left(\gamma^f_{[a}n_{{c}]}\gamma^g_{[{b}}n_{d]}\right)\,
    n^e\left(\nabla_eK_{fg}\right)\;,
    \nonumber\\*&
    + 4\,a_{[a}n_{c]}a_{[b}n_{d]}
    - 4\,n_{[b}\left(D_{d]}a_{[a}\right)n_{c]}
    \;.
\end{align}
Projecting the decomposition onto the respective temporal and spatial indices (and specifying to the $(3+1)$ coordinate conventions in~\cref{eq:coordinate-conventions}) results in the Gauss-, Codazzi-, and Ricci-equation, respectively, i.e.,
\begin{align}
    \gamma_a^e \gamma_b^f \gamma_c^g \gamma_d^h R_{efgh} &=
    {}^{(3)}\!R_{acbd}
    +2\,K_{a[c}K_{d]b}\;,
    \\[0.4em]
    \gamma_a^e \gamma_b^f \gamma_c^g n^d R_{efgd} &= 
    -2\,D_{[a}K_{b]c}\;,
    \\
    \gamma_b^e \gamma_d^f n^a n^c R_{aecf} &= 
    \mathcal{L}_nK_{bd} 
    + K^e_b K_{de} 
    + \frac{1}{\alpha}D_bD_d\alpha\;.
\end{align}
Other contractions with two normal vectors are either equivalent (by the symmetries of the Riemann tensor) to the above or vanish. All contractions with more than two normal vectors also vanish.

\section{Decomposition of $\nabla_a\nabla_b\mathcal{R}$}
\label{app:DDR}

Here, we detail the split of $\nabla_a\nabla_b \mathcal{R}$ into spatial and temporal part. We start from
\begin{align}
    \nabla_a\nabla_b \mathcal{R} &= g^c_a \nabla_c (g^d_b \nabla_d \mathcal{R}) \nonumber \\
    &= (\gamma^c_a - n_a n^c) \nabla_c [(\gamma^d_b - n_b n^d) \nabla_d \mathcal{R}] \nonumber \\
    &= 
    + \underbrace{\gamma^c_a \nabla_c (\gamma^d_b \nabla_d \mathcal{R})}_\text{(I)}
    + \underbrace{n_a n^c \nabla_c (n_b n^d \nabla_d \mathcal{R})}_\text{(II)}
    \nonumber\\*&\quad
    - \underbrace{\gamma^c_a \nabla_c(n_b n^d \nabla_d \mathcal{R})}_\text{(III)}
    - \underbrace{n_a n^c \nabla_c (\gamma^d_b \nabla_d \mathcal{R})}_\text{(IV)} 
    \nonumber\;,
\end{align}
and look at each term individually, i.e.,
\begin{align}
    \text{(I)} &= \gamma^c_a \nabla_c (\gamma^d_b \nabla_d \mathcal{R}) 
    \equiv 
    D_a D_b \mathcal{R}
    \; ,
    \nonumber\\[0.5em]
    \text{(II)} &= 
    n_a n^c \nabla_c (n_b n^d \nabla_d \mathcal{R})
    \nonumber\\*&
    =
    -n_a n_b \left(n^c \nabla_c \hat{\mathcal{R}}\right)
    -n_a a_b \hat{\mathcal{R}}
    \;,
    \nonumber\\[0.5em]
    \text{(III)} 
    &= 
    \gamma^c_a \nabla_c(n_b n^d \nabla_d \mathcal{R}) 
    =
    - n_b D_a \hat{\mathcal{R}}
    - \gamma^c_a (\nabla_cn_b) \hat{\mathcal{R}}
    \nonumber\\*&
    =
    - n_b D_a \hat{\mathcal{R}} 
    + K_{ab} \hat{\mathcal{R}}
    \nonumber\; ,
\end{align}
where we have introduced the acceleration $a_b\equiv n^c\nabla_c n_b$ and inserted the definition of $\hat{\mathcal{R}}\equiv -n^a\nabla_a\mathcal{R}$.

Finally, term (IV) can be rewritten by commuting covariant derivatives, i.e.,
\begin{align}
    \text{(IV)}
    &= 
    n_an^c \nabla_c (\gamma^d_b \nabla_d \mathcal{R})
    \nonumber\\*&
    =
    n_a(n^c \nabla_c \gamma^d_b)(\nabla_d \mathcal{R})
    + n_a\gamma^d_b\,n^c \nabla_c\nabla_d \mathcal{R}
    \nonumber\\*&
    =
    n_a n^c (n^d\nabla_cn_b + n_b\nabla_cn^d)(\nabla_d \mathcal{R})
    +n_a\gamma^d_b\,n^c \nabla_d\nabla_c \mathcal{R}
    \nonumber\\*&
    =
    - n_a\,a_b\,\hat{\mathcal{R}}
    - n_a n_b\,a_c D^c\mathcal{R}
    + n_a\gamma^d_b\,\nabla_d(n^c\nabla_c \mathcal{R})
    \nonumber\\*&\quad
    - \gamma^d_b\,(n_a\nabla_d n^c)(\nabla_c \mathcal{R})
    \nonumber\\*&
    =
    - n_a\,a_b\,\hat{\mathcal{R}}
    - n_a n_b\,a_c D^c\mathcal{R}
    - n_aD_b\hat{\mathcal{R}}
    \nonumber\\*&\quad
    + \gamma^d_b \,(n^c\nabla_d n_a)(\nabla_c \mathcal{R})
    - \gamma^d_b\,(\nabla_d \gamma_a^c)(\nabla_c \mathcal{R})
    \nonumber\\*&
    =
    - n_a\,a_b\,\hat{\mathcal{R}}
    - n_a n_b\,a_c D^c\mathcal{R}
    - n_aD_b\hat{\mathcal{R}}
    + K_{ab}\hat{\mathcal{R}}
    \nonumber\;,
\end{align}
where we have twice used that $0=\nabla_d g^c_a = \nabla_d(\gamma^c_a - n_an^c) = \nabla_d\gamma^c_a - n_a\nabla_dn^c -n^c\nabla_dn_a$\;.
Note that there are no remaining temporal derivatives in any of these terms. Collecting results, we find
\begin{align}
    \nabla_a\nabla_b \mathcal{R} &=
    D_a D_b \mathcal{R}
    + 2\,n_{(a}D_{b)}\hat{\mathcal{R}}
    - 2\,K_{ab} \hat{\mathcal{R}}
    \nonumber\\*&\quad
    - n_a n_b \left(n^c \nabla_c \hat{\mathcal{R}}\right)
    + n_a n_b\,a_c D^c\mathcal{R}\;,
\end{align}
which is also given in the main text.

\section{Decomposition of $\Box\widetilde{\mathcal{R}}_{ab}$}
\label{app:boxRab}

Here, we detail the split of $\Box\widetilde{\mathcal{R}}_{ab}$ into spatial and temporal part. We start from
\begin{align}
    \Box \widetilde{\mathcal{R}}_{ab} &=  
    - \underbrace{
        \gamma\indices{^d_c} \nabla_d (n^c n^e \nabla_e \widetilde{\mathcal{R}}_{ab})
    }_\text{(I)}
    + \underbrace{
        n_c n^d \nabla_d ( n^c n^e \nabla_e \widetilde{\mathcal{R}}_{ab})
    }_\text{(II)}
    \nonumber\\*&\quad
    + \underbrace{
        n_c n^d \nabla_d (\gamma^{ce} \nabla_e \widetilde{\mathcal{R}}_{ab})
    }_\text{(III)}
    - \underbrace{
        \gamma\indices{^d_c} \nabla_d (\gamma^{ce} \nabla_e \widetilde{\mathcal{R}}_{ab})
    }_\text{(IV)}
    \nonumber\;,
\end{align}
project the covariant derivatives onto spatial and temporal part, and look at each term individually. In the first two terms, we can introduce the first-order fiducial variable $\widetilde{V}_{ab} = - n^c \nabla_c \widetilde{\mathcal{R}}_{ab}$ to find
\begin{align}
    \text{(I)} &=
    \gamma\indices{^d_c} \nabla_d (n^c n^e \nabla_e \widetilde{\mathcal{R}}_{ab}) =
    - \gamma\indices{^d_c} \nabla_d \left(n^c \widetilde{V}_{ab}\right) = 
    K \widetilde{V}_{ab}
    \;,
    \nonumber\\
    \text{(II)} &=
    n_c n^d \nabla_d ( n^c n^e \nabla_e \widetilde{\mathcal{R}}_{ab}) = 
    - n_c n^d \nabla_d ( n^c \widetilde{V}_{ab})
    \nonumber\\*&=
    n^d\nabla_d V_{ab}
    \nonumber\;,
\end{align}
For the third term, we find
\begin{align}
    \text{(III)} &= n_c n^d \nabla_d (\gamma^{ce} \nabla_e \widetilde{\mathcal{R}}_{ab}) = 
    n_c n^d (\nabla_d \gamma^{ce}) (\nabla_e \widetilde{\mathcal{R}}_{ab})
    \nonumber\\*&=
    n_c n^d (\nabla_d n^c n^e) (\nabla_e \widetilde{\mathcal{R}}_{ab}) = a^e\nabla_e\widetilde{\mathcal{R}}_{ab}
    \nonumber\\*&=
    a^e\gamma_e^c\nabla_c\widetilde{\mathcal{R}}_{ab} \equiv
    a^c D_c\widetilde{\mathcal{R}}_{ab}
    \nonumber\;.
\end{align}
Here, as well as in the fourth term,
\begin{align}
    \text{(IV)} &= 
    \gamma\indices{^d_c} \nabla_d (\gamma^{ce} \nabla_e \widetilde{\mathcal{R}}_{ab}) \equiv
    D_cD^c\widetilde{\mathcal{R}}_{ab}
    \nonumber\;,
\end{align}
the spatial covariant derivatives should be understood as a shorthand notation and not yet as a purely spatial quantity.
This is important since $\widetilde{\mathcal{R}}_{ab}$ is not yet projected and thus contains temporal components.
With this subtlety in mind, we collect results and find
\begin{align}
    \Box \widetilde{\mathcal{R}}_{ab} &=  
    n^c \nabla_c \widetilde{V}_{ab}
    + (D_c + a_c)D^c\widetilde{\mathcal{R}}_{ab}
    - K \widetilde{V}_{ab}\;,
\end{align}
which is also given in the main text.

\section{Projections of $(n^c\nabla_c\widetilde{\mathcal{R}}_{ab})$ and $(n^c\nabla_c \widetilde{V}_{ab})$}
\label{app:LHSprojections}

Here, we project the left-hand side of the covariant traceless evolutions equations, i.e., $(n^c\nabla_c\widetilde{\mathcal{R}}_{ab})$ and $(n^c\nabla_c \widetilde{V}_{ab})$, onto spatial and temporal parts. In the following, we go through the $\widetilde{\mathcal{R}}_{ab}$-case, but the $\widetilde{V}_{ab}$-case proceeds analogously.

For the spatial projection, we derive
\begin{align}
    \gamma_i^a\gamma_j^b&\left(n^c\nabla_c\widetilde{\mathcal{R}}_{ab}\right) =
    \left(n^c\nabla_c\gamma_i^a\gamma_j^b\widetilde{\mathcal{R}}_{ab}\right) 
    - \left(n^c\nabla_c\gamma_i^a\gamma_j^b\right)\widetilde{\mathcal{R}}_{ab}
    \nonumber\\*&=
    \left(n^c\nabla_c\mathcal{A}_{ij}\right) 
    + \frac{1}{3}\left(n^c\nabla_c\gamma_{ij}\mathcal{A}\right)
    - 2\left(n^c\nabla_c\gamma_{(i}^a\right)\gamma_{j)}^b\widetilde{\mathcal{R}}_{ab}
    \nonumber\\*&=
    \left(n^c\nabla_c\mathcal{A}_{ij}\right) 
    + \frac{1}{3}\gamma_{ij}\left(n^c\nabla_c\mathcal{A}\right)
    + \frac{1}{3}\mathcal{A}\left(n^c\nabla_c\gamma_{ij}\right)
    \nonumber\\*&\quad\;
    - 2\left(n^c\nabla_c n^a n_{(i}\right)\gamma_{j)}^b\widetilde{\mathcal{R}}_{ab}
    \nonumber\\*&=
    \left(n^c\nabla_c\mathcal{A}_{ij}\right) 
    + \frac{1}{3}\gamma_{ij}\left(n^c\nabla_c\mathcal{A}\right)
    - \frac{2}{3}\mathcal{A}\left(D_{(i}n_{j)} + K_{ij}\right)
    \nonumber\\*&\quad\;
    - 2\left(a^a n_{(i} + n^a a_{(i}\right)\gamma_{j)}^b\widetilde{\mathcal{R}}_{ab}\nonumber\\*&=
    \left(n^c\nabla_c\mathcal{A}_{ij}\right) 
    + \frac{1}{3}\gamma_{ij}\left(n^c\nabla_c\mathcal{A}\right)
    - \frac{2}{3}\mathcal{A}\left(D_{(i}n_{j)} + K_{ij}\right)
    \nonumber\\*&\quad\;
    - 2\,a^c\left(\mathcal{A}_{c(i}n_{j)} + \frac{1}{3}\gamma_{c(i}n_{j)}\mathcal{A}\right) - 2a_{(i}\mathcal{C}_{j)}
    \;,
    \label{eq:derivation_spatial-projection_Rab}
\end{align}
where we have used the decomposition of $\widetilde{\mathcal{R}}_{ab}$ (cf.~\cref{eq:decomp_RabVab}) in the second and last equality; the evolution equation for the spatial metric (cf.~\cref{eq:evol-gamma-York}) in the fourth equality; and throughout, the decomposition of the metric itself (cf.~\cref{eq:decomp_metric}).

We can independently derive the spatial trace as
\begin{align}
    \gamma^{ab}\left(n^c\nabla_c\widetilde{\mathcal{R}}_{ab}\right) &=
    \left(n^c\nabla_c\gamma^{ab}\widetilde{\mathcal{R}}_{ab}\right) - \left(n^c\nabla_c\gamma^{ab}\right)\widetilde{\mathcal{R}}_{ab}
    \nonumber\\*&=
    \left(n^c\nabla_c\mathcal{A}\right) - \left(n^c\nabla_c n^a n^b\right)\widetilde{\mathcal{R}}_{ab}
    \nonumber\\*&=
    \left(n^c\nabla_c\mathcal{A}\right) - 2\,n^{(a} a^{b)}\widetilde{\mathcal{R}}_{ab}
    \nonumber\\*&=
    \left(n^c\nabla_c\mathcal{A}\right) - 2\,a^c\mathcal{C}_c
    \;,
    \label{eq:derivation_trace-projection_Rab}
\end{align}
which serves as a crosscheck and agrees with the trace of \cref{eq:derivation_spatial-projection_Rab}.

Analogously, we derive the mixed projection, i.e.,
\begin{align}
    n^a\gamma^b_d\left(n^c\nabla_c\widetilde{\mathcal{R}}_{ab}\right) &=
    \left(n^c\nabla_c n^a\gamma^b_d\widetilde{\mathcal{R}}_{ab}\right) 
    - \left(n^c\nabla_c n^a\right)\gamma^b_d\widetilde{\mathcal{R}}_{ab}
    \nonumber\\*&\quad\;
    - n^a\left(n^c\nabla_c\gamma^b_d\right)\widetilde{\mathcal{R}}_{ab}
    \nonumber\\*&=
    \left(n^c\nabla_c \mathcal{C}_d\right) 
    - a^a\left(\mathcal{A}_{ad} - \frac{1}{3}\gamma_{ad}\mathcal{A}\right)
    \nonumber\\*&\quad\;
    - n^a\left(n^c\nabla_c n^b n_d\right)\widetilde{\mathcal{R}}_{ab}
    \nonumber\\*&=
    \left(n^c\nabla_c \mathcal{C}_d\right) 
    - a^a\left(\mathcal{A}_{ad} - \frac{1}{3}\gamma_{ad}\mathcal{A}\right)
    \nonumber\\*&\quad\;
    - n^a\left(a^b n_d + n^b a_d\right)\widetilde{\mathcal{R}}_{ab}
    \nonumber\\*&=
    \left(n^c\nabla_c \mathcal{C}_d\right) 
    - a^a\left(\mathcal{A}_{ad} + \frac{2}{3}\gamma_{ad}\mathcal{A}
    \right)
    \nonumber\\*&\quad\;
    - n_d a^c \mathcal{C}_c
    \;,
    \label{eq:derivation_mixed-projection_Rab}
\end{align}
and the temporal projection (which -- by construction -- agrees with the spatial trace, cf.~\cref{eq:derivation_trace-projection_Rab}, such that the 4D trace vanishes), i.e.,
\begin{align}
    n^an^b\left(n^c\nabla_c\widetilde{\mathcal{R}}_{ab}\right) &=
    \left(n^c\nabla_c n^a n^b\widetilde{\mathcal{R}}_{ab}\right) 
    - \left(n^c\nabla_c n^a n^b\right)\widetilde{\mathcal{R}}_{ab}
    \nonumber\\*&=
    \left(n^c\nabla_c\mathcal{A}\right)
    - 2\,a^c\mathcal{C}_c
    \;.
    \label{eq:derivation_temporal-projection_Rab}
\end{align}

\section{BSSN equations}
\label{app:BSSN}

For completeness, we provide the implemented BSSN equations that we use to evolve the metric sector of the theory. Our conventions agree with~\cite{Baumgarte:2010ndz}. With a split of the conformal metric and the extrinsic curvature into trace and traceless part, i.e.,
\begin{align}
	\tilde{\gamma}_{ij} &= e^{-4\phi}\gamma_{ij}\;, \quad\text{with}\quad \phi = \frac{\ln(\gamma)}{12}\;,
	\\*
	\tilde{A}_{ij} &= e^{-4\phi}\left(K_{ij} - \frac{1}{3}\gamma_{ij}K\right)\;,
\end{align}
the York-variant of the ADM equations (cf.~\cref{eq:evol-gamma-York,eq:evol-extrinsic-York}) can be recast into BSSN form, i.e.,
\begin{align}
	\label{eq:evol-phi-BSSN}
	\partial_t \phi =& - \frac{1}{6} \alpha K + \beta^i \partial_i \phi
	+ \frac{1}{6} \partial_i \beta^i
	\\[0.5em]
	\label{eq:evol-K-BSSN}
	\partial_t K =& - \gamma^{ij} D_j D_i \alpha + 
	\alpha(\tilde A_{ij} \tilde A^{ij}
	+ \frac{1}{3} K^2) 
	\nonumber\\*&\quad
	+ \frac{1}{2\MPl^2} (\widetilde{\rho} + \widetilde{S}) 
	+ \beta^i \partial_i K \;,
	\\[0.5em]
	\label{eq:evol-gamma-BSSN}
	\partial_t \tilde \gamma_{ij} =&
	- 2 \alpha \tilde A_{ij} 
	+ \beta^k \partial_k \tilde \gamma_{ij} 
	+ \tilde \gamma_{ik} \partial_j \beta^k
	+ \tilde \gamma_{kj} \partial_i \beta^k
	\nonumber\\*&\quad
	- \frac{2}{3} \tilde \gamma_{ij} \partial_k \beta^k \;,
	\\[0.5em]
	\label{eq:evol-A-BSSN}
	\partial_t \tilde A_{ij} = & \;e^{- 4 \phi} \left[ 
	    - ( D_i D_j \alpha )^\text{TF} 
	    +\alpha \left({}^{(3)}\!R_{ij}^\text{TF} - \frac{1}{\MPl^2} \widetilde{S}_{ij}^\text{TF} \right) 
	\right]
	\nonumber\\*&
	+ \beta^k \partial_k \tilde A_{ij} 
	+ \tilde A_{ik} \partial_j \beta^k
	+ \tilde A_{kj} \partial_i \beta^k
	- \frac{2}{3} \tilde A_{ij} \partial_k \beta^k
	\nonumber\\*&
	+ \alpha (K \tilde A_{ij} - 2 \tilde A_{il} \tilde A^l_{~j}) \;,
	\\[0.5em]
	\label{eq:evol-Gamma-BSSN}
	\partial_t \tilde \Gamma^i = &\;
	2\alpha \left(
	    \tilde \Gamma^i_{jk} \tilde A^{kj} 
	    - \frac{2}{3} \tilde \gamma^{ij} \partial_j K
	    - \frac{1}{\MPl^2} \tilde{\gamma}^{ij} \widetilde{S}_j 
	    + 6 \tilde A^{ij} \partial_j \phi
	\right)
	\nonumber\\*&
	- 2 \tilde A^{ij} \partial_j \alpha 
	+ \beta^j \partial_j \tilde \Gamma^i 
	- \tilde \Gamma^j \partial_j \beta^i
	\nonumber\\*&
	+ \frac{2}{3} \tilde \Gamma^i \partial_j \beta^j 
	+ \frac{1}{3} \tilde \gamma^{li} \partial_l\partial_j\beta^j
	+ \tilde \gamma^{lj} \partial_j\partial_l\beta^i \;.
\end{align}
\cref{eq:evol-phi-BSSN,eq:evol-K-BSSN} (as well as \cref{eq:evol-gamma-BSSN,eq:evol-A-BSSN}) evolve the trace part (as well as the traceless part) of the metric and extrinsic curvature. They are obtained from the York-ADM equations by tracing and subtracting the trace, respectively. Superscripts $^\text{TF}$ denote trace-free parts. \cref{eq:evol-Gamma-BSSN} is introduced to remove $2^\text{nd}$-order mixed spatial derivatives in $R_{ij}^\text{TF}$ of \cref{eq:evol-A-BSSN} by extending the system. The explicit expression for $R_{ij}^\text{TF}$ in terms of the conformal connection functions $\tilde{\Gamma}^i$ can be found, e.g., in \cite{Baumgarte:1998te}. Their definition
\begin{align}
	\label{eq:conformal-constraint}
	\tilde{\Gamma}^i \equiv \tilde{\gamma}^{jk}\tilde{\Gamma}^i_{jk} = - \partial_j\tilde{\gamma}^{ij}
\end{align}
serves as an additional constraint. Initial data is physical only if it also obeys \cref{eq:conformal-constraint}. Finally -- and crucially with regards to numerical stability -- the shift constraint has been used in \cref{eq:evol-Gamma-BSSN} to remove spatial derivatives of~$\tilde{A}_{ij}$.

\section{Convergence Tests}
\label{app:convergence}

All simulations were performed under LANL supercomupter Darwin. Darwin is a very heterogeneous cluster with a wide variety of hardware available, including x86, Power PC and ARM CPU architectures, systems with terabytes of memory, and a variety of GPUs and other accelerators. In particular, we choose \texttt{x86\_64 Intel CPUs} partition which has dual socket 2.1 GHz 18 core Intel Broadwell E5 2695v4 processor with 45MB of cache and 128GB of RAM on each node.

We perform standard convergence tests. To be specific, the self-convergence ratio is given by
\begin{align}
    \label{eqn:SelfConv_ratio}
    \mathcal{C}_{\textrm{self}} = \log_2 \frac{||\mathbf{F}_{h_i} - \mathbf{F}_{h_{i+1}}||_q}{||\mathbf{F}_{h_{i+1}} - \mathbf{F}_{h_{i+2}}||_q} ,
\end{align}
where $\mathbf{F}$ is the state vector for all evolution variables, and $|| \cdot ||_q$ is a general expression for different norms. Convergence tests have to be performed with respect to a specific norm which is suitable for given system of evolution of equations.
In the following, we denote with $|| \cdot ||_{H_1}$ the $H_1$ norm. This norm is computed in a discrete approximation that replaces the respective continuum norm~\cite{Giannakopoulos2020}.

Similarly, the exact convergence ratio, with $\mathbf{F}_\textrm{exact} = 0$, can be computed
\begin{align}
    \label{eqn:ExactConv_ratio}
    \mathcal{C}_{\textrm{exact}} = 
    \log_2 \frac{||\mathbf{F}_{h_i} - \mathbf{F}_\textrm{exact}||_q}{||\mathbf{F}_{h_{i+1}} - \mathbf{F}_\textrm{exact}||_q} = 
    \log_2 \frac{||\mathbf{F}_{h_i} ||_q}{||\mathbf{F}_{h_{i+1}} ||_q} \; .
\end{align}
Given the employed fourth-order scheme, the expected convergence rate is four, in both cases.
A more detailed discussions of convergence tests is given in~\cite{Held2021}.

\begin{figure}[!t]
    \centering
    \includegraphics[trim={0cm 0cm 0cm 0cm},clip,width=\columnwidth]{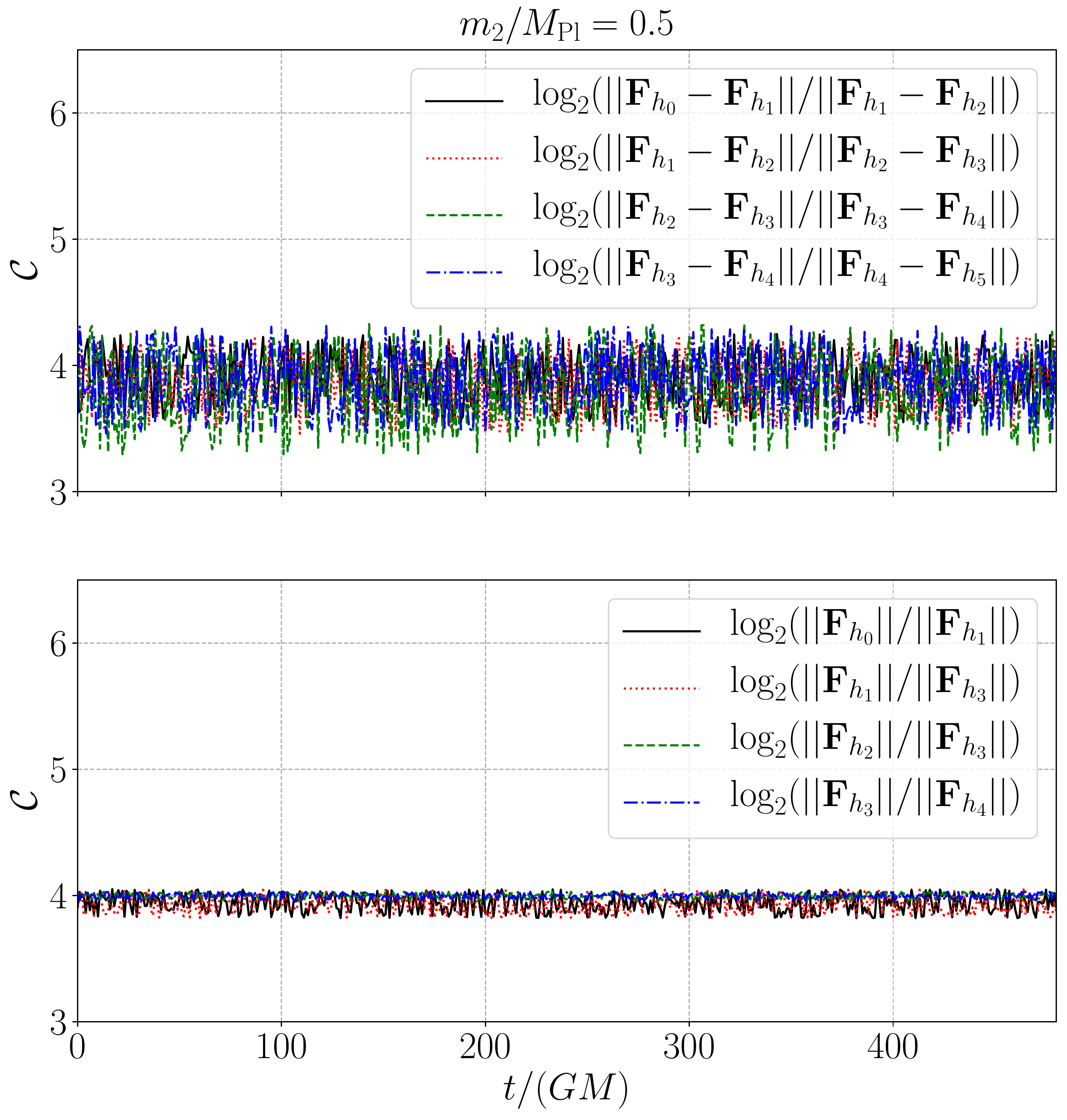}
\caption{
\label{fig:ConvRatio}
   Self-convergence test for Schwarzschild spacetime (upper panel) and exact convergence test (lower panel) as a function of physical time. Both cases exhibit the expected convergence ratio.
}
\end{figure}

In \cref{fig:ConvRatio}, we show the self-convergence test for Schwarzschild spacetime (upper) and the exact convergence test. In both cases, we find the expected fourth-order convergence ratio which matches the implemented fourth-order discretization scheme.
\\

\begin{figure}[!t]
    \centering
    \includegraphics[width=\linewidth]{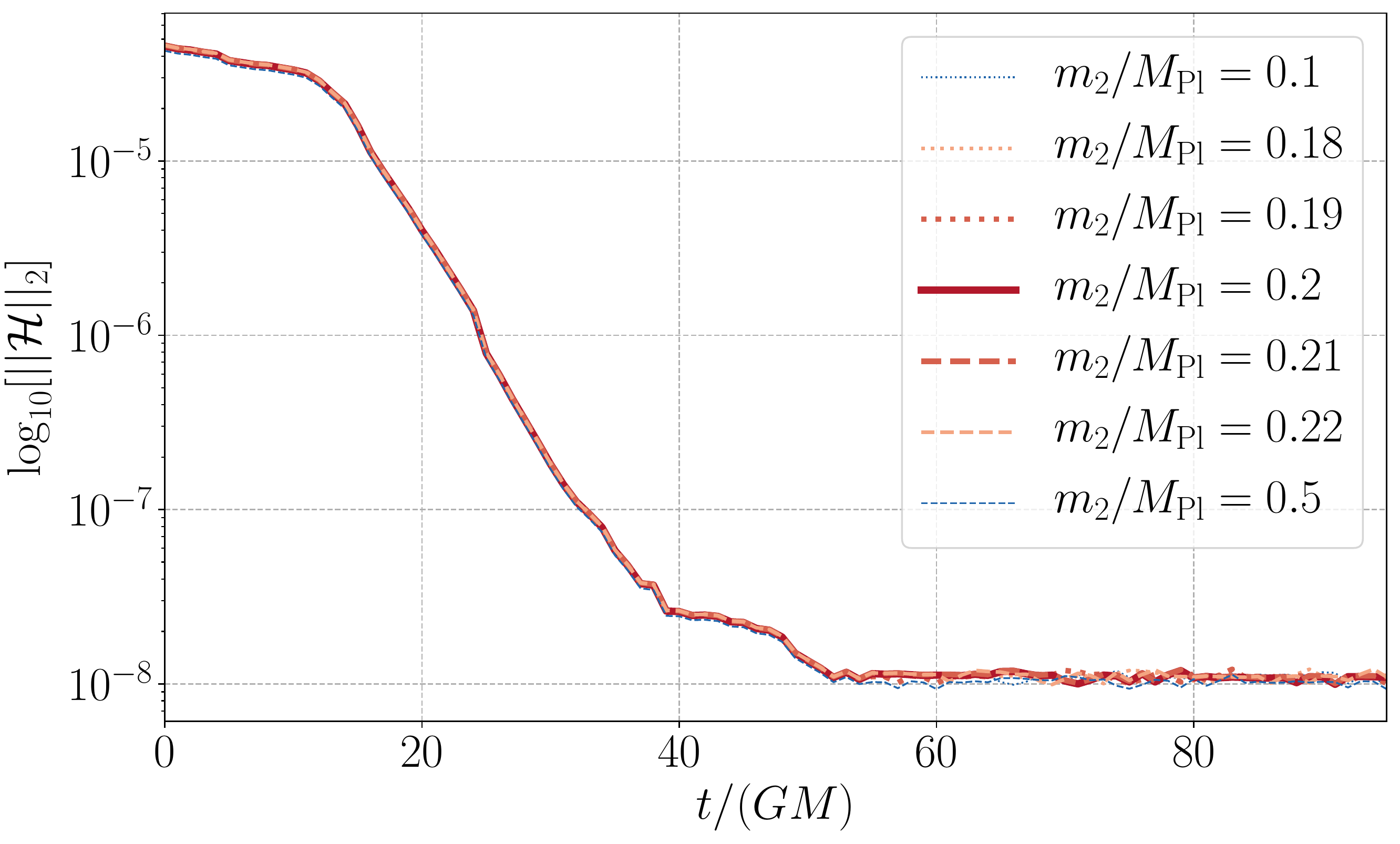}
\caption{
    We show the constraint evolution corresponding to the simulations in \cref{fig:GL}.
}
\label{fig:constraint_GL}
\end{figure}

\begin{figure}[!t]
    \centering
    \includegraphics[width=\linewidth]{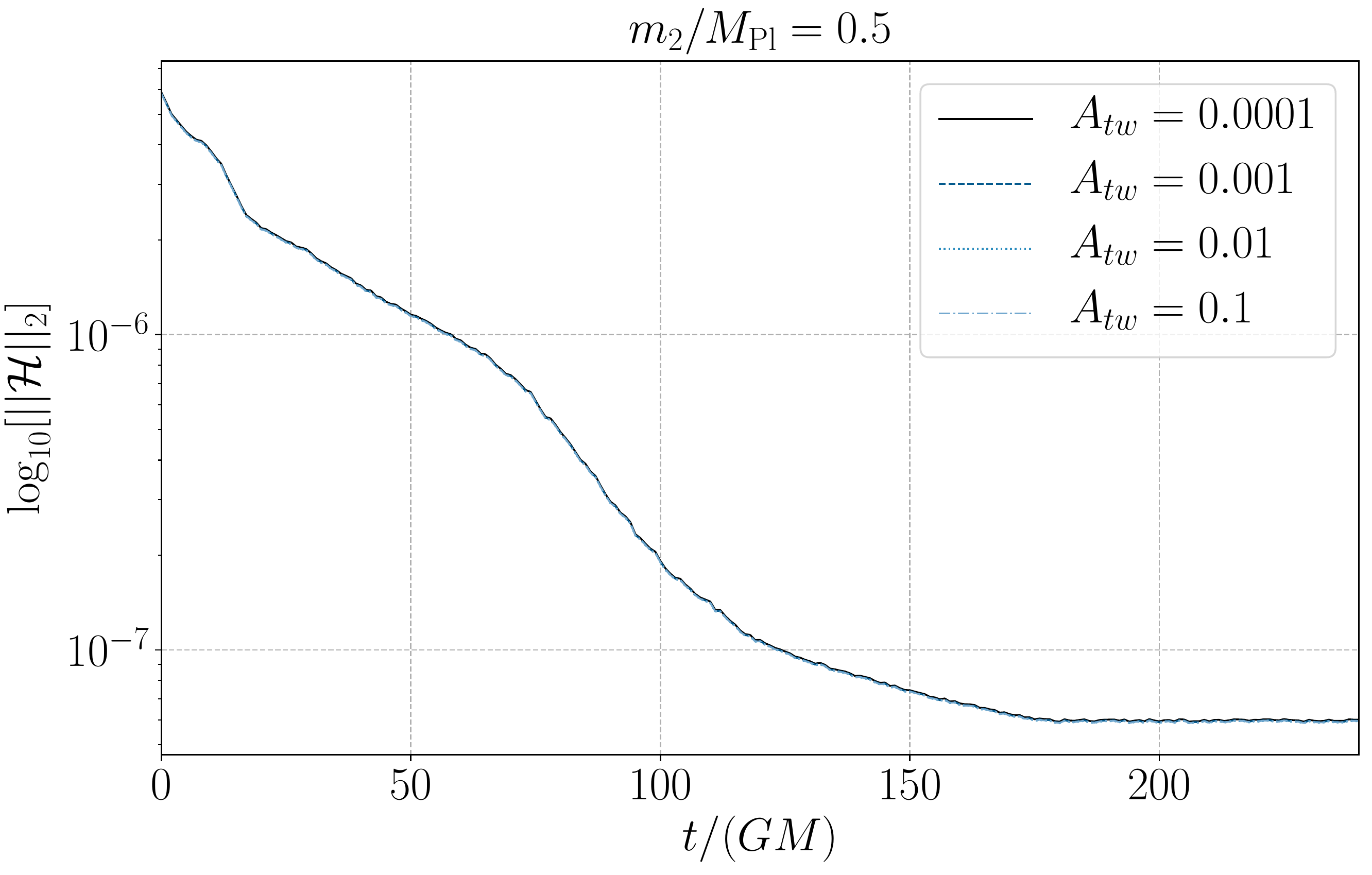}
\caption{We show the constraint evolution corresponding to the simulation in \cref{fig:tw_bh_ricci}.}
\label{fig:constraint_tw}
\end{figure}

\begin{figure}[!t]
    \centering
        \includegraphics[width=\linewidth]{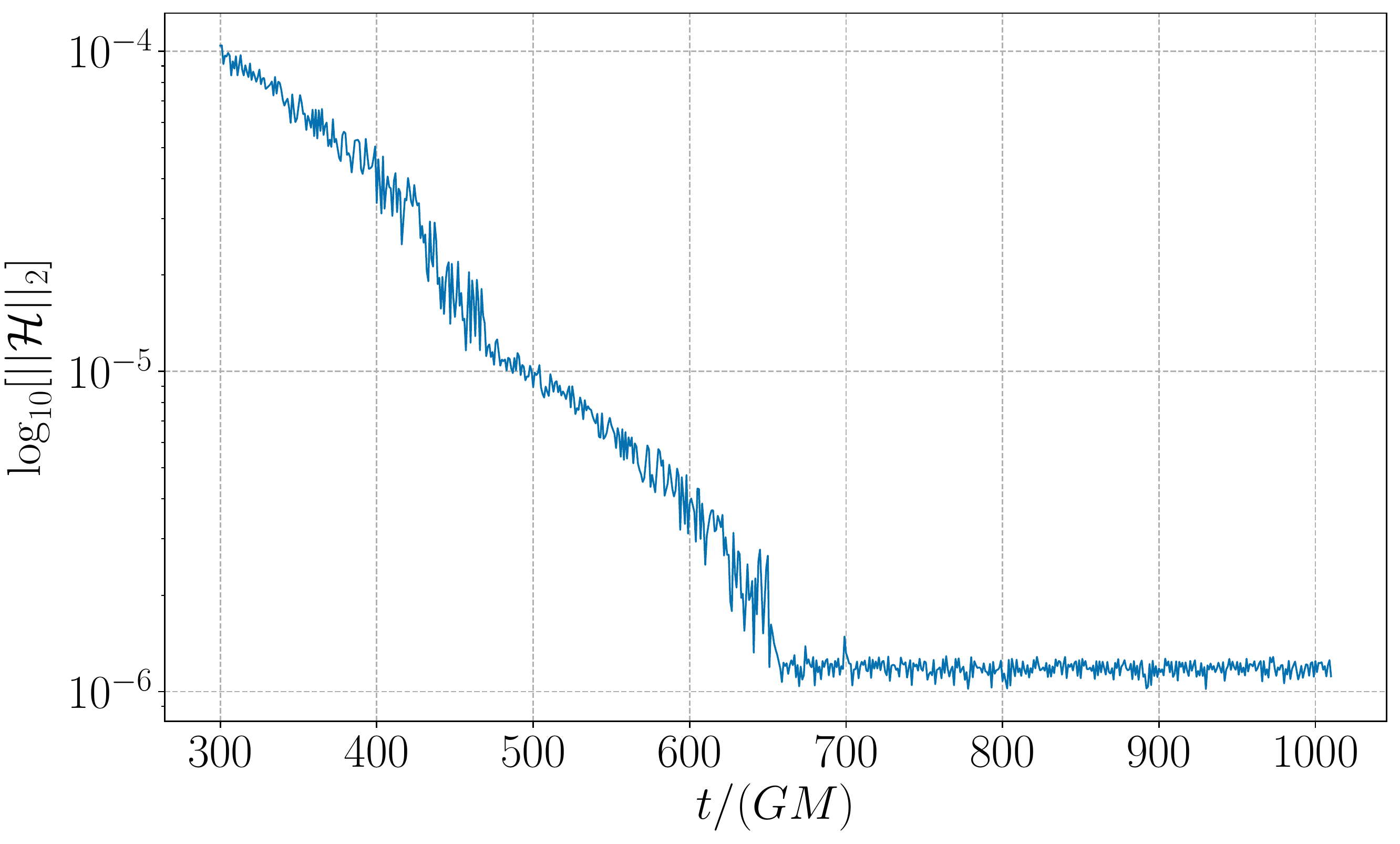}
\caption{We show the constraint evolution corresponding to the beyond-GR simulation in \cref{fig:bbh}. }
\label{fig:constraint_bbh}
\end{figure}

For completeness, we show plots of the constraint (i.e., the $l2$-norm of the Hamiltonian constraint in \cref{eq:Hamiltonian-constraint-York}) for all of our numerical simulations: \cref{fig:constraint_GL} refers to the linear instability in \cref{sec:Gregory-Laflamme}, see also \cref{fig:GL}; \cref{fig:constraint_tw} refers to the Teukolsky wave test in \cref{sec:Teukolsky}, see also \cref{fig:tw_bh_ricci}; \cref{fig:constraint_bbh} refers to the 
linear instability in \cref{sec:binary}, see also \cref{fig:bbh}. Clearly, in all cases, the constrain violations remain small and even decay.


\bibliography{References}

\begin{thebibliography}{99}%
\makeatletter
\providecommand \@ifxundefined [1]{%
 \@ifx{#1\undefined}
}%
\providecommand \@ifnum [1]{%
 \ifnum #1\expandafter \@firstoftwo
 \else \expandafter \@secondoftwo
 \fi
}%
\providecommand \@ifx [1]{%
 \ifx #1\expandafter \@firstoftwo
 \else \expandafter \@secondoftwo
 \fi
}%
\providecommand \natexlab [1]{#1}%
\providecommand \enquote  [1]{``#1''}%
\providecommand \bibnamefont  [1]{#1}%
\providecommand \bibfnamefont [1]{#1}%
\providecommand \citenamefont [1]{#1}%
\providecommand \href@noop [0]{\@secondoftwo}%
\providecommand \href [0]{\begingroup \@sanitize@url \@href}%
\providecommand \@href[1]{\@@startlink{#1}\@@href}%
\providecommand \@@href[1]{\endgroup#1\@@endlink}%
\providecommand \@sanitize@url [0]{\catcode `\\12\catcode `\$12\catcode
  `\&12\catcode `\#12\catcode `\^12\catcode `\_12\catcode `\%12\relax}%
\providecommand \@@startlink[1]{}%
\providecommand \@@endlink[0]{}%
\providecommand \url  [0]{\begingroup\@sanitize@url \@url }%
\providecommand \@url [1]{\endgroup\@href {#1}{\urlprefix }}%
\providecommand \urlprefix  [0]{URL }%
\providecommand \Eprint [0]{\href }%
\providecommand \doibase [0]{http://dx.doi.org/}%
\providecommand \selectlanguage [0]{\@gobble}%
\providecommand \bibinfo  [0]{\@secondoftwo}%
\providecommand \bibfield  [0]{\@secondoftwo}%
\providecommand \translation [1]{[#1]}%
\providecommand \BibitemOpen [0]{}%
\providecommand \bibitemStop [0]{}%
\providecommand \bibitemNoStop [0]{.\EOS\space}%
\providecommand \EOS [0]{\spacefactor3000\relax}%
\providecommand \BibitemShut  [1]{\csname bibitem#1\endcsname}%
\let\auto@bib@innerbib\@empty
\bibitem [{\citenamefont {Abbott}\ \emph {et~al.}(2016)\citenamefont {Abbott}
  \emph {et~al.}}]{LIGOScientific:2016aoc}%
  \BibitemOpen
  \bibfield  {author} {\bibinfo {author} {\bibfnamefont {B.~P.}\ \bibnamefont
  {Abbott}} \emph {et~al.} (\bibinfo {collaboration} {LIGO Scientific,
  Virgo}),\ }\href {\doibase 10.1103/PhysRevLett.116.061102} {\bibfield
  {journal} {\bibinfo  {journal} {Phys. Rev. Lett.}\ }\textbf {\bibinfo
  {volume} {116}},\ \bibinfo {pages} {061102} (\bibinfo {year} {2016})},\
  \Eprint {http://arxiv.org/abs/1602.03837} {arXiv:1602.03837 [gr-qc]}
  \BibitemShut {NoStop}%
\bibitem [{\citenamefont {Akiyama}\ \emph {et~al.}(2019)\citenamefont {Akiyama}
  \emph {et~al.}}]{EventHorizonTelescope:2019dse}%
  \BibitemOpen
  \bibfield  {author} {\bibinfo {author} {\bibfnamefont {K.}~\bibnamefont
  {Akiyama}} \emph {et~al.} (\bibinfo {collaboration} {Event Horizon
  Telescope}),\ }\href {\doibase 10.3847/2041-8213/ab0ec7} {\bibfield
  {journal} {\bibinfo  {journal} {Astrophys. J. Lett.}\ }\textbf {\bibinfo
  {volume} {875}},\ \bibinfo {pages} {L1} (\bibinfo {year} {2019})},\ \Eprint
  {http://arxiv.org/abs/1906.11238} {arXiv:1906.11238 [astro-ph.GA]}
  \BibitemShut {NoStop}%
\bibitem [{\citenamefont {Bertone}\ \emph {et~al.}(2005)\citenamefont
  {Bertone}, \citenamefont {Hooper},\ and\ \citenamefont
  {Silk}}]{Bertone:2004pz}%
  \BibitemOpen
  \bibfield  {author} {\bibinfo {author} {\bibfnamefont {G.}~\bibnamefont
  {Bertone}}, \bibinfo {author} {\bibfnamefont {D.}~\bibnamefont {Hooper}}, \
  and\ \bibinfo {author} {\bibfnamefont {J.}~\bibnamefont {Silk}},\ }\href
  {\doibase 10.1016/j.physrep.2004.08.031} {\bibfield  {journal} {\bibinfo
  {journal} {Phys. Rept.}\ }\textbf {\bibinfo {volume} {405}},\ \bibinfo
  {pages} {279} (\bibinfo {year} {2005})},\ \Eprint
  {http://arxiv.org/abs/hep-ph/0404175} {arXiv:hep-ph/0404175} \BibitemShut
  {NoStop}%
\bibitem [{\citenamefont {Barack}\ \emph {et~al.}(2019)\citenamefont {Barack}
  \emph {et~al.}}]{Barack:2018yly}%
  \BibitemOpen
  \bibfield  {author} {\bibinfo {author} {\bibfnamefont {L.}~\bibnamefont
  {Barack}} \emph {et~al.},\ }\href {\doibase 10.1088/1361-6382/ab0587}
  {\bibfield  {journal} {\bibinfo  {journal} {Class. Quant. Grav.}\ }\textbf
  {\bibinfo {volume} {36}},\ \bibinfo {pages} {143001} (\bibinfo {year}
  {2019})},\ \Eprint {http://arxiv.org/abs/1806.05195} {arXiv:1806.05195
  [gr-qc]} \BibitemShut {NoStop}%
\bibitem [{\citenamefont {Peebles}\ and\ \citenamefont
  {Ratra}(2003)}]{Peebles:2002gy}%
  \BibitemOpen
  \bibfield  {author} {\bibinfo {author} {\bibfnamefont {P.~J.~E.}\
  \bibnamefont {Peebles}}\ and\ \bibinfo {author} {\bibfnamefont
  {B.}~\bibnamefont {Ratra}},\ }\href {\doibase 10.1103/RevModPhys.75.559}
  {\bibfield  {journal} {\bibinfo  {journal} {Rev. Mod. Phys.}\ }\textbf
  {\bibinfo {volume} {75}},\ \bibinfo {pages} {559} (\bibinfo {year} {2003})},\
  \Eprint {http://arxiv.org/abs/astro-ph/0207347} {arXiv:astro-ph/0207347}
  \BibitemShut {NoStop}%
\bibitem [{\citenamefont {Penrose}(1965)}]{Penrose:1964wq}%
  \BibitemOpen
  \bibfield  {author} {\bibinfo {author} {\bibfnamefont {R.}~\bibnamefont
  {Penrose}},\ }\href {\doibase 10.1103/PhysRevLett.14.57} {\bibfield
  {journal} {\bibinfo  {journal} {Phys. Rev. Lett.}\ }\textbf {\bibinfo
  {volume} {14}},\ \bibinfo {pages} {57} (\bibinfo {year} {1965})}\BibitemShut
  {NoStop}%
\bibitem [{\citenamefont {Hawking}(1965)}]{Hawking:1965mf}%
  \BibitemOpen
  \bibfield  {author} {\bibinfo {author} {\bibfnamefont {S.}~\bibnamefont
  {Hawking}},\ }\href {\doibase 10.1103/PhysRevLett.15.689} {\bibfield
  {journal} {\bibinfo  {journal} {Phys. Rev. Lett.}\ }\textbf {\bibinfo
  {volume} {15}},\ \bibinfo {pages} {689} (\bibinfo {year} {1965})}\BibitemShut
  {NoStop}%
\bibitem [{\citenamefont {Geroch}(1966)}]{Geroch:1966ur}%
  \BibitemOpen
  \bibfield  {author} {\bibinfo {author} {\bibfnamefont {R.~P.}\ \bibnamefont
  {Geroch}},\ }\href {\doibase 10.1103/PhysRevLett.17.445} {\bibfield
  {journal} {\bibinfo  {journal} {Phys. Rev. Lett.}\ }\textbf {\bibinfo
  {volume} {17}},\ \bibinfo {pages} {445} (\bibinfo {year} {1966})}\BibitemShut
  {NoStop}%
\bibitem [{\citenamefont {Hawking}\ and\ \citenamefont
  {Penrose}(1970)}]{Hawking:1969sw}%
  \BibitemOpen
  \bibfield  {author} {\bibinfo {author} {\bibfnamefont {S.~W.}\ \bibnamefont
  {Hawking}}\ and\ \bibinfo {author} {\bibfnamefont {R.}~\bibnamefont
  {Penrose}},\ }\href {\doibase 10.1098/rspa.1970.0021} {\bibfield  {journal}
  {\bibinfo  {journal} {Proc. Roy. Soc. Lond.}\ }\textbf {\bibinfo {volume}
  {A314}},\ \bibinfo {pages} {529} (\bibinfo {year} {1970})}\BibitemShut
  {NoStop}%
\bibitem [{\citenamefont {{'t Hooft}}\ and\ \citenamefont
  {Veltman}(1974)}]{tHooft:1974toh}%
  \BibitemOpen
  \bibfield  {author} {\bibinfo {author} {\bibfnamefont {G.}~\bibnamefont {{'t
  Hooft}}}\ and\ \bibinfo {author} {\bibfnamefont {M.~J.~G.}\ \bibnamefont
  {Veltman}},\ }\href@noop {} {\bibfield  {journal} {\bibinfo  {journal} {Ann.
  Inst. H. Poincare Phys. Theor.}\ }\textbf {\bibinfo {volume} {A20}},\
  \bibinfo {pages} {69} (\bibinfo {year} {1974})}\BibitemShut {NoStop}%
\bibitem [{\citenamefont {Stelle}(1977)}]{Stelle:1976gc}%
  \BibitemOpen
  \bibfield  {author} {\bibinfo {author} {\bibfnamefont {K.~S.}\ \bibnamefont
  {Stelle}},\ }\href {\doibase 10.1103/PhysRevD.16.953} {\bibfield  {journal}
  {\bibinfo  {journal} {Phys. Rev. D}\ }\textbf {\bibinfo {volume} {16}},\
  \bibinfo {pages} {953} (\bibinfo {year} {1977})}\BibitemShut {NoStop}%
\bibitem [{\citenamefont {Goroff}\ and\ \citenamefont
  {Sagnotti}(1985)}]{Goroff:1985sz}%
  \BibitemOpen
  \bibfield  {author} {\bibinfo {author} {\bibfnamefont {M.~H.}\ \bibnamefont
  {Goroff}}\ and\ \bibinfo {author} {\bibfnamefont {A.}~\bibnamefont
  {Sagnotti}},\ }\href {\doibase 10.1016/0370-2693(85)91470-4} {\bibfield
  {journal} {\bibinfo  {journal} {Phys. Lett.}\ }\textbf {\bibinfo {volume}
  {160B}},\ \bibinfo {pages} {81} (\bibinfo {year} {1985})}\BibitemShut
  {NoStop}%
\bibitem [{\citenamefont {Avramidi}\ and\ \citenamefont
  {Barvinsky}(1985)}]{Avramidi:1985ki}%
  \BibitemOpen
  \bibfield  {author} {\bibinfo {author} {\bibfnamefont {I.~G.}\ \bibnamefont
  {Avramidi}}\ and\ \bibinfo {author} {\bibfnamefont {A.~O.}\ \bibnamefont
  {Barvinsky}},\ }\href {\doibase 10.1016/0370-2693(85)90248-5} {\bibfield
  {journal} {\bibinfo  {journal} {Phys. Lett. B}\ }\textbf {\bibinfo {volume}
  {159}},\ \bibinfo {pages} {269} (\bibinfo {year} {1985})}\BibitemShut
  {NoStop}%
\bibitem [{\citenamefont {van~de Ven}(1992)}]{vandeVen:1991gw}%
  \BibitemOpen
  \bibfield  {author} {\bibinfo {author} {\bibfnamefont {A.~E.~M.}\
  \bibnamefont {van~de Ven}},\ }\bibfield  {booktitle} {\emph {\bibinfo
  {booktitle} {{Conference on Strings and Symmetries Stony Brook, New York, May
  20-25, 1991}}},\ }\href {\doibase 10.1016/0550-3213(92)90011-Y} {\bibfield
  {journal} {\bibinfo  {journal} {Nucl. Phys.}\ }\textbf {\bibinfo {volume}
  {B378}},\ \bibinfo {pages} {309} (\bibinfo {year} {1992})}\BibitemShut
  {NoStop}%
\bibitem [{\citenamefont {Ohta}\ \emph {et~al.}(2018)\citenamefont {Ohta},
  \citenamefont {Percacci},\ and\ \citenamefont {Pereira}}]{Ohta:2018sze}%
  \BibitemOpen
  \bibfield  {author} {\bibinfo {author} {\bibfnamefont {N.}~\bibnamefont
  {Ohta}}, \bibinfo {author} {\bibfnamefont {R.}~\bibnamefont {Percacci}}, \
  and\ \bibinfo {author} {\bibfnamefont {A.~D.}\ \bibnamefont {Pereira}},\
  }\href {\doibase 10.1103/PhysRevD.97.104039} {\bibfield  {journal} {\bibinfo
  {journal} {Phys. Rev. D}\ }\textbf {\bibinfo {volume} {97}},\ \bibinfo
  {pages} {104039} (\bibinfo {year} {2018})},\ \Eprint
  {http://arxiv.org/abs/1804.01608} {arXiv:1804.01608 [hep-th]} \BibitemShut
  {NoStop}%
\bibitem [{\citenamefont {Loll}(2020)}]{Loll:2019rdj}%
  \BibitemOpen
  \bibfield  {author} {\bibinfo {author} {\bibfnamefont {R.}~\bibnamefont
  {Loll}},\ }\href {\doibase 10.1088/1361-6382/ab57c7} {\bibfield  {journal}
  {\bibinfo  {journal} {Class. Quant. Grav.}\ }\textbf {\bibinfo {volume}
  {37}},\ \bibinfo {pages} {013002} (\bibinfo {year} {2020})},\ \Eprint
  {http://arxiv.org/abs/1905.08669} {arXiv:1905.08669 [hep-th]} \BibitemShut
  {NoStop}%
\bibitem [{\citenamefont {Ashtekar}\ \emph {et~al.}(1989)\citenamefont
  {Ashtekar}, \citenamefont {Balachandran},\ and\ \citenamefont
  {Jo}}]{Ashtekar:1988sw}%
  \BibitemOpen
  \bibfield  {author} {\bibinfo {author} {\bibfnamefont {A.}~\bibnamefont
  {Ashtekar}}, \bibinfo {author} {\bibfnamefont {A.~P.}\ \bibnamefont
  {Balachandran}}, \ and\ \bibinfo {author} {\bibfnamefont {S.}~\bibnamefont
  {Jo}},\ }\href {\doibase 10.1142/S0217751X89000649} {\bibfield  {journal}
  {\bibinfo  {journal} {Int. J. Mod. Phys. A}\ }\textbf {\bibinfo {volume}
  {4}},\ \bibinfo {pages} {1493} (\bibinfo {year} {1989})}\BibitemShut
  {NoStop}%
\bibitem [{\citenamefont {Boulware}\ and\ \citenamefont
  {Deser}(1985)}]{Boulware:1985wk}%
  \BibitemOpen
  \bibfield  {author} {\bibinfo {author} {\bibfnamefont {D.~G.}\ \bibnamefont
  {Boulware}}\ and\ \bibinfo {author} {\bibfnamefont {S.}~\bibnamefont
  {Deser}},\ }\href {\doibase 10.1103/PhysRevLett.55.2656} {\bibfield
  {journal} {\bibinfo  {journal} {Phys. Rev. Lett.}\ }\textbf {\bibinfo
  {volume} {55}},\ \bibinfo {pages} {2656} (\bibinfo {year}
  {1985})}\BibitemShut {NoStop}%
\bibitem [{\citenamefont {Zwiebach}(1985)}]{Zwiebach:1985uq}%
  \BibitemOpen
  \bibfield  {author} {\bibinfo {author} {\bibfnamefont {B.}~\bibnamefont
  {Zwiebach}},\ }\href {\doibase 10.1016/0370-2693(85)91616-8} {\bibfield
  {journal} {\bibinfo  {journal} {Phys. Lett.}\ }\textbf {\bibinfo {volume}
  {156B}},\ \bibinfo {pages} {315} (\bibinfo {year} {1985})}\BibitemShut
  {NoStop}%
\bibitem [{\citenamefont {Benedetti}\ \emph {et~al.}(2009)\citenamefont
  {Benedetti}, \citenamefont {Machado},\ and\ \citenamefont
  {Saueressig}}]{Benedetti:2009rx}%
  \BibitemOpen
  \bibfield  {author} {\bibinfo {author} {\bibfnamefont {D.}~\bibnamefont
  {Benedetti}}, \bibinfo {author} {\bibfnamefont {P.~F.}\ \bibnamefont
  {Machado}}, \ and\ \bibinfo {author} {\bibfnamefont {F.}~\bibnamefont
  {Saueressig}},\ }\href {\doibase 10.1142/S0217732309031521} {\bibfield
  {journal} {\bibinfo  {journal} {Mod. Phys. Lett. A}\ }\textbf {\bibinfo
  {volume} {24}},\ \bibinfo {pages} {2233} (\bibinfo {year} {2009})},\ \Eprint
  {http://arxiv.org/abs/0901.2984} {arXiv:0901.2984 [hep-th]} \BibitemShut
  {NoStop}%
\bibitem [{\citenamefont {Ohta}\ and\ \citenamefont
  {Percacci}(2016)}]{Ohta:2015zwa}%
  \BibitemOpen
  \bibfield  {author} {\bibinfo {author} {\bibfnamefont {N.}~\bibnamefont
  {Ohta}}\ and\ \bibinfo {author} {\bibfnamefont {R.}~\bibnamefont
  {Percacci}},\ }\href {\doibase 10.1088/0264-9381/33/3/035001} {\bibfield
  {journal} {\bibinfo  {journal} {Class. Quant. Grav.}\ }\textbf {\bibinfo
  {volume} {33}},\ \bibinfo {pages} {035001} (\bibinfo {year} {2016})},\
  \Eprint {http://arxiv.org/abs/1506.05526} {arXiv:1506.05526 [hep-th]}
  \BibitemShut {NoStop}%
\bibitem [{\citenamefont {Knorr}(2021)}]{Knorr:2021slg}%
  \BibitemOpen
  \bibfield  {author} {\bibinfo {author} {\bibfnamefont {B.}~\bibnamefont
  {Knorr}},\ }\href {\doibase 10.21468/SciPostPhysCore.4.3.020} {\bibfield
  {journal} {\bibinfo  {journal} {SciPost Phys. Core}\ }\textbf {\bibinfo
  {volume} {4}},\ \bibinfo {pages} {020} (\bibinfo {year} {2021})},\ \Eprint
  {http://arxiv.org/abs/2104.11336} {arXiv:2104.11336 [hep-th]} \BibitemShut
  {NoStop}%
\bibitem [{\citenamefont {Baldazzi}\ and\ \citenamefont
  {Falls}(2021)}]{Baldazzi:2021orb}%
  \BibitemOpen
  \bibfield  {author} {\bibinfo {author} {\bibfnamefont {A.}~\bibnamefont
  {Baldazzi}}\ and\ \bibinfo {author} {\bibfnamefont {K.}~\bibnamefont
  {Falls}},\ }\href {\doibase 10.3390/universe7080294} {\bibfield  {journal}
  {\bibinfo  {journal} {Universe}\ }\textbf {\bibinfo {volume} {7}},\ \bibinfo
  {pages} {294} (\bibinfo {year} {2021})},\ \Eprint
  {http://arxiv.org/abs/2107.00671} {arXiv:2107.00671 [hep-th]} \BibitemShut
  {NoStop}%
\bibitem [{\citenamefont {Stelle}(1978)}]{Stelle:1977ry}%
  \BibitemOpen
  \bibfield  {author} {\bibinfo {author} {\bibfnamefont {K.~S.}\ \bibnamefont
  {Stelle}},\ }\href {\doibase 10.1007/BF00760427} {\bibfield  {journal}
  {\bibinfo  {journal} {Gen. Rel. Grav.}\ }\textbf {\bibinfo {volume} {9}},\
  \bibinfo {pages} {353} (\bibinfo {year} {1978})}\BibitemShut {NoStop}%
\bibitem [{\citenamefont {Kanti}\ \emph {et~al.}(1996)\citenamefont {Kanti},
  \citenamefont {Mavromatos}, \citenamefont {Rizos}, \citenamefont {Tamvakis},\
  and\ \citenamefont {Winstanley}}]{Kanti:1995vq}%
  \BibitemOpen
  \bibfield  {author} {\bibinfo {author} {\bibfnamefont {P.}~\bibnamefont
  {Kanti}}, \bibinfo {author} {\bibfnamefont {N.~E.}\ \bibnamefont
  {Mavromatos}}, \bibinfo {author} {\bibfnamefont {J.}~\bibnamefont {Rizos}},
  \bibinfo {author} {\bibfnamefont {K.}~\bibnamefont {Tamvakis}}, \ and\
  \bibinfo {author} {\bibfnamefont {E.}~\bibnamefont {Winstanley}},\ }\href
  {\doibase 10.1103/PhysRevD.54.5049} {\bibfield  {journal} {\bibinfo
  {journal} {Phys. Rev. D}\ }\textbf {\bibinfo {volume} {54}},\ \bibinfo
  {pages} {5049} (\bibinfo {year} {1996})},\ \Eprint
  {http://arxiv.org/abs/hep-th/9511071} {arXiv:hep-th/9511071} \BibitemShut
  {NoStop}%
\bibitem [{\citenamefont {Alexander}\ and\ \citenamefont
  {Yunes}(2009)}]{Alexander:2009tp}%
  \BibitemOpen
  \bibfield  {author} {\bibinfo {author} {\bibfnamefont {S.}~\bibnamefont
  {Alexander}}\ and\ \bibinfo {author} {\bibfnamefont {N.}~\bibnamefont
  {Yunes}},\ }\href {\doibase 10.1016/j.physrep.2009.07.002} {\bibfield
  {journal} {\bibinfo  {journal} {Phys. Rept.}\ }\textbf {\bibinfo {volume}
  {480}},\ \bibinfo {pages} {1} (\bibinfo {year} {2009})},\ \Eprint
  {http://arxiv.org/abs/0907.2562} {arXiv:0907.2562 [hep-th]} \BibitemShut
  {NoStop}%
\bibitem [{\citenamefont {Ruhdorfer}\ \emph {et~al.}(2020)\citenamefont
  {Ruhdorfer}, \citenamefont {Serra},\ and\ \citenamefont
  {Weiler}}]{Ruhdorfer:2019qmk}%
  \BibitemOpen
  \bibfield  {author} {\bibinfo {author} {\bibfnamefont {M.}~\bibnamefont
  {Ruhdorfer}}, \bibinfo {author} {\bibfnamefont {J.}~\bibnamefont {Serra}}, \
  and\ \bibinfo {author} {\bibfnamefont {A.}~\bibnamefont {Weiler}},\ }\href
  {\doibase 10.1007/JHEP05(2020)083} {\bibfield  {journal} {\bibinfo  {journal}
  {JHEP}\ }\textbf {\bibinfo {volume} {05}},\ \bibinfo {pages} {083} (\bibinfo
  {year} {2020})},\ \Eprint {http://arxiv.org/abs/1908.08050} {arXiv:1908.08050
  [hep-ph]} \BibitemShut {NoStop}%
\bibitem [{\citenamefont {Burgess}(2004)}]{Burgess:2003jk}%
  \BibitemOpen
  \bibfield  {author} {\bibinfo {author} {\bibfnamefont {C.~P.}\ \bibnamefont
  {Burgess}},\ }\href {\doibase 10.12942/lrr-2004-5} {\bibfield  {journal}
  {\bibinfo  {journal} {Living Rev. Rel.}\ }\textbf {\bibinfo {volume} {7}},\
  \bibinfo {pages} {5} (\bibinfo {year} {2004})},\ \Eprint
  {http://arxiv.org/abs/gr-qc/0311082} {arXiv:gr-qc/0311082} \BibitemShut
  {NoStop}%
\bibitem [{\citenamefont {Endlich}\ \emph {et~al.}(2017)\citenamefont
  {Endlich}, \citenamefont {Gorbenko}, \citenamefont {Huang},\ and\
  \citenamefont {Senatore}}]{Endlich:2017tqa}%
  \BibitemOpen
  \bibfield  {author} {\bibinfo {author} {\bibfnamefont {S.}~\bibnamefont
  {Endlich}}, \bibinfo {author} {\bibfnamefont {V.}~\bibnamefont {Gorbenko}},
  \bibinfo {author} {\bibfnamefont {J.}~\bibnamefont {Huang}}, \ and\ \bibinfo
  {author} {\bibfnamefont {L.}~\bibnamefont {Senatore}},\ }\href {\doibase
  10.1007/JHEP09(2017)122} {\bibfield  {journal} {\bibinfo  {journal} {JHEP}\
  }\textbf {\bibinfo {volume} {09}},\ \bibinfo {pages} {122} (\bibinfo {year}
  {2017})},\ \Eprint {http://arxiv.org/abs/1704.01590} {arXiv:1704.01590
  [gr-qc]} \BibitemShut {NoStop}%
\bibitem [{\citenamefont {Penrose}(1969)}]{Penrose:1969pc}%
  \BibitemOpen
  \bibfield  {author} {\bibinfo {author} {\bibfnamefont {R.}~\bibnamefont
  {Penrose}},\ }\href {\doibase 10.1023/A:1016578408204} {\bibfield  {journal}
  {\bibinfo  {journal} {Riv. Nuovo Cim.}\ }\textbf {\bibinfo {volume} {1}},\
  \bibinfo {pages} {252} (\bibinfo {year} {1969})}\BibitemShut {NoStop}%
\bibitem [{\citenamefont {Wald}(1997)}]{Wald:1997wa}%
  \BibitemOpen
  \bibfield  {author} {\bibinfo {author} {\bibfnamefont {R.~M.}\ \bibnamefont
  {Wald}}\ }(\bibinfo {year} {1997})\ pp.\ \bibinfo {pages} {69--85},\ \Eprint
  {http://arxiv.org/abs/gr-qc/9710068} {arXiv:gr-qc/9710068} \BibitemShut
  {NoStop}%
\bibitem [{\citenamefont {Choptuik}(1993)}]{Choptuik:1992jv}%
  \BibitemOpen
  \bibfield  {author} {\bibinfo {author} {\bibfnamefont {M.~W.}\ \bibnamefont
  {Choptuik}},\ }\href {\doibase 10.1103/PhysRevLett.70.9} {\bibfield
  {journal} {\bibinfo  {journal} {Phys. Rev. Lett.}\ }\textbf {\bibinfo
  {volume} {70}},\ \bibinfo {pages} {9} (\bibinfo {year} {1993})}\BibitemShut
  {NoStop}%
\bibitem [{\citenamefont {Gundlach}\ and\ \citenamefont
  {Martin-Garcia}(2007)}]{Gundlach:2007gc}%
  \BibitemOpen
  \bibfield  {author} {\bibinfo {author} {\bibfnamefont {C.}~\bibnamefont
  {Gundlach}}\ and\ \bibinfo {author} {\bibfnamefont {J.~M.}\ \bibnamefont
  {Martin-Garcia}},\ }\href {\doibase 10.12942/lrr-2007-5} {\bibfield
  {journal} {\bibinfo  {journal} {Living Rev. Rel.}\ }\textbf {\bibinfo
  {volume} {10}},\ \bibinfo {pages} {5} (\bibinfo {year} {2007})},\ \Eprint
  {http://arxiv.org/abs/0711.4620} {arXiv:0711.4620 [gr-qc]} \BibitemShut
  {NoStop}%
\bibitem [{\citenamefont {Abbott}\ \emph {et~al.}(2019)\citenamefont {Abbott}
  \emph {et~al.}}]{LIGOScientific:2018mvr}%
  \BibitemOpen
  \bibfield  {author} {\bibinfo {author} {\bibfnamefont {B.~P.}\ \bibnamefont
  {Abbott}} \emph {et~al.} (\bibinfo {collaboration} {LIGO Scientific,
  Virgo}),\ }\href {\doibase 10.1103/PhysRevX.9.031040} {\bibfield  {journal}
  {\bibinfo  {journal} {Phys. Rev. X}\ }\textbf {\bibinfo {volume} {9}},\
  \bibinfo {pages} {031040} (\bibinfo {year} {2019})},\ \Eprint
  {http://arxiv.org/abs/1811.12907} {arXiv:1811.12907 [astro-ph.HE]}
  \BibitemShut {NoStop}%
\bibitem [{\citenamefont {Abbott}\ \emph
  {et~al.}(2021{\natexlab{a}})\citenamefont {Abbott} \emph
  {et~al.}}]{LIGOScientific:2020ibl}%
  \BibitemOpen
  \bibfield  {author} {\bibinfo {author} {\bibfnamefont {R.}~\bibnamefont
  {Abbott}} \emph {et~al.} (\bibinfo {collaboration} {LIGO Scientific,
  Virgo}),\ }\href {\doibase 10.1103/PhysRevX.11.021053} {\bibfield  {journal}
  {\bibinfo  {journal} {Phys. Rev. X}\ }\textbf {\bibinfo {volume} {11}},\
  \bibinfo {pages} {021053} (\bibinfo {year} {2021}{\natexlab{a}})},\ \Eprint
  {http://arxiv.org/abs/2010.14527} {arXiv:2010.14527 [gr-qc]} \BibitemShut
  {NoStop}%
\bibitem [{\citenamefont {Abbott}\ \emph
  {et~al.}(2021{\natexlab{b}})\citenamefont {Abbott} \emph
  {et~al.}}]{LIGOScientific:2021djp}%
  \BibitemOpen
  \bibfield  {author} {\bibinfo {author} {\bibfnamefont {R.}~\bibnamefont
  {Abbott}} \emph {et~al.} (\bibinfo {collaboration} {LIGO Scientific, VIRGO,
  KAGRA}),\ }\href@noop {} {\  (\bibinfo {year} {2021}{\natexlab{b}})},\
  \Eprint {http://arxiv.org/abs/2111.03606} {arXiv:2111.03606 [gr-qc]}
  \BibitemShut {NoStop}%
\bibitem [{\citenamefont {Abbott}\ \emph
  {et~al.}(2021{\natexlab{c}})\citenamefont {Abbott} \emph
  {et~al.}}]{LIGOScientific:2021sio}%
  \BibitemOpen
  \bibfield  {author} {\bibinfo {author} {\bibfnamefont {R.}~\bibnamefont
  {Abbott}} \emph {et~al.} (\bibinfo {collaboration} {LIGO Scientific, VIRGO,
  KAGRA}),\ }\href@noop {} {\  (\bibinfo {year} {2021}{\natexlab{c}})},\
  \Eprint {http://arxiv.org/abs/2112.06861} {arXiv:2112.06861 [gr-qc]}
  \BibitemShut {NoStop}%
\bibitem [{\citenamefont {Shibata}\ and\ \citenamefont
  {Nakamura}(1995)}]{Shibata:1995we}%
  \BibitemOpen
  \bibfield  {author} {\bibinfo {author} {\bibfnamefont {M.}~\bibnamefont
  {Shibata}}\ and\ \bibinfo {author} {\bibfnamefont {T.}~\bibnamefont
  {Nakamura}},\ }\href {\doibase 10.1103/PhysRevD.52.5428} {\bibfield
  {journal} {\bibinfo  {journal} {Phys. Rev. D}\ }\textbf {\bibinfo {volume}
  {52}},\ \bibinfo {pages} {5428} (\bibinfo {year} {1995})}\BibitemShut
  {NoStop}%
\bibitem [{\citenamefont {Baumgarte}\ and\ \citenamefont
  {Shapiro}(1998)}]{Baumgarte:1998te}%
  \BibitemOpen
  \bibfield  {author} {\bibinfo {author} {\bibfnamefont {T.~W.}\ \bibnamefont
  {Baumgarte}}\ and\ \bibinfo {author} {\bibfnamefont {S.~L.}\ \bibnamefont
  {Shapiro}},\ }\href {\doibase 10.1103/PhysRevD.59.024007} {\bibfield
  {journal} {\bibinfo  {journal} {Phys. Rev. D}\ }\textbf {\bibinfo {volume}
  {59}},\ \bibinfo {pages} {024007} (\bibinfo {year} {1998})},\ \Eprint
  {http://arxiv.org/abs/gr-qc/9810065} {arXiv:gr-qc/9810065} \BibitemShut
  {NoStop}%
\bibitem [{\citenamefont {Pretorius}(2005{\natexlab{a}})}]{Pretorius2005}%
  \BibitemOpen
  \bibfield  {author} {\bibinfo {author} {\bibfnamefont {F.}~\bibnamefont
  {Pretorius}},\ }\href {\doibase 10.1088/0264-9381/22/2/014} {\bibfield
  {journal} {\bibinfo  {journal} {Classical and Quantum Gravity}\ }\textbf
  {\bibinfo {volume} {22}},\ \bibinfo {pages} {425} (\bibinfo {year}
  {2005}{\natexlab{a}})}\BibitemShut {NoStop}%
\bibitem [{\citenamefont {Pretorius}(2005{\natexlab{b}})}]{Pretorius:2005gq}%
  \BibitemOpen
  \bibfield  {author} {\bibinfo {author} {\bibfnamefont {F.}~\bibnamefont
  {Pretorius}},\ }\href {\doibase 10.1103/PhysRevLett.95.121101} {\bibfield
  {journal} {\bibinfo  {journal} {Phys. Rev. Lett.}\ }\textbf {\bibinfo
  {volume} {95}},\ \bibinfo {pages} {121101} (\bibinfo {year}
  {2005}{\natexlab{b}})},\ \Eprint {http://arxiv.org/abs/gr-qc/0507014}
  {arXiv:gr-qc/0507014} \BibitemShut {NoStop}%
\bibitem [{\citenamefont {Sarbach}\ and\ \citenamefont
  {Tiglio}(2012)}]{Sarbach2012}%
  \BibitemOpen
  \bibfield  {author} {\bibinfo {author} {\bibfnamefont {O.}~\bibnamefont
  {Sarbach}}\ and\ \bibinfo {author} {\bibfnamefont {M.}~\bibnamefont
  {Tiglio}},\ }\href {\doibase 10.12942/lrr-2012-9} {\bibfield  {journal}
  {\bibinfo  {journal} {Living Reviews in Relativity}\ }\textbf {\bibinfo
  {volume} {15}},\ \bibinfo {pages} {9} (\bibinfo {year} {2012})}\BibitemShut
  {NoStop}%
\bibitem [{\citenamefont {Isenberg}(2014)}]{Isenberg2014}%
  \BibitemOpen
  \bibfield  {author} {\bibinfo {author} {\bibfnamefont {J.}~\bibnamefont
  {Isenberg}},\ }\enquote {\bibinfo {title} {The initial value problem in
  general relativity},}\ in\ \href {\doibase 10.1007/978-3-642-41992-8_16}
  {\emph {\bibinfo {booktitle} {Springer Handbook of Spacetime}}},\ \bibinfo
  {editor} {edited by\ \bibinfo {editor} {\bibfnamefont {A.}~\bibnamefont
  {Ashtekar}}\ and\ \bibinfo {editor} {\bibfnamefont {V.}~\bibnamefont
  {Petkov}}}\ (\bibinfo  {publisher} {Springer Berlin Heidelberg},\ \bibinfo
  {address} {Berlin, Heidelberg},\ \bibinfo {year} {2014})\ pp.\ \bibinfo
  {pages} {303--321}\BibitemShut {NoStop}%
\bibitem [{\citenamefont {Witek}\ \emph {et~al.}(2019)\citenamefont {Witek},
  \citenamefont {Gualtieri}, \citenamefont {Pani},\ and\ \citenamefont
  {Sotiriou}}]{Witek2019}%
  \BibitemOpen
  \bibfield  {author} {\bibinfo {author} {\bibfnamefont {H.}~\bibnamefont
  {Witek}}, \bibinfo {author} {\bibfnamefont {L.}~\bibnamefont {Gualtieri}},
  \bibinfo {author} {\bibfnamefont {P.}~\bibnamefont {Pani}}, \ and\ \bibinfo
  {author} {\bibfnamefont {T.~P.}\ \bibnamefont {Sotiriou}},\ }\href {\doibase
  10.1103/PhysRevD.99.064035} {\bibfield  {journal} {\bibinfo  {journal} {Phys.
  Rev. D}\ }\textbf {\bibinfo {volume} {99}},\ \bibinfo {pages} {064035}
  (\bibinfo {year} {2019})}\BibitemShut {NoStop}%
\bibitem [{\citenamefont {Okounkova}\ \emph {et~al.}(2019)\citenamefont
  {Okounkova}, \citenamefont {Stein}, \citenamefont {Scheel},\ and\
  \citenamefont {Teukolsky}}]{Okounkova2019}%
  \BibitemOpen
  \bibfield  {author} {\bibinfo {author} {\bibfnamefont {M.}~\bibnamefont
  {Okounkova}}, \bibinfo {author} {\bibfnamefont {L.~C.}\ \bibnamefont
  {Stein}}, \bibinfo {author} {\bibfnamefont {M.~A.}\ \bibnamefont {Scheel}}, \
  and\ \bibinfo {author} {\bibfnamefont {S.~A.}\ \bibnamefont {Teukolsky}},\
  }\href {\doibase 10.1103/PhysRevD.100.104026} {\bibfield  {journal} {\bibinfo
   {journal} {Phys. Rev. D}\ }\textbf {\bibinfo {volume} {100}},\ \bibinfo
  {pages} {104026} (\bibinfo {year} {2019})}\BibitemShut {NoStop}%
\bibitem [{\citenamefont {Okounkova}\ \emph {et~al.}(2020)\citenamefont
  {Okounkova}, \citenamefont {Stein}, \citenamefont {Moxon}, \citenamefont
  {Scheel},\ and\ \citenamefont {Teukolsky}}]{Okounkova2020}%
  \BibitemOpen
  \bibfield  {author} {\bibinfo {author} {\bibfnamefont {M.}~\bibnamefont
  {Okounkova}}, \bibinfo {author} {\bibfnamefont {L.~C.}\ \bibnamefont
  {Stein}}, \bibinfo {author} {\bibfnamefont {J.}~\bibnamefont {Moxon}},
  \bibinfo {author} {\bibfnamefont {M.~A.}\ \bibnamefont {Scheel}}, \ and\
  \bibinfo {author} {\bibfnamefont {S.~A.}\ \bibnamefont {Teukolsky}},\ }\href
  {\doibase 10.1103/PhysRevD.101.104016} {\bibfield  {journal} {\bibinfo
  {journal} {Phys. Rev. D}\ }\textbf {\bibinfo {volume} {101}},\ \bibinfo
  {pages} {104016} (\bibinfo {year} {2020})}\BibitemShut {NoStop}%
\bibitem [{\citenamefont {Witek}\ \emph {et~al.}(2020)\citenamefont {Witek},
  \citenamefont {Gualtieri},\ and\ \citenamefont {Pani}}]{Witek2020}%
  \BibitemOpen
  \bibfield  {author} {\bibinfo {author} {\bibfnamefont {H.}~\bibnamefont
  {Witek}}, \bibinfo {author} {\bibfnamefont {L.}~\bibnamefont {Gualtieri}}, \
  and\ \bibinfo {author} {\bibfnamefont {P.}~\bibnamefont {Pani}},\ }\href
  {\doibase 10.1103/PhysRevD.101.124055} {\bibfield  {journal} {\bibinfo
  {journal} {Phys. Rev. D}\ }\textbf {\bibinfo {volume} {101}},\ \bibinfo
  {pages} {124055} (\bibinfo {year} {2020})}\BibitemShut {NoStop}%
\bibitem [{\citenamefont {Ripley}\ and\ \citenamefont
  {Pretorius}(2020{\natexlab{a}})}]{Ripley2020a}%
  \BibitemOpen
  \bibfield  {author} {\bibinfo {author} {\bibfnamefont {J.~L.}\ \bibnamefont
  {Ripley}}\ and\ \bibinfo {author} {\bibfnamefont {F.}~\bibnamefont
  {Pretorius}},\ }\href {\doibase 10.1088/1361-6382/ab9bbb} {\bibfield
  {journal} {\bibinfo  {journal} {Classical and Quantum Gravity}\ }\textbf
  {\bibinfo {volume} {37}},\ \bibinfo {pages} {155003} (\bibinfo {year}
  {2020}{\natexlab{a}})}\BibitemShut {NoStop}%
\bibitem [{\citenamefont {Ripley}\ and\ \citenamefont
  {Pretorius}(2020{\natexlab{b}})}]{Ripley2020b}%
  \BibitemOpen
  \bibfield  {author} {\bibinfo {author} {\bibfnamefont {J.~L.}\ \bibnamefont
  {Ripley}}\ and\ \bibinfo {author} {\bibfnamefont {F.}~\bibnamefont
  {Pretorius}},\ }\href {\doibase 10.1103/PhysRevD.101.044015} {\bibfield
  {journal} {\bibinfo  {journal} {Phys. Rev. D}\ }\textbf {\bibinfo {volume}
  {101}},\ \bibinfo {pages} {044015} (\bibinfo {year}
  {2020}{\natexlab{b}})}\BibitemShut {NoStop}%
\bibitem [{\citenamefont {Okounkova}(2020)}]{Okounkova:2020rqw}%
  \BibitemOpen
  \bibfield  {author} {\bibinfo {author} {\bibfnamefont {M.}~\bibnamefont
  {Okounkova}},\ }\href {\doibase 10.1103/PhysRevD.102.084046} {\bibfield
  {journal} {\bibinfo  {journal} {Phys. Rev. D}\ }\textbf {\bibinfo {volume}
  {102}},\ \bibinfo {pages} {084046} (\bibinfo {year} {2020})},\ \Eprint
  {http://arxiv.org/abs/2001.03571} {arXiv:2001.03571 [gr-qc]} \BibitemShut
  {NoStop}%
\bibitem [{\citenamefont {East}\ and\ \citenamefont
  {Ripley}(2021{\natexlab{a}})}]{East2021}%
  \BibitemOpen
  \bibfield  {author} {\bibinfo {author} {\bibfnamefont {W.~E.}\ \bibnamefont
  {East}}\ and\ \bibinfo {author} {\bibfnamefont {J.~L.}\ \bibnamefont
  {Ripley}},\ }\href {\doibase 10.1103/PhysRevD.103.044040} {\bibfield
  {journal} {\bibinfo  {journal} {Phys. Rev. D}\ }\textbf {\bibinfo {volume}
  {103}},\ \bibinfo {pages} {044040} (\bibinfo {year}
  {2021}{\natexlab{a}})}\BibitemShut {NoStop}%
\bibitem [{\citenamefont {Silva}\ \emph {et~al.}(2020)\citenamefont {Silva},
  \citenamefont {Witek}, \citenamefont {Elley},\ and\ \citenamefont
  {Yunes}}]{Silva:2020omi}%
  \BibitemOpen
  \bibfield  {author} {\bibinfo {author} {\bibfnamefont {H.~O.}\ \bibnamefont
  {Silva}}, \bibinfo {author} {\bibfnamefont {H.}~\bibnamefont {Witek}},
  \bibinfo {author} {\bibfnamefont {M.}~\bibnamefont {Elley}}, \ and\ \bibinfo
  {author} {\bibfnamefont {N.}~\bibnamefont {Yunes}},\ }\href@noop {} {\
  (\bibinfo {year} {2020})},\ \Eprint {http://arxiv.org/abs/2012.10436}
  {arXiv:2012.10436 [gr-qc]} \BibitemShut {NoStop}%
\bibitem [{\citenamefont {Figueras}\ and\ \citenamefont
  {Fran\c{c}a}(2020)}]{Figueras:2020dzx}%
  \BibitemOpen
  \bibfield  {author} {\bibinfo {author} {\bibfnamefont {P.}~\bibnamefont
  {Figueras}}\ and\ \bibinfo {author} {\bibfnamefont {T.}~\bibnamefont
  {Fran\c{c}a}},\ }\href {\doibase 10.1088/1361-6382/abb693} {\bibfield
  {journal} {\bibinfo  {journal} {Class. Quant. Grav.}\ }\textbf {\bibinfo
  {volume} {37}},\ \bibinfo {pages} {225009} (\bibinfo {year} {2020})},\
  \Eprint {http://arxiv.org/abs/2006.09414} {arXiv:2006.09414 [gr-qc]}
  \BibitemShut {NoStop}%
\bibitem [{\citenamefont {Kov\'acs}\ and\ \citenamefont
  {Reall}(2020{\natexlab{a}})}]{Kovacs:2020pns}%
  \BibitemOpen
  \bibfield  {author} {\bibinfo {author} {\bibfnamefont {A.~D.}\ \bibnamefont
  {Kov\'acs}}\ and\ \bibinfo {author} {\bibfnamefont {H.~S.}\ \bibnamefont
  {Reall}},\ }\href {\doibase 10.1103/PhysRevLett.124.221101} {\bibfield
  {journal} {\bibinfo  {journal} {Phys. Rev. Lett.}\ }\textbf {\bibinfo
  {volume} {124}},\ \bibinfo {pages} {221101} (\bibinfo {year}
  {2020}{\natexlab{a}})},\ \Eprint {http://arxiv.org/abs/2003.04327}
  {arXiv:2003.04327 [gr-qc]} \BibitemShut {NoStop}%
\bibitem [{\citenamefont {Kov\'acs}\ and\ \citenamefont
  {Reall}(2020{\natexlab{b}})}]{Kovacs:2020ywu}%
  \BibitemOpen
  \bibfield  {author} {\bibinfo {author} {\bibfnamefont {A.~D.}\ \bibnamefont
  {Kov\'acs}}\ and\ \bibinfo {author} {\bibfnamefont {H.~S.}\ \bibnamefont
  {Reall}},\ }\href {\doibase 10.1103/PhysRevD.101.124003} {\bibfield
  {journal} {\bibinfo  {journal} {Phys. Rev. D}\ }\textbf {\bibinfo {volume}
  {101}},\ \bibinfo {pages} {124003} (\bibinfo {year} {2020}{\natexlab{b}})},\
  \Eprint {http://arxiv.org/abs/2003.08398} {arXiv:2003.08398 [gr-qc]}
  \BibitemShut {NoStop}%
\bibitem [{\citenamefont {Cayuso}\ and\ \citenamefont
  {Lehner}(2020)}]{Cayuso:2020lca}%
  \BibitemOpen
  \bibfield  {author} {\bibinfo {author} {\bibfnamefont {R.}~\bibnamefont
  {Cayuso}}\ and\ \bibinfo {author} {\bibfnamefont {L.}~\bibnamefont
  {Lehner}},\ }\href {\doibase 10.1103/PhysRevD.102.084008} {\bibfield
  {journal} {\bibinfo  {journal} {Phys. Rev. D}\ }\textbf {\bibinfo {volume}
  {102}},\ \bibinfo {pages} {084008} (\bibinfo {year} {2020})},\ \Eprint
  {http://arxiv.org/abs/2005.13720} {arXiv:2005.13720 [gr-qc]} \BibitemShut
  {NoStop}%
\bibitem [{\citenamefont {Cayuso}\ \emph {et~al.}(2023)\citenamefont {Cayuso},
  \citenamefont {Figueras}, \citenamefont {Fran\c{c}a},\ and\ \citenamefont
  {Lehner}}]{Cayuso:2023aht}%
  \BibitemOpen
  \bibfield  {author} {\bibinfo {author} {\bibfnamefont {R.}~\bibnamefont
  {Cayuso}}, \bibinfo {author} {\bibfnamefont {P.}~\bibnamefont {Figueras}},
  \bibinfo {author} {\bibfnamefont {T.}~\bibnamefont {Fran\c{c}a}}, \ and\
  \bibinfo {author} {\bibfnamefont {L.}~\bibnamefont {Lehner}},\ }\href@noop {}
  {\  (\bibinfo {year} {2023})},\ \Eprint {http://arxiv.org/abs/2303.07246}
  {arXiv:2303.07246 [gr-qc]} \BibitemShut {NoStop}%
\bibitem [{\citenamefont {Cayuso}\ \emph {et~al.}(2017)\citenamefont {Cayuso},
  \citenamefont {Ortiz},\ and\ \citenamefont {Lehner}}]{Cayuso:2017iqc}%
  \BibitemOpen
  \bibfield  {author} {\bibinfo {author} {\bibfnamefont {J.}~\bibnamefont
  {Cayuso}}, \bibinfo {author} {\bibfnamefont {N.}~\bibnamefont {Ortiz}}, \
  and\ \bibinfo {author} {\bibfnamefont {L.}~\bibnamefont {Lehner}},\ }\href
  {\doibase 10.1103/PhysRevD.96.084043} {\bibfield  {journal} {\bibinfo
  {journal} {Phys. Rev. D}\ }\textbf {\bibinfo {volume} {96}},\ \bibinfo
  {pages} {084043} (\bibinfo {year} {2017})},\ \Eprint
  {http://arxiv.org/abs/1706.07421} {arXiv:1706.07421 [gr-qc]} \BibitemShut
  {NoStop}%
\bibitem [{\citenamefont {Held}\ and\ \citenamefont {Lim}(2021)}]{Held2021}%
  \BibitemOpen
  \bibfield  {author} {\bibinfo {author} {\bibfnamefont {A.}~\bibnamefont
  {Held}}\ and\ \bibinfo {author} {\bibfnamefont {H.}~\bibnamefont {Lim}},\
  }\href {\doibase 10.1103/PhysRevD.104.084075} {\bibfield  {journal} {\bibinfo
   {journal} {Phys. Rev. D}\ }\textbf {\bibinfo {volume} {104}},\ \bibinfo
  {pages} {084075} (\bibinfo {year} {2021})}\BibitemShut {NoStop}%
\bibitem [{\citenamefont {{Noakes}}(1983)}]{Noakes:1983}%
  \BibitemOpen
  \bibfield  {author} {\bibinfo {author} {\bibfnamefont {D.~R.}\ \bibnamefont
  {{Noakes}}},\ }\href {\doibase 10.1063/1.525906} {\bibfield  {journal}
  {\bibinfo  {journal} {Journal of Mathematical Physics}\ }\textbf {\bibinfo
  {volume} {24}},\ \bibinfo {pages} {1846} (\bibinfo {year}
  {1983})}\BibitemShut {NoStop}%
\bibitem [{\citenamefont {Morales}\ and\ \citenamefont
  {Santillán}(2019)}]{Morales:2018imi}%
  \BibitemOpen
  \bibfield  {author} {\bibinfo {author} {\bibfnamefont {J.~O.}\ \bibnamefont
  {Morales}}\ and\ \bibinfo {author} {\bibfnamefont {O.~P.}\ \bibnamefont
  {Santillán}},\ }\href {\doibase 10.1088/1475-7516/2019/03/026} {\bibfield
  {journal} {\bibinfo  {journal} {JCAP}\ }\textbf {\bibinfo {volume} {03}},\
  \bibinfo {pages} {026} (\bibinfo {year} {2019})},\ \Eprint
  {http://arxiv.org/abs/1811.07869} {arXiv:1811.07869 [hep-th]} \BibitemShut
  {NoStop}%
\bibitem [{\citenamefont {Hindawi}\ \emph
  {et~al.}(1996{\natexlab{a}})\citenamefont {Hindawi}, \citenamefont {Ovrut},\
  and\ \citenamefont {Waldram}}]{Hindawi:1995an}%
  \BibitemOpen
  \bibfield  {author} {\bibinfo {author} {\bibfnamefont {A.}~\bibnamefont
  {Hindawi}}, \bibinfo {author} {\bibfnamefont {B.~A.}\ \bibnamefont {Ovrut}},
  \ and\ \bibinfo {author} {\bibfnamefont {D.}~\bibnamefont {Waldram}},\ }\href
  {\doibase 10.1103/PhysRevD.53.5583} {\bibfield  {journal} {\bibinfo
  {journal} {Phys. Rev. D}\ }\textbf {\bibinfo {volume} {53}},\ \bibinfo
  {pages} {5583} (\bibinfo {year} {1996}{\natexlab{a}})},\ \Eprint
  {http://arxiv.org/abs/hep-th/9509142} {arXiv:hep-th/9509142} \BibitemShut
  {NoStop}%
\bibitem [{\citenamefont {Hinterbichler}\ and\ \citenamefont
  {Saravani}(2016)}]{Hinterbichler:2015soa}%
  \BibitemOpen
  \bibfield  {author} {\bibinfo {author} {\bibfnamefont {K.}~\bibnamefont
  {Hinterbichler}}\ and\ \bibinfo {author} {\bibfnamefont {M.}~\bibnamefont
  {Saravani}},\ }\href {\doibase 10.1103/PhysRevD.93.065006} {\bibfield
  {journal} {\bibinfo  {journal} {Phys. Rev. D}\ }\textbf {\bibinfo {volume}
  {93}},\ \bibinfo {pages} {065006} (\bibinfo {year} {2016})},\ \Eprint
  {http://arxiv.org/abs/1508.02401} {arXiv:1508.02401 [hep-th]} \BibitemShut
  {NoStop}%
\bibitem [{\citenamefont {L\"u}\ \emph {et~al.}(2015)\citenamefont {L\"u},
  \citenamefont {Perkins}, \citenamefont {Pope},\ and\ \citenamefont
  {Stelle}}]{Lu:2015psa}%
  \BibitemOpen
  \bibfield  {author} {\bibinfo {author} {\bibfnamefont {H.}~\bibnamefont
  {L\"u}}, \bibinfo {author} {\bibfnamefont {A.}~\bibnamefont {Perkins}},
  \bibinfo {author} {\bibfnamefont {C.~N.}\ \bibnamefont {Pope}}, \ and\
  \bibinfo {author} {\bibfnamefont {K.~S.}\ \bibnamefont {Stelle}},\ }\href
  {\doibase 10.1103/PhysRevD.92.124019} {\bibfield  {journal} {\bibinfo
  {journal} {Phys. Rev. D}\ }\textbf {\bibinfo {volume} {92}},\ \bibinfo
  {pages} {124019} (\bibinfo {year} {2015})},\ \Eprint
  {http://arxiv.org/abs/1508.00010} {arXiv:1508.00010 [hep-th]} \BibitemShut
  {NoStop}%
\bibitem [{\citenamefont {Held}\ and\ \citenamefont
  {Zhang}(2023)}]{Held:2022abx}%
  \BibitemOpen
  \bibfield  {author} {\bibinfo {author} {\bibfnamefont {A.}~\bibnamefont
  {Held}}\ and\ \bibinfo {author} {\bibfnamefont {J.}~\bibnamefont {Zhang}},\
  }\href {\doibase 10.1103/PhysRevD.107.064060} {\bibfield  {journal} {\bibinfo
   {journal} {Phys. Rev. D}\ }\textbf {\bibinfo {volume} {107}},\ \bibinfo
  {pages} {064060} (\bibinfo {year} {2023})},\ \Eprint
  {http://arxiv.org/abs/2209.01867} {arXiv:2209.01867 [gr-qc]} \BibitemShut
  {NoStop}%
\bibitem [{\citenamefont {York}(2014)}]{York:2014}%
  \BibitemOpen
  \bibfield  {author} {\bibinfo {author} {\bibfnamefont {J.~W.}\ \bibnamefont
  {York}},\ }in\ \href@noop {} {\emph {\bibinfo {booktitle} {{Workshop on
  Sources of Gravitational Radiation}}}}\ (\bibinfo {year} {2014})\ pp.\
  \bibinfo {pages} {83--126}\BibitemShut {NoStop}%
\bibitem [{\citenamefont {Martin-Garcia}\ \emph {et~al.}(2007)\citenamefont
  {Martin-Garcia}, \citenamefont {Portugal},\ and\ \citenamefont
  {Manssur}}]{Martin-Garcia:2007bqa}%
  \BibitemOpen
  \bibfield  {author} {\bibinfo {author} {\bibfnamefont {J.~M.}\ \bibnamefont
  {Martin-Garcia}}, \bibinfo {author} {\bibfnamefont {R.}~\bibnamefont
  {Portugal}}, \ and\ \bibinfo {author} {\bibfnamefont {L.~R.~U.}\ \bibnamefont
  {Manssur}},\ }\href {\doibase 10.1016/j.cpc.2007.05.015} {\bibfield
  {journal} {\bibinfo  {journal} {Comput. Phys. Commun.}\ }\textbf {\bibinfo
  {volume} {177}},\ \bibinfo {pages} {640} (\bibinfo {year} {2007})},\ \Eprint
  {http://arxiv.org/abs/0704.1756} {arXiv:0704.1756 [cs.SC]} \BibitemShut
  {NoStop}%
\bibitem [{\citenamefont {Alic}\ \emph {et~al.}(2012)\citenamefont {Alic},
  \citenamefont {Bona-Casas}, \citenamefont {Bona}, \citenamefont {Rezzolla},\
  and\ \citenamefont {Palenzuela}}]{Alic2012}%
  \BibitemOpen
  \bibfield  {author} {\bibinfo {author} {\bibfnamefont {D.}~\bibnamefont
  {Alic}}, \bibinfo {author} {\bibfnamefont {C.}~\bibnamefont {Bona-Casas}},
  \bibinfo {author} {\bibfnamefont {C.}~\bibnamefont {Bona}}, \bibinfo {author}
  {\bibfnamefont {L.}~\bibnamefont {Rezzolla}}, \ and\ \bibinfo {author}
  {\bibfnamefont {C.}~\bibnamefont {Palenzuela}},\ }\href {\doibase
  10.1103/PhysRevD.85.064040} {\bibfield  {journal} {\bibinfo  {journal} {Phys.
  Rev. D}\ }\textbf {\bibinfo {volume} {85}},\ \bibinfo {pages} {064040}
  (\bibinfo {year} {2012})}\BibitemShut {NoStop}%
\bibitem [{\citenamefont {Alic}\ \emph {et~al.}(2013)\citenamefont {Alic},
  \citenamefont {Kastaun},\ and\ \citenamefont {Rezzolla}}]{Alic2013}%
  \BibitemOpen
  \bibfield  {author} {\bibinfo {author} {\bibfnamefont {D.}~\bibnamefont
  {Alic}}, \bibinfo {author} {\bibfnamefont {W.}~\bibnamefont {Kastaun}}, \
  and\ \bibinfo {author} {\bibfnamefont {L.}~\bibnamefont {Rezzolla}},\ }\href
  {\doibase 10.1103/PhysRevD.88.064049} {\bibfield  {journal} {\bibinfo
  {journal} {Phys. Rev. D}\ }\textbf {\bibinfo {volume} {88}},\ \bibinfo
  {pages} {064049} (\bibinfo {year} {2013})}\BibitemShut {NoStop}%
\bibitem [{\citenamefont {Loffler}\ \emph {et~al.}(2012)\citenamefont {Loffler}
  \emph {et~al.}}]{Loffler2011ay}%
  \BibitemOpen
  \bibfield  {author} {\bibinfo {author} {\bibfnamefont {F.}~\bibnamefont
  {Loffler}} \emph {et~al.},\ }\href {\doibase 10.1088/0264-9381/29/11/115001}
  {\bibfield  {journal} {\bibinfo  {journal} {Class. Quant. Grav.}\ }\textbf
  {\bibinfo {volume} {29}},\ \bibinfo {pages} {115001} (\bibinfo {year}
  {2012})},\ \Eprint {http://arxiv.org/abs/1111.3344} {arXiv:1111.3344 [gr-qc]}
  \BibitemShut {NoStop}%
\bibitem [{\citenamefont {Clough}\ \emph {et~al.}(2015)\citenamefont {Clough},
  \citenamefont {Figueras}, \citenamefont {Finkel}, \citenamefont {Kunesch},
  \citenamefont {Lim},\ and\ \citenamefont {Tunyasuvunakool}}]{Clough2015}%
  \BibitemOpen
  \bibfield  {author} {\bibinfo {author} {\bibfnamefont {K.}~\bibnamefont
  {Clough}}, \bibinfo {author} {\bibfnamefont {P.}~\bibnamefont {Figueras}},
  \bibinfo {author} {\bibfnamefont {H.}~\bibnamefont {Finkel}}, \bibinfo
  {author} {\bibfnamefont {M.}~\bibnamefont {Kunesch}}, \bibinfo {author}
  {\bibfnamefont {E.~A.}\ \bibnamefont {Lim}}, \ and\ \bibinfo {author}
  {\bibfnamefont {S.}~\bibnamefont {Tunyasuvunakool}},\ }\href {\doibase
  10.1088/0264-9381/32/24/245011} {\bibfield  {journal} {\bibinfo  {journal}
  {Classical and Quantum Gravity}\ }\textbf {\bibinfo {volume} {32}},\ \bibinfo
  {pages} {245011} (\bibinfo {year} {2015})}\BibitemShut {NoStop}%
\bibitem [{Spe()}]{SpecCode}%
  \BibitemOpen
  \href {https://www.black-holes.org/code/SpEC.html} {\enquote {\bibinfo
  {title} {Spectral einstein code},}\ }\BibitemShut {NoStop}%
\bibitem [{\citenamefont {Fernando}\ \emph {et~al.}(2019)\citenamefont
  {Fernando}, \citenamefont {Neilsen}, \citenamefont {Lim}, \citenamefont
  {Hirschmann},\ and\ \citenamefont {Sundar}}]{Fernando2019}%
  \BibitemOpen
  \bibfield  {author} {\bibinfo {author} {\bibfnamefont {M.}~\bibnamefont
  {Fernando}}, \bibinfo {author} {\bibfnamefont {D.}~\bibnamefont {Neilsen}},
  \bibinfo {author} {\bibfnamefont {H.}~\bibnamefont {Lim}}, \bibinfo {author}
  {\bibfnamefont {E.}~\bibnamefont {Hirschmann}}, \ and\ \bibinfo {author}
  {\bibfnamefont {H.}~\bibnamefont {Sundar}},\ }\href@noop {} {\bibfield
  {journal} {\bibinfo  {journal} {SIAM J. Sci. Comput.}\ }\textbf {\bibinfo
  {volume} {41}},\ \bibinfo {pages} {C97} (\bibinfo {year} {2019})}\BibitemShut
  {NoStop}%
\bibitem [{\citenamefont {Alcubierre}\ \emph {et~al.}(2003)\citenamefont
  {Alcubierre}, \citenamefont {Br\"ugmann}, \citenamefont {Diener},
  \citenamefont {Koppitz}, \citenamefont {Pollney}, \citenamefont {Seidel},\
  and\ \citenamefont {Takahashi}}]{Alcubierre2003}%
  \BibitemOpen
  \bibfield  {author} {\bibinfo {author} {\bibfnamefont {M.}~\bibnamefont
  {Alcubierre}}, \bibinfo {author} {\bibfnamefont {B.}~\bibnamefont
  {Br\"ugmann}}, \bibinfo {author} {\bibfnamefont {P.}~\bibnamefont {Diener}},
  \bibinfo {author} {\bibfnamefont {M.}~\bibnamefont {Koppitz}}, \bibinfo
  {author} {\bibfnamefont {D.}~\bibnamefont {Pollney}}, \bibinfo {author}
  {\bibfnamefont {E.}~\bibnamefont {Seidel}}, \ and\ \bibinfo {author}
  {\bibfnamefont {R.}~\bibnamefont {Takahashi}},\ }\href {\doibase
  10.1103/PhysRevD.67.084023} {\bibfield  {journal} {\bibinfo  {journal} {Phys.
  Rev. D}\ }\textbf {\bibinfo {volume} {67}},\ \bibinfo {pages} {084023}
  (\bibinfo {year} {2003})}\BibitemShut {NoStop}%
\bibitem [{\citenamefont {Beyer}\ and\ \citenamefont
  {Sarbach}(2004)}]{Beyer:2004sv}%
  \BibitemOpen
  \bibfield  {author} {\bibinfo {author} {\bibfnamefont {H.~R.}\ \bibnamefont
  {Beyer}}\ and\ \bibinfo {author} {\bibfnamefont {O.}~\bibnamefont
  {Sarbach}},\ }\href {\doibase 10.1103/PhysRevD.70.104004} {\bibfield
  {journal} {\bibinfo  {journal} {Phys. Rev. D}\ }\textbf {\bibinfo {volume}
  {70}},\ \bibinfo {pages} {104004} (\bibinfo {year} {2004})},\ \Eprint
  {http://arxiv.org/abs/gr-qc/0406003} {arXiv:gr-qc/0406003} \BibitemShut
  {NoStop}%
\bibitem [{\citenamefont {{Courant}}\ \emph {et~al.}(1967)\citenamefont
  {{Courant}}, \citenamefont {{Friedrichs}},\ and\ \citenamefont
  {{Lewy}}}]{Courant:1967}%
  \BibitemOpen
  \bibfield  {author} {\bibinfo {author} {\bibfnamefont {R.}~\bibnamefont
  {{Courant}}}, \bibinfo {author} {\bibfnamefont {K.}~\bibnamefont
  {{Friedrichs}}}, \ and\ \bibinfo {author} {\bibfnamefont {H.}~\bibnamefont
  {{Lewy}}},\ }\href {\doibase 10.1147/rd.112.0215} {\bibfield  {journal}
  {\bibinfo  {journal} {IBM Journal of Research and Development}\ }\textbf
  {\bibinfo {volume} {11}},\ \bibinfo {pages} {215} (\bibinfo {year}
  {1967})}\BibitemShut {NoStop}%
\bibitem [{\citenamefont {Brandt}\ and\ \citenamefont
  {Br\"ugmann}(1997)}]{Steven1997}%
  \BibitemOpen
  \bibfield  {author} {\bibinfo {author} {\bibfnamefont {S.}~\bibnamefont
  {Brandt}}\ and\ \bibinfo {author} {\bibfnamefont {B.}~\bibnamefont
  {Br\"ugmann}},\ }\href {\doibase 10.1103/PhysRevLett.78.3606} {\bibfield
  {journal} {\bibinfo  {journal} {Phys. Rev. Lett.}\ }\textbf {\bibinfo
  {volume} {78}},\ \bibinfo {pages} {3606} (\bibinfo {year}
  {1997})}\BibitemShut {NoStop}%
\bibitem [{\citenamefont {Brito}\ \emph {et~al.}(2013)\citenamefont {Brito},
  \citenamefont {Cardoso},\ and\ \citenamefont {Pani}}]{Brito:2013wya}%
  \BibitemOpen
  \bibfield  {author} {\bibinfo {author} {\bibfnamefont {R.}~\bibnamefont
  {Brito}}, \bibinfo {author} {\bibfnamefont {V.}~\bibnamefont {Cardoso}}, \
  and\ \bibinfo {author} {\bibfnamefont {P.}~\bibnamefont {Pani}},\ }\href
  {\doibase 10.1103/PhysRevD.88.023514} {\bibfield  {journal} {\bibinfo
  {journal} {Phys. Rev. D}\ }\textbf {\bibinfo {volume} {88}},\ \bibinfo
  {pages} {023514} (\bibinfo {year} {2013})},\ \Eprint
  {http://arxiv.org/abs/1304.6725} {arXiv:1304.6725 [gr-qc]} \BibitemShut
  {NoStop}%
\bibitem [{\citenamefont {Gregory}\ and\ \citenamefont
  {Laflamme}(1993)}]{Gregory:1993vy}%
  \BibitemOpen
  \bibfield  {author} {\bibinfo {author} {\bibfnamefont {R.}~\bibnamefont
  {Gregory}}\ and\ \bibinfo {author} {\bibfnamefont {R.}~\bibnamefont
  {Laflamme}},\ }\href {\doibase 10.1103/PhysRevLett.70.2837} {\bibfield
  {journal} {\bibinfo  {journal} {Phys. Rev. Lett.}\ }\textbf {\bibinfo
  {volume} {70}},\ \bibinfo {pages} {2837} (\bibinfo {year} {1993})},\ \Eprint
  {http://arxiv.org/abs/hep-th/9301052} {arXiv:hep-th/9301052} \BibitemShut
  {NoStop}%
\bibitem [{\citenamefont {Brito}\ \emph {et~al.}(2020)\citenamefont {Brito},
  \citenamefont {Grillo},\ and\ \citenamefont {Pani}}]{Brito:2020lup}%
  \BibitemOpen
  \bibfield  {author} {\bibinfo {author} {\bibfnamefont {R.}~\bibnamefont
  {Brito}}, \bibinfo {author} {\bibfnamefont {S.}~\bibnamefont {Grillo}}, \
  and\ \bibinfo {author} {\bibfnamefont {P.}~\bibnamefont {Pani}},\ }\href
  {\doibase 10.1103/PhysRevLett.124.211101} {\bibfield  {journal} {\bibinfo
  {journal} {Phys. Rev. Lett.}\ }\textbf {\bibinfo {volume} {124}},\ \bibinfo
  {pages} {211101} (\bibinfo {year} {2020})},\ \Eprint
  {http://arxiv.org/abs/2002.04055} {arXiv:2002.04055 [gr-qc]} \BibitemShut
  {NoStop}%
\bibitem [{\citenamefont {Teukolsky}(1982)}]{Teukolsky:waves}%
  \BibitemOpen
  \bibfield  {author} {\bibinfo {author} {\bibfnamefont {S.~A.}\ \bibnamefont
  {Teukolsky}},\ }\href {\doibase 10.1103/PhysRevD.26.745} {\bibfield
  {journal} {\bibinfo  {journal} {Phys. Rev. D}\ }\textbf {\bibinfo {volume}
  {26}},\ \bibinfo {pages} {745} (\bibinfo {year} {1982})}\BibitemShut
  {NoStop}%
\bibitem [{\citenamefont {{Brill}}(1959)}]{1959AnPhy...7..466B}%
  \BibitemOpen
  \bibfield  {author} {\bibinfo {author} {\bibfnamefont {D.~R.}\ \bibnamefont
  {{Brill}}},\ }\href {\doibase 10.1016/0003-4916(59)90055-7} {\bibfield
  {journal} {\bibinfo  {journal} {Annals of Physics}\ }\textbf {\bibinfo
  {volume} {7}},\ \bibinfo {pages} {466} (\bibinfo {year} {1959})}\BibitemShut
  {NoStop}%
\bibitem [{\citenamefont {Baumgarte}\ and\ \citenamefont
  {Shapiro}(2010)}]{Baumgarte:2010ndz}%
  \BibitemOpen
  \bibfield  {author} {\bibinfo {author} {\bibfnamefont {T.~W.}\ \bibnamefont
  {Baumgarte}}\ and\ \bibinfo {author} {\bibfnamefont {S.~L.}\ \bibnamefont
  {Shapiro}},\ }\href {\doibase 10.1017/CBO9781139193344} {\emph {\bibinfo
  {title} {{Numerical Relativity: Solving Einstein's Equations on the
  Computer}}}}\ (\bibinfo  {publisher} {Cambridge University Press},\ \bibinfo
  {year} {2010})\BibitemShut {NoStop}%
\bibitem [{\citenamefont {Brewin}(2017)}]{Leo2017}%
  \BibitemOpen
  \bibfield  {author} {\bibinfo {author} {\bibfnamefont {L.}~\bibnamefont
  {Brewin}},\ }\href {\doibase 10.1103/PhysRevD.96.024037} {\bibfield
  {journal} {\bibinfo  {journal} {Phys. Rev. D}\ }\textbf {\bibinfo {volume}
  {96}},\ \bibinfo {pages} {024037} (\bibinfo {year} {2017})}\BibitemShut
  {NoStop}%
\bibitem [{\citenamefont {Fern\'andez}\ \emph {et~al.}(2021)\citenamefont
  {Fern\'andez}, \citenamefont {Baumgarte},\ and\ \citenamefont
  {Hilditch}}]{Fernandez:2021ckg}%
  \BibitemOpen
  \bibfield  {author} {\bibinfo {author} {\bibfnamefont {I.~S.}\ \bibnamefont
  {Fern\'andez}}, \bibinfo {author} {\bibfnamefont {T.~W.}\ \bibnamefont
  {Baumgarte}}, \ and\ \bibinfo {author} {\bibfnamefont {D.}~\bibnamefont
  {Hilditch}},\ }\href {\doibase 10.1103/PhysRevD.104.124036} {\bibfield
  {journal} {\bibinfo  {journal} {Phys. Rev. D}\ }\textbf {\bibinfo {volume}
  {104}},\ \bibinfo {pages} {124036} (\bibinfo {year} {2021})},\ \Eprint
  {http://arxiv.org/abs/2111.04752} {arXiv:2111.04752 [gr-qc]} \BibitemShut
  {NoStop}%
\bibitem [{\citenamefont {East}\ and\ \citenamefont
  {Ripley}(2021{\natexlab{b}})}]{East2021b}%
  \BibitemOpen
  \bibfield  {author} {\bibinfo {author} {\bibfnamefont {W.~E.}\ \bibnamefont
  {East}}\ and\ \bibinfo {author} {\bibfnamefont {J.~L.}\ \bibnamefont
  {Ripley}},\ }\href {\doibase 10.1103/PhysRevLett.127.101102} {\bibfield
  {journal} {\bibinfo  {journal} {Phys. Rev. Lett.}\ }\textbf {\bibinfo
  {volume} {127}},\ \bibinfo {pages} {101102} (\bibinfo {year}
  {2021}{\natexlab{b}})}\BibitemShut {NoStop}%
\bibitem [{\citenamefont {Corman}\ \emph {et~al.}(2023)\citenamefont {Corman},
  \citenamefont {Ripley},\ and\ \citenamefont {East}}]{Corman2023}%
  \BibitemOpen
  \bibfield  {author} {\bibinfo {author} {\bibfnamefont {M.}~\bibnamefont
  {Corman}}, \bibinfo {author} {\bibfnamefont {J.~L.}\ \bibnamefont {Ripley}},
  \ and\ \bibinfo {author} {\bibfnamefont {W.~E.}\ \bibnamefont {East}},\
  }\href {\doibase 10.1103/PhysRevD.107.024014} {\bibfield  {journal} {\bibinfo
   {journal} {Phys. Rev. D}\ }\textbf {\bibinfo {volume} {107}},\ \bibinfo
  {pages} {024014} (\bibinfo {year} {2023})}\BibitemShut {NoStop}%
\bibitem [{\citenamefont {Bowen}(1979)}]{bowen1979general}%
  \BibitemOpen
  \bibfield  {author} {\bibinfo {author} {\bibfnamefont {J.~M.}\ \bibnamefont
  {Bowen}},\ }\href@noop {} {\bibfield  {journal} {\bibinfo  {journal} {General
  Relativity and Gravitation}\ }\textbf {\bibinfo {volume} {11}},\ \bibinfo
  {pages} {227} (\bibinfo {year} {1979})}\BibitemShut {NoStop}%
\bibitem [{\citenamefont {{York}}(1989)}]{1989fnr..book...89Y}%
  \BibitemOpen
  \bibfield  {author} {\bibinfo {author} {\bibfnamefont {J.}~\bibnamefont
  {{York}}, \bibfnamefont {J.~W.}},\ }in\ \href@noop {} {\emph {\bibinfo
  {booktitle} {Frontiers in Numerical Relativity}}},\ \bibinfo {editor} {edited
  by\ \bibinfo {editor} {\bibfnamefont {C.~R.}\ \bibnamefont {{Evans}}},
  \bibinfo {editor} {\bibfnamefont {L.~S.}\ \bibnamefont {{Finn}}}, \ and\
  \bibinfo {editor} {\bibfnamefont {D.~W.}\ \bibnamefont {{Hobill}}}}\
  (\bibinfo {year} {1989})\ pp.\ \bibinfo {pages} {89--109}\BibitemShut
  {NoStop}%
\bibitem [{\citenamefont {Abbott}\ and\ \citenamefont {et.
  al}(2016)}]{Abbott2016}%
  \BibitemOpen
  \bibfield  {author} {\bibinfo {author} {\bibfnamefont {B.~P.}\ \bibnamefont
  {Abbott}}\ and\ \bibinfo {author} {\bibnamefont {et. al}} (\bibinfo
  {collaboration} {LIGO Scientific Collaboration and Virgo Collaboration}),\
  }\href {\doibase 10.1103/PhysRevLett.116.241102} {\bibfield  {journal}
  {\bibinfo  {journal} {Phys. Rev. Lett.}\ }\textbf {\bibinfo {volume} {116}},\
  \bibinfo {pages} {241102} (\bibinfo {year} {2016})}\BibitemShut {NoStop}%
\bibitem [{\citenamefont {Wardell}\ \emph {et~al.}(2016)\citenamefont
  {Wardell}, \citenamefont {Hinder},\ and\ \citenamefont
  {Bentivegna}}]{Wardell2016}%
  \BibitemOpen
  \bibfield  {author} {\bibinfo {author} {\bibfnamefont {B.}~\bibnamefont
  {Wardell}}, \bibinfo {author} {\bibfnamefont {I.}~\bibnamefont {Hinder}}, \
  and\ \bibinfo {author} {\bibfnamefont {E.}~\bibnamefont {Bentivegna}},\
  }\href {\doibase 10.5281/zenodo.155394} {\enquote {\bibinfo {title}
  {{Simulation of GW150914 binary black hole merger using the Einstein
  Toolkit}},}\ } (\bibinfo {year} {2016})\BibitemShut {NoStop}%
\bibitem [{\citenamefont {Choptuik}\ \emph {et~al.}(2003)\citenamefont
  {Choptuik}, \citenamefont {Lehner}, \citenamefont {Olabarrieta},
  \citenamefont {Petryk}, \citenamefont {Pretorius},\ and\ \citenamefont
  {Villegas}}]{Choptuik:2003qd}%
  \BibitemOpen
  \bibfield  {author} {\bibinfo {author} {\bibfnamefont {M.~W.}\ \bibnamefont
  {Choptuik}}, \bibinfo {author} {\bibfnamefont {L.}~\bibnamefont {Lehner}},
  \bibinfo {author} {\bibfnamefont {I.}~\bibnamefont {Olabarrieta}}, \bibinfo
  {author} {\bibfnamefont {R.}~\bibnamefont {Petryk}}, \bibinfo {author}
  {\bibfnamefont {F.}~\bibnamefont {Pretorius}}, \ and\ \bibinfo {author}
  {\bibfnamefont {H.}~\bibnamefont {Villegas}},\ }\href {\doibase
  10.1103/PhysRevD.68.044001} {\bibfield  {journal} {\bibinfo  {journal} {Phys.
  Rev. D}\ }\textbf {\bibinfo {volume} {68}},\ \bibinfo {pages} {044001}
  (\bibinfo {year} {2003})},\ \Eprint {http://arxiv.org/abs/gr-qc/0304085}
  {arXiv:gr-qc/0304085} \BibitemShut {NoStop}%
\bibitem [{\citenamefont {Lehner}\ and\ \citenamefont
  {Pretorius}(2010)}]{Lehner:2010pn}%
  \BibitemOpen
  \bibfield  {author} {\bibinfo {author} {\bibfnamefont {L.}~\bibnamefont
  {Lehner}}\ and\ \bibinfo {author} {\bibfnamefont {F.}~\bibnamefont
  {Pretorius}},\ }\href {\doibase 10.1103/PhysRevLett.105.101102} {\bibfield
  {journal} {\bibinfo  {journal} {Phys. Rev. Lett.}\ }\textbf {\bibinfo
  {volume} {105}},\ \bibinfo {pages} {101102} (\bibinfo {year} {2010})},\
  \Eprint {http://arxiv.org/abs/1006.5960} {arXiv:1006.5960 [hep-th]}
  \BibitemShut {NoStop}%
\bibitem [{\citenamefont {Figueras}\ \emph {et~al.}(2023)\citenamefont
  {Figueras}, \citenamefont {Fran\c{c}a}, \citenamefont {Gu},\ and\
  \citenamefont {Andrade}}]{Figueras:2022zkg}%
  \BibitemOpen
  \bibfield  {author} {\bibinfo {author} {\bibfnamefont {P.}~\bibnamefont
  {Figueras}}, \bibinfo {author} {\bibfnamefont {T.}~\bibnamefont
  {Fran\c{c}a}}, \bibinfo {author} {\bibfnamefont {C.}~\bibnamefont {Gu}}, \
  and\ \bibinfo {author} {\bibfnamefont {T.}~\bibnamefont {Andrade}},\ }\href
  {\doibase 10.1103/PhysRevD.107.044028} {\bibfield  {journal} {\bibinfo
  {journal} {Phys. Rev. D}\ }\textbf {\bibinfo {volume} {107}},\ \bibinfo
  {pages} {044028} (\bibinfo {year} {2023})},\ \Eprint
  {http://arxiv.org/abs/2210.13501} {arXiv:2210.13501 [hep-th]} \BibitemShut
  {NoStop}%
\bibitem [{\citenamefont {Hindawi}\ \emph
  {et~al.}(1996{\natexlab{b}})\citenamefont {Hindawi}, \citenamefont {Ovrut},\
  and\ \citenamefont {Waldram}}]{Hindawi:1995cu}%
  \BibitemOpen
  \bibfield  {author} {\bibinfo {author} {\bibfnamefont {A.}~\bibnamefont
  {Hindawi}}, \bibinfo {author} {\bibfnamefont {B.~A.}\ \bibnamefont {Ovrut}},
  \ and\ \bibinfo {author} {\bibfnamefont {D.}~\bibnamefont {Waldram}},\ }\href
  {\doibase 10.1103/PhysRevD.53.5597} {\bibfield  {journal} {\bibinfo
  {journal} {Phys. Rev. D}\ }\textbf {\bibinfo {volume} {53}},\ \bibinfo
  {pages} {5597} (\bibinfo {year} {1996}{\natexlab{b}})},\ \Eprint
  {http://arxiv.org/abs/hep-th/9509147} {arXiv:hep-th/9509147} \BibitemShut
  {NoStop}%
\bibitem [{\citenamefont {Deffayet}\ \emph {et~al.}(2022)\citenamefont
  {Deffayet}, \citenamefont {Mukohyama},\ and\ \citenamefont
  {Vikman}}]{Deffayet:2021nnt}%
  \BibitemOpen
  \bibfield  {author} {\bibinfo {author} {\bibfnamefont {C.}~\bibnamefont
  {Deffayet}}, \bibinfo {author} {\bibfnamefont {S.}~\bibnamefont {Mukohyama}},
  \ and\ \bibinfo {author} {\bibfnamefont {A.}~\bibnamefont {Vikman}},\ }\href
  {\doibase 10.1103/PhysRevLett.128.041301} {\bibfield  {journal} {\bibinfo
  {journal} {Phys. Rev. Lett.}\ }\textbf {\bibinfo {volume} {128}},\ \bibinfo
  {pages} {041301} (\bibinfo {year} {2022})},\ \Eprint
  {http://arxiv.org/abs/2108.06294} {arXiv:2108.06294 [gr-qc]} \BibitemShut
  {NoStop}%
\bibitem [{\citenamefont {Deffayet}\ \emph {et~al.}(2023)\citenamefont
  {Deffayet}, \citenamefont {Held}, \citenamefont {Mukohyama},\ and\
  \citenamefont {Vikman}}]{Deffayet:2023wdg}%
  \BibitemOpen
  \bibfield  {author} {\bibinfo {author} {\bibfnamefont {C.}~\bibnamefont
  {Deffayet}}, \bibinfo {author} {\bibfnamefont {A.}~\bibnamefont {Held}},
  \bibinfo {author} {\bibfnamefont {S.}~\bibnamefont {Mukohyama}}, \ and\
  \bibinfo {author} {\bibfnamefont {A.}~\bibnamefont {Vikman}},\ }\href@noop {}
  {\  (\bibinfo {year} {2023})},\ \Eprint {http://arxiv.org/abs/2305.09631}
  {arXiv:2305.09631 [gr-qc]} \BibitemShut {NoStop}%
\bibitem [{\citenamefont {Bueno}\ \emph {et~al.}(2017)\citenamefont {Bueno},
  \citenamefont {Cano}, \citenamefont {Min},\ and\ \citenamefont
  {Visser}}]{Bueno:2016ypa}%
  \BibitemOpen
  \bibfield  {author} {\bibinfo {author} {\bibfnamefont {P.}~\bibnamefont
  {Bueno}}, \bibinfo {author} {\bibfnamefont {P.~A.}\ \bibnamefont {Cano}},
  \bibinfo {author} {\bibfnamefont {V.~S.}\ \bibnamefont {Min}}, \ and\
  \bibinfo {author} {\bibfnamefont {M.~R.}\ \bibnamefont {Visser}},\ }\href
  {\doibase 10.1103/PhysRevD.95.044010} {\bibfield  {journal} {\bibinfo
  {journal} {Phys. Rev. D}\ }\textbf {\bibinfo {volume} {95}},\ \bibinfo
  {pages} {044010} (\bibinfo {year} {2017})},\ \Eprint
  {http://arxiv.org/abs/1610.08519} {arXiv:1610.08519 [hep-th]} \BibitemShut
  {NoStop}%
\bibitem [{\citenamefont {Giannakopoulos}\ \emph {et~al.}(2020)\citenamefont
  {Giannakopoulos}, \citenamefont {Hilditch},\ and\ \citenamefont
  {Zilh\~ao}}]{Giannakopoulos2020}%
  \BibitemOpen
  \bibfield  {author} {\bibinfo {author} {\bibfnamefont {T.}~\bibnamefont
  {Giannakopoulos}}, \bibinfo {author} {\bibfnamefont {D.}~\bibnamefont
  {Hilditch}}, \ and\ \bibinfo {author} {\bibfnamefont {M.}~\bibnamefont
  {Zilh\~ao}},\ }\href {\doibase 10.1103/PhysRevD.102.064035} {\bibfield
  {journal} {\bibinfo  {journal} {Phys. Rev. D}\ }\textbf {\bibinfo {volume}
  {102}},\ \bibinfo {pages} {064035} (\bibinfo {year} {2020})}\BibitemShut
  {NoStop}%
\end{thebibliography}%

\end{document}